\documentclass[twocolumn, tighten, times]{aastex63}

\usepackage{amsmath}
\usepackage{lipsum}

\shorttitle{Cosmogenic component in blazar SEDs}
\shortauthors{Das et al.}

\begin{document}
\vspace*{-1.5cm}

\title{Ultrahigh-energy cosmic ray interactions as the origin of very high energy $\gamma-$rays from BL Lacs}

\correspondingauthor{Saikat Das}
\email{saikatdas@rri.res.in}

\author[0000-0001-5796-225X]{Saikat Das}
\affiliation{Astronomy \& Astrophysics Group, Raman Research Institute, Bengaluru 560080, Karnataka, India}

\author[0000-0002-1188-7503]{Nayantara Gupta}
\affiliation{Astronomy \& Astrophysics Group, Raman Research Institute, Bengaluru 560080, Karnataka, India}
\email{nayan@rri.res.in}

\author[0000-0002-0130-2460]{Soebur Razzaque}
\affiliation{Centre for Astro-Particle Physics (CAPP) and Department of Physics, University of Johannesburg, PO Box 524, Auckland Park 2006, South Africa}
\email{srazzaque@uj.ac.za}




\begin{abstract}
We explain the observed multiwavelength photon spectrum of a number of BL Lac objects detected at very high energy (VHE, $E \gtrsim 30$ GeV), using a lepto-hadronic emission model. The one-zone leptonic emission is employed to fit the synchrotron peak. Subsequently, the SSC spectrum is calculated, such that it extends up to the highest energy possible for the jet parameters considered. The data points beyond this energy, and also in the entire VHE range are well explained using a hadronic emission model. The ultrahigh-energy cosmic rays (UHECRs, $E\gtrsim 0.1$ EeV) escaping from the source interact with the extragalactic background light (EBL) during propagation over cosmological distances to initiate electromagnetic cascade down to $\sim1$ GeV energies. The resulting photon spectrum peaks at $\sim1$ TeV energies. We consider a random turbulent extragalactic magnetic field (EGMF) with a Kolmogorov power spectrum to find the survival rate of UHECRs within 0.1 degrees of the direction of propagation in which the observer is situated. We restrict ourselves to an RMS value of EGMF, $B_{\rm rms}\sim 10^{-5}$ nG, for a significant contribution to the photon spectral energy distribution (SED) from UHECR interactions. We found that UHECR interactions on the EBL and secondary cascade emission can fit gamma-ray data from the BL Lacs we considered at the highest energies. The required luminosity in UHECRs and corresponding jet power are below the Eddington luminosities of the super-massive black holes in these BL Lacs.

\end{abstract}


\section{Introduction} \label{sec:introduction}

The far-reaching progress made in $\gamma-$ray observations in the last decade and the considerable amount of data collected by Fermi-LAT \citep{Fermi_3FGL, Fermi_3LAC}, MAGIC \citep{MAGIC_19}, HESS \citep{HESS_06}, VERITAS \citep{VERITAS_18}, HAWC \citep{HAWC_17a, HAWC_17b} experiments, etc., have evolved the understanding of our universe in the high-energy regime. The precision measurement of photon fluxes from a wide variety of astrophysical objects provides an impetus to study the $\gamma-$ray sky as a probe to various hadronic interaction processes. It also puts to test the speculations on the origin of ultrahigh energy cosmic rays (UHECRs), i.e., particles with energies beyond $10^{18}$ eV and extending up to a few times $10^{20}$ eV. 

Active galactic nuclei (AGN) are a class of astrophysical objects which are considered to be the potential acceleration sites of UHECRs. The collimated beam of outflow from AGN contains a tangled magnetic field and a relativistic population of electrons and protons, assumed to be accelerated by the same physical process. These jets transport energy and momentum over large distances by hosting a variety of interactions and radiative emission processes, thus injecting high energy particles into the universe. A particular subclass of radio-loud AGN called blazars have their relativistic jets oriented closely along the line of sight of the observer. The non-thermal continuum emission from blazars span over a wide range of frequencies, from radio to very-high energy (VHE) $\gamma-$rays ($E>30$ GeV), and exhibit variability at diverse timescales which may or may not be coherent in all wavebands. The spectral energy distribution (SED) of blazars show two distinct components \textemdash a low energy peak between optical to X-ray energies and a high energy peak at $\gamma-$ray regime. While the former is attributed to synchrotron emission of relativistic electrons accelerated in the jet, several propositions exist for the latter. The most prevalent explanation in the light of leptonic model invokes the inverse Compton (IC) scattering of synchrotron photons (SSC) or external photons (accretion disk, dusty torus, or broad-line region) by the same population of relativistic electrons. This allows the detection of VHE $\gamma-$ rays due to relativistic beaming of the emitted photons from the jet. Alternately, the hadronic models predict that high energy protons inside the jet can produce electron-positron pairs ($\mathrm{e^+e^-}$), charged/neutral pions ($\pi^0$, $\pi^+$) by photohadronic interactions. The $\mathrm{e^+e^-}$ pairs can undergo synchrotron emission losses and the neutral pions decay to produce $\gamma$ photons. The charged pions and muons produced in these photohadronic interactions can also radiate by synchrotron process before decaying into electrons. The mass of pions and muons being two orders of magnitude higher than electrons, the efficiency of such loss is lower than electron synchrotron radiation in general. Also possible is the proton-synchrotron emission peaking at multi-TeV energies \citep{Aharonian_02}. However, the latter requires very high jet power, that may surpass the Eddington luminosity ($L_{\rm Edd}$) in some cases \citep{Zdziarski_15}. Interpreting the multiwavelength spectrum using lepto-hadronic models require a minimum jet power, which is crucial to understand the interactions and physical properties inside the jet \citep{Bottcher_13}. 

High-frequency peaked BL Lacs (HBLs) have their peak synchrotron emission in the UV to X-ray energy range and their spectrum extends up to multi-TeV energies. They comprise the majority of extragalactic VHE $\gamma-$ray sources. At these extreme energies, the high energy photons if produced inside the jet can be absorbed by intrinsic $\gamma\gamma$ pair production with target photons originating from broad-line region (BLR), dusty torus (DT), or accretion disk. They will also suffer attenuation losses due to $\gamma\gamma_{bg}$ collision with the EBL photons while propagating over cosmological distances. Hence the intrinsic spectrum must be harder than the observed spectrum. For some HBLs, the intrinsic spectrum at TeV energies is too hard to be explained by one-zone leptonic model due to suppression of IC emission through Klein-Nishina (KN) effect and intrinsic/extrinsic pair production losses. The TeV spectra can be explained by another scenario, in which the UHECRs from the sources interact with the cosmic background photons to produce electromagnetic (EM) particles, viz., electrons, positrons, photons \citep[see][]{Essey_10a}. These secondary particles can initiate electromagnetic cascade and produce a resultant $\gamma-$ray spectra at VHE range \citep{Essey_10b, Essey_11a, Razzaque_12, Murase_12, Kalashev_13, Takami_13, Supanitsky_13, Tavecchio_14}. Such a scenario is applicable for HBLs at redshifts greater than the mean interaction length of UHECR protons for photohadronic interactions, and lesser than the distance where a majority of the photons produced are either scattered off from the line of sight or absorbed completely by the EBL photons. These cascade photons can contribute significantly to the spectrum at the highest observed energies. Resonant photopion production on CMB occurs for protons with energies $E\geqslant50$ EeV. At lower energies interactions with the EBL photon dominates. 

In this paper, we consider a representative sample of high-energy BL Lacs, for which the TeV emission is prominent, non-variable, and the source parameters are suitable for UHECR acceleration. More than 3130 of the identified or associated sources in the 50 MeV $-$ 1 TeV range, listed in the Fermi-LAT 8 years source catalog (4FGL), are blazars \citep{Fermi_4FGL}. The 4LAC catalog, which is the companion of 4FGL focused on AGN, splits the blazar candidates based on their spectral properties into 650 flat-spectrum radio quasars (FSRQs) and 1052 BL Lacs with 1092 blazars of unknown type \citep{Fermi_4LAC}. While majority of FSRQs are low-synchrotron peaked, the BL Lacs are fairly evenly distributed between low-synchrotron peaked, intermediate-synchrotron peaked and high-synchrotron peaked subclasses. The redshift is unknown for the majority of BL Lacs. However, the redshifts of the sources considered in this work are well studied. We model the multiwavelength SED in the quiescent state by a one-zone leptonic synchrotron/SSC model that extends up to the highest energies allowed by the leptonic model. Next, we fit the data points in the VHE range by secondary photons originating from UHECR interactions while propagating in the CMB and EBL close to the line of sight. The requirement of this additional component is compelling for the SED of HBLs considered in this analysis. During the propagation of UHECR protons, they are subjected to deflections by the EGMF. The EGMF is not known to high precision. We set constraints on the RMS value of the EGMF, such that a significant fraction of UHECR protons are constricted near to the line of sight, respecting the luminosity budget. We calculate the required kinetic power in electrons and protons to compare it with the Eddington luminosity of the AGN. 

The hindrance due to EGMF, in addition to the EBL attenuation of photon fluxes originating from the jet increases the required total jet power to explain the SED. The plausibility of hadronic interactions inside the jet has been studied earlier in great details \citep{Sahu_12, Bottcher_13, Petropoulou_15a, Xue_19a, Sahu_19}. However, the efficiency of photon production in such a scenario can be correlated directly with the internal pair production, owing to the same seed photon distribution in both processes \citep{Bottcher_16}. The interaction efficiency of photohadronic interactions in the high energy limit is $10^{-3}$ times smaller than the peak $\gamma\gamma$ opacity, leading to a substantial reduction in the escaping photon flux \citep{Murase_18}. In our model, we calculate the opacity of $p\gamma$ interactions inside the jet to find that the contribution from such interactions is insignificant and escape by diffusion dominates for protons upto ultrahigh energies.

Also, if neutrinos are produced in the same UHECR interactions as photons, they must contribute to the observed neutrino flux at the highest energies. The upper limits to the neutrino flux comes from measurements at IceCube \citep{IceCube_18} and Pierre Auger \citep{PAO_15} experiments. We find an estimate for the neutrino flux from these high energy TeV blazars and show that a detection is difficult in near future owing to the extremely low flux value compared to the current and upcoming detector sensitivities. We discuss the theoretical framework and the model used for the relativistic jet and UHECR propagation in Sec.~\ref{sec:framework}. We show our results from modeling the multiwavelength SEDs of HBLs, and the resultant UHECR and neutrino event rates in Sec.~\ref{sec:result}. We discuss the results obtained from our analysis in Sec.~\ref{sec:discussions} and draw our conclusions in Sec.~\ref{sec:conclusions}. 


\section{Theoretical Framework} \label{sec:framework}

\subsection{Leptonic Modeling} \label{subsec:leptons}

We consider the emission region inside the jet consists of a relativistic plasma of electrons and protons moving through a uniform magnetic field $B$ in a spherical blob of radius $R$. We use a one-zone leptonic model to fit the observed broadband SED and consider the constant spectrum of electrons injected into the system in the comoving frame of the jet as,
\begin{equation}
Q_e(E_e) = A_e \bigg(\dfrac{E_e}{E_0}\bigg)^{-\alpha}\exp\bigg(-\dfrac{E_e}{E_{\rm e,cut}}\bigg) \label{eqn:electron_inj}
\end{equation}
where $E_0=0.5$ GeV is a reference energy. The cutoff energy ($E_{\rm e,cut}$), the injection spectral index ($\alpha$), and the minimum electron energy ($E_{e, \rm min}$) are found from modeling the synchrotron spectrum. The resulting IC spectrum is then adjusted to find the highest energy up to which it can extend. Increasing $E_{\rm e,cut}$ beyond this value may worsen the synchrotron fit or turns out to be ineffective because of KN effect. Electrons and positrons are cooled radiatively by energy loss via synchrotron emission and synchrotron self Compton (SSC) emission processes. We use the open-source code GAMERA{\protect\footnote{\url{http://libgamera.github.io/GAMERA/docs/main_page.html}}} to model the SED up to the highest energies possible using the leptonic model. It solves the 1D transport equation,
\begin{equation}
\dfrac{\partial N_e}{\partial t} = Q_e(E_e,t) - \dfrac{\partial}{\partial E}(bN_e) - \dfrac{N_e}{t^e_{\rm esc}}
\end{equation}
to calculate the spectrum of particles $N_e(E_e,t)$ at a time $t$, where $b=b(E_e,t)$ is the energy loss rate of leptons and $t^e_{\rm esc}$ is the timescale on which the leptons escape from the system.

We assume the electrons escape over dynamical timescale, $t_{\rm esc}^e= t_{\rm dyn} = R/c$, which is constant in energy and time. Since the cooling rate for electrons is significantly higher than protons due to radiative losses inside the blob, the escape timescale will be higher and maybe assumed in a simplistic approach to be equivalent to the dynamical timescale $R/c$. Thus invoking diffusion loss is not important. For a continuous lepton injection $Q_e(E_e)$, we calculate the resultant spectrum of photons from $N_e(E_e,t)$ at a time much greater than that required to attain the steady state. This ensures that the obtained luminosities are that of the quiescent state and the variabilities, if any, are averaged out.

The photons from the jet are Doppler boosted due to relativistic beaming by a factor of $\delta_D = [\Gamma(1-\beta\cos\theta)]^{-1}$. $\Gamma$ is the bulk Lorentz factor, and $\beta c$ is the velocity of the emitting plasma. $\theta\lesssim 1/\Gamma$ is the viewing angle of the blazar with respect to the line of sight. The synchrotron and SSC luminosities are doppler boosted by $\delta_D^4$ in the observer frame \citep{Celotti_08}. The luminosity in electrons required in the AGN frame is given as,
\begin{equation}
L_{\rm e} = \pi R^2 \Gamma^2 c u'_e 
\end{equation}
where $u'_e = (1/V) \times \int_{E_{e, \rm min}}^{E_{e, \rm max}} dE_e \ Q(E_e) E_e$ is the energy density of electrons in the comoving frame of the jet and V is the volume of the blob. The luminosity in magnetic field, i.e., the power carried as Poynting flux is given by,
\begin{equation}
L_{\rm B} = \dfrac{1}{8} R^2 \Gamma^2 \beta c B^2 
\end{equation}

It may be possible to explain the entire SED by employing multiple emission zones, and/or other processes involving the geometry of the HBLs. Nonetheless, we restrict ourselves to a single-zone leptonic model only, to accommodate the hadronic component originating from UHECR interactions.

\subsection{Hadronic Modeling} \label{subsec:hadrons}
Protons are accelerated inside the jet up to maximum energies which can be estimated from the Hillas condition \citep{Hillas_84},
\begin{equation}
E_{\rm p,max} \sim 2\beta c Ze B R \label{eqn:Hillas}
\end{equation} 

Protons, being heavier than the electrons, are not cooled sufficiently inside the jet and one can simply write $N_p(E_p) = t_{\rm dyn}Q_p(E_p)$. We find that escape dominates over energy losses inside the jet up to ultrahigh energies. We assume protons are injected into the ISM following a power-law injection of the form,
\begin{equation}
N_p(E_p)=\dfrac{dN}{dE_p} = A_p E_{p}^{-\alpha} \label{eqn:proton_injection}
\end{equation} 
where we take $E_{\rm p,min}=0.1$ EeV and $E_{\rm p,max}=10$ EeV. The choice of $E_{\rm p,max}$ is explained later (cf. Sec.~\ref{sec:result}). We assume the same spectral index for the injection of leptons and hadrons, implying the same inherent acceleration process for both. 

For the sake of simplicity, we consider protons (Z=1) as the only UHECRs for this study, and a steep cutoff instead of an exponential one. The signatures of an exponential term is greatly obscured in the resulting photon spectrum, by virtue of electromagnetic cascade which is primarily driven by CMB and/or EBL photon density, and EGMF. The resonant photopion production of UHECRs with CMB occurs at $E_{\rm p}=50$ EeV. In our study, owing to a lower value of $E_{\rm p,max}$, the dominant contribution to pion decay photons comes from UHECR interactions on EBL. 

The timescales of photohadronic interactions on synchrotron and SSC photons inside the jet are given by,
\begin{equation}
\dfrac{1}{t_{\rm p\gamma}} = \dfrac{c}{2\gamma_p^2}\int_{\epsilon_{th}/2\gamma_p}^\infty d\epsilon_\gamma'\dfrac{n(\epsilon'_\gamma)}{\epsilon_\gamma'^2}\int_{\epsilon_{th}}^{2\epsilon\gamma_p}d\epsilon_r \sigma(\epsilon_r) K(\epsilon_r) \label{eqn:efficiency_ph}
\end{equation}
where $\sigma(\epsilon_r)$ and $K(\epsilon_r)$ are the cross-section and inelasticity respectively of photopion production or Bethe-Heitler pair production as a function of photon energy $\epsilon_r$ in the proton rest frame \citep{Berezinskii_88, Chodorowski_92, Mucke_00, Berezinsky_06}. n($\epsilon_\gamma'$) is the seed photon density per unit energy interval as a function of photon energy $\epsilon_\gamma'$ in the comoving jet frame. It is related to the observed photon spectrum by the relation \citep{Ghisellini_93, Joshi_13},
\begin{equation}
4\pi R^2 c \ \epsilon_\gamma'^2 n(\epsilon_\gamma') = 4 \pi d_L^2 \delta_D^{-p} \bigg(\epsilon_\gamma^{2}\dfrac{dN_\gamma}{d\epsilon_\gamma dt dA}\bigg)_{\rm obs}
\label{eqn:frame_change}
\end{equation}
where $p=n+2$ and $n=$ 2, 3 for continuous jet and a moving sphere respectively. The subscript `obs' in Eqn.~(\ref{eqn:frame_change}) represents the quantity in the observer frame. The photon energy transforms as $\epsilon_\gamma = \delta_D\epsilon_\gamma'/(1+z)$. For a source at redshift $z$ from the observer, $d_L=(1+z)d_c$ is the luminosity distance and $d_c$ is the comoving distance. The escape timescale of protons from the source is given as,
\begin{equation}
t^p_{\rm esc} = \dfrac{R^2}{4D} \label{eqn:escape_timescale}
\end{equation}  
where we write the diffusion coefficient as, $D=D_0 (E/E_0)^{2-q}$, where $q$ is the turbulence spectral index and is taken to be $q=3/2$ for the Kraichnan model. The acceleration timescale is calculated from,
\begin{equation}
t^p_{\rm acc} \simeq \dfrac{20\eta}{3} \dfrac{r_L}{c} \simeq \dfrac{20\eta}{3} \dfrac{\gamma_p m_pc}{eB}
\label{eqn:acceleration}
\end{equation}
$\eta$ is the ratio of the mean magnetic field energy density to the turbulent magnetic field energy density \citep{Lagage_83}. We consider $\eta=1$ in our calculations. 

The escaping protons propagate through extragalactic distances interacting on CMB and EBL, producing electrons, positrons, $\gamma-$rays, and neutrinos through $\Delta-$resonance or Bethe-Heitler pair production given as,
\begin{align}
p+\gamma_{bg} &\rightarrow p + \mathrm{e^+e^-} \\
p+\gamma_{bg} &\rightarrow \Delta^+ \rightarrow
\begin{cases}
n+ \pi^+\\
p+ \pi^0
\end{cases}
\end{align}
The neutral pions decay to produce $\gamma$ photons ($\pi^0\rightarrow\gamma\gamma$) and the charged pions decay to produce neutrinos ($\pi^+\rightarrow\mu^+ + \nu_\mu \rightarrow \mathrm{e^+} + \nu_e + \overline{\nu}_\mu + \nu_\mu$). The secondary photons and electrons can initiate electromagnetic cascades down to GeV energies. The secondary $\gamma$ photons interact with cosmic background radiations and universal radio background (URB) leading to Breit-Wheeler pair production or double pair production.  The relativistic cascade electrons and positrons also lose energy by synchrotron radiation on deflections in EGMF, triplet pair production, and IC scattering of background photons. The neutrinos, on the other hand, once produced propagate rectilinearly unhindered by EGMF or interactions. The proton luminosity required in the AGN frame is calculated as,
\begin{equation}
L_{\rm p} = \pi R^2 \Gamma^2 c u'_p
\end{equation}

We use the public astrophysical simulation framework \textsc{CRPropa 3} to propagate the UHECR protons from the source to observer \citep{Batista_16}. \textsc{CRPropa 3} provides two external codes, DINT \citep{Lee_98} and EleCa \citep{Settimo_15}, to calculate the development of EM cascades. DINT solves the transport equations to produce the observed spectrum and is thus computationally more efficient. EleCa, on the other hand, does a full Monte Carlo tracking of individual particles. For 1D simulations, DINT can also be combined with EleCa or \textsc{CRPropa}
\citep{Heiter_18}. We propagate the EM particles produced from UHECR interactions using DINT only. We consider the Gilmore et al. EBL model \citep{Gilmore_12} and a random turbulent EGMF with a Kolmogorov power spectrum of RMS field strength $B_{\rm rms}=10^{-5}$ nG. 


The total luminosity of the HBL is obtained as $L_{\rm jet} = L_{\rm e} + L_{\rm p} + L_{\rm B}$. We compare the obtained value of $L_{\rm jet}$ with the Eddington luminosity $L_{\rm Edd}$ of the supermassive black hole. Thus, we can check whether a scenario that invokes UHECR interactions can account for the origin of $\gamma-$rays observed at the highest energies from the HBLs.

\section{Results\protect} \label{sec:result}


\begin{figure}
\centering
\includegraphics[width = 0.49\textwidth]{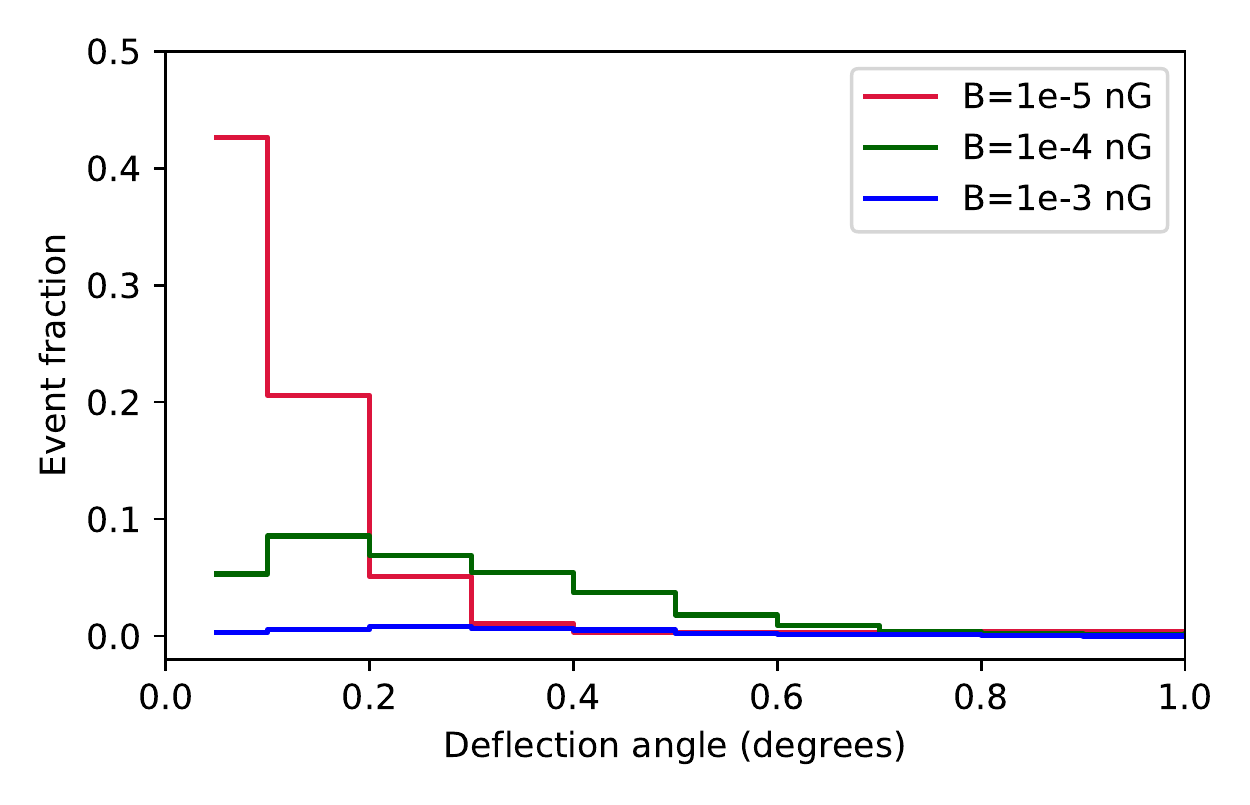}
\caption{\small{Distribution of propagated UHECRs as a function of deflection angle in a random turbulent magnetic field.}}
\label{fig:deflection}
\end{figure}

We explore the effect of our choice of the extragalactic magnetic field (EGMF). The root mean square deflection in the trajectories of cosmic rays on propagation over a distance $D$ in a random turbulent magnetic field with coherence length $l_{c}$ can be approximated as \citep{Dermer_09},
\begin{equation}
\Phi_{\rm rms} \approx 4^\circ\dfrac{60 \text{ EeV}}{E/Z} \dfrac{B_{\rm rms}}{10^{-9} \text{ G}}\sqrt{\dfrac{D}{100\text{ Mpc}}}\sqrt{\dfrac{l_c}{1 \text{ Mpc}}} \label{eqn:phi_rms}
\end{equation}

The constraints on turbulent EGMF models derived from the correlation between observed UHECR arrival directions and potential AGN sources yield the condition $B_{\rm rms}\sqrt{l_c}\leq 10^{-9} \text{G}\sqrt{\text{Mpc}}$ \citep{PAO_08}. One finds from Eqn.~(\ref{eqn:phi_rms}), the deflection of a proton of 1 EeV traversing a distance of 1 Gpc from the source to the observer through an EGMF with $B_{\rm rms}=10^{-5}$ nG to be $\approx 0.0076^\circ$. In realistic scenarios, the emitted UHECRs are not monoenergetic, but follows a distribution according to Eqn.~(\ref{eqn:proton_injection}). We run 3D test simulations in \textsc{CRPropa 3} for the propagation of UHECRs from a source at a distance of 1000 Mpc away from the observer to check the effect of varying $B_{\rm rms}$. We consider a magnetic field with Fourier modes taken from a Gaussian distribution with $\langle|B(\mathbf{k})|^2\rangle$ given by a Kolmogorov power spectrum and random polarization. In all cases, we set the value of the turbulent correlation length $l_c=1$ Mpc and thus take the observer to be a sphere of radius $R_{\rm obs} = 1$ Mpc. The proton injection spectrum is assumed to be a power law, $dN/dE_p \propto E^{-\alpha}$ with $\alpha=2$ and energies in the range $0.1-10$ EeV. All energy loss processes, viz., photopion production, Bethe-Heitler interaction, nuclear decay and adiabatic expansion of the universe are accounted for in the simulation. The emission from the source is such that almost all of the emitted protons are initially directed towards the observer. The deflections of the UHECRs on arrival at the surface of the observer sphere is calculated from initial and final momentum vector directions, and the distribution of observed events as a function of deflection angle is shown in Fig.~\ref{fig:deflection}. Here the events are binned over deflection angle with respect to the direction along which the UHECRs are emitted, in bin-widths of $0.1^\circ$. The event fraction within $0-0.1^\circ$ indicates the survival rate of UHECRs close to the initial direction of propagation.

\begin{figure}
\centering
\includegraphics[width = 0.49\textwidth, trim={2.0cm 1.0cm 2.5cm 3.0cm},clip]{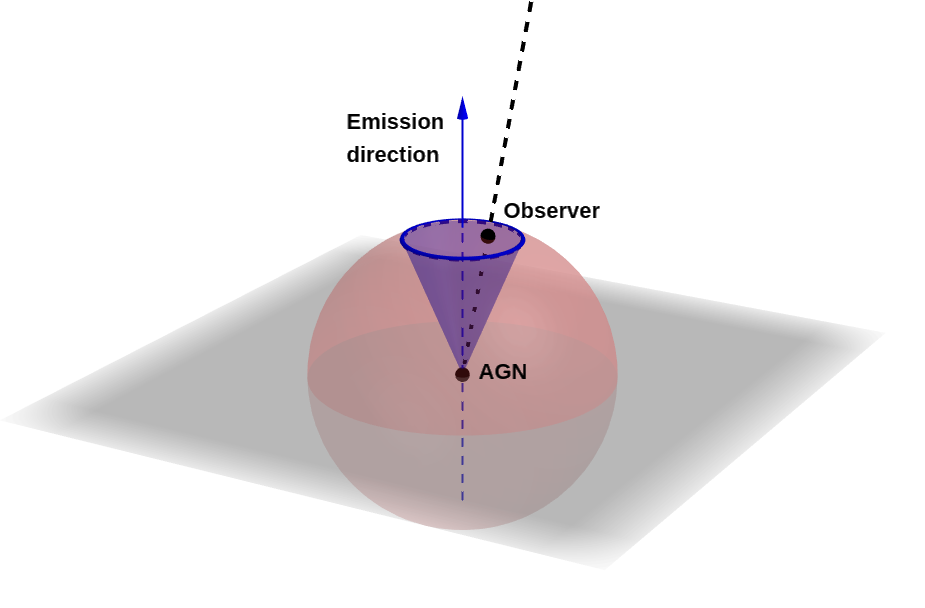}
\caption{\small{The emission from the blazar centered on the sphere is assumed to be constricted within the blue cone. The black dashed line represents the line of sight towards the center of the sphere for an observer on the surface.}}
\label{fig:geometry}
\end{figure}

The survival rate of UHECRs increases with decreasing $B_{\rm rms}$, as well as, with higher containment angles. Since we are interested in the calculation of contribution from UHECR interactions to the observed $\gamma-$ray SEDs of HBLs, we find the survival rate of UHECRs within $0.1^\circ$ of the direction of propagation in which the observer is positioned and denote it by $\xi_B$. Accordingly, from the results of the 3D simulations, we consider $B_{\rm rms}=10^{-5}$ nG for the rest of our study, to increase the fraction of events close to the direction of jet emission axis of HBLs. In some studies, a lower bound of $\sim 10^{-6}$ nG is found for the EGMF \citep{Neronov_10}; while others have estimated it to be, as low as, $10^{-7}$ nG \citep{Razzaque_12} or $10^{-8}$ nG \citep{Essey_11b}. An increase in the survival rate along the direction of emission for blazars will essentially increase the detection rate along the line of sight of the observer. The angular resolution of Fermi-LAT for observation of a single photon at $E>10$ GeV is $\sim0.15^\circ$ \citep{Fermi_4FGL}.

\begin{table}
\caption{\label{tab:hadronic} UHECR model parameters}
 \begin{ruledtabular}
 \begin{tabular}{lcrccc}
 HBL & $z$ & $d_{L}$ & $\xi_B$ & $f_{\rm CR}$ & $f_\nu$ \\
 \hline
 1ES 1011+496 & 0.212 & 1085 Mpc & 0.45 & 0.084 & 0.00064 \\
 1ES 0229+200 & 0.140 & 687 Mpc  & 0.46 & 0.052 & 0.00039 \\
 1ES 1101--232 & 0.186 & 938 Mpc & 0.48 & 0.079 & 0.00058 \\
 1ES 0414+009 & 0.287 & 1529 Mpc & 0.39 & 0.085 & 0.00077 \\
 \end{tabular}
 \end{ruledtabular}
\end{table}

Fig.~\ref{fig:geometry} shows the jet emission axis by a vertical solid line directed upwards. The sphere has a radius $d_L$ equivalent to the luminosity distance from the source to the observer. The source is at the center of the sphere. The cone is the extrapolation of the jet emission from the HBL, with a semi-apex angle equal to the jet opening angle $\theta_{\rm jet}$. The observer is at the position where the dashed line pierces through the surface of the sphere. The line of sight is thus directed towards the center of the sphere along the line passing through the observer, and may also lie outside the emission cone. When the line of sight lies outside the jet opening angle, the collimation of outflow along the observer's direction becomes poor and thus the required jet power increases. However, we do not include any angular dependence for varying observer position in our analysis and assume the jet emission stays collimated along the line of sight. The observer sphere is not shown in this diagram since $R_{\rm obs}\ll d_L$. Assuming the angle between the emission direction and line of sight to be a few degrees, the viewing angle $\theta\lesssim 1/\Gamma$. Now, the flux of secondary cascade photons will be distributed across the area of the spherical cap subtended on the surface of the sphere by the emission cone.

DINT solves the transport equation in 1D for the propagation and electromagnetic cascade of secondary $\mathrm{e^+}$, $\mathrm{e^-}$ and $\gamma$ photons produced from UHECR interactions on CMB and EBL. Hence, we run 1D simulations in CRPropa for the propagation of UHECRs, producing EM particles, to calculate the secondary photon flux at the Earth. To find the flux intercepted by the observer's line of sight in the presence of an EGMF, we multiply this photon flux by $\xi_B$. Although $\xi_B$ is calculated for an observer sphere of radius $R_{\rm obs}=1$ Mpc, this introduces little error as $l_c=1$ Mpc and also, the mean free path for energy loss of UHECRs in the energy range considered is much greater than this value \citep{Dermer_09, Heiter_18}. In this process, we already reject the contribution to secondary photons from all UHECRs outside of $0.1^\circ$ w.r.t the propagation direction along which the observer is located.

Thus, the luminosity required in UHECR protons, $L_{\rm UHECR}$ can be calculated from the expression,
\begin{equation}
L_{\rm UHECR} = \dfrac{2\pi d_L^2(1-\cos\theta_{\rm jet})}{\xi_B f_{\rm CR}} \int_{\epsilon_{\gamma,\rm min}}^{\epsilon_{\gamma,\rm max}} \epsilon_\gamma\dfrac{dN}{d\epsilon_\gamma dAdt}d\epsilon_\gamma \label{eqn:luminosity}
\end{equation}
where $2\pi d_L^2(1-\cos\theta_{\rm jet})$ is the area of the spherical cap region. We consider typical values of $\theta_{\rm jet}\sim 0.1$ radians \citep{Pushkarev_09, Finke_19}. $f_{\rm CR}$ is the ratio of the power in produced secondary photons $L_{\gamma,\rm p}$ due to propagation of UHECRs and cascade of resulting EM particles, to the injected UHECR power $L_{\rm UHECR}$. Both $\xi_B$ and $f_{CR}$ will be a function of propagation distance. The values of these quantities obtained for the sources in our study, while modeling the VHE $\gamma-$ray component, are listed in Table~\ref{tab:hadronic}. The integral is over the flux of secondary photons of hadronic origin, required to fit the observed blazar SED. We obtain the secondary EM particles and cascade photons from 1D simulations, and account for the conical distribution by calculating the luminosity according to Eqn.~(\ref{eqn:luminosity}).

\begin{figure*}
\centering
\includegraphics[width = 0.49\textwidth]{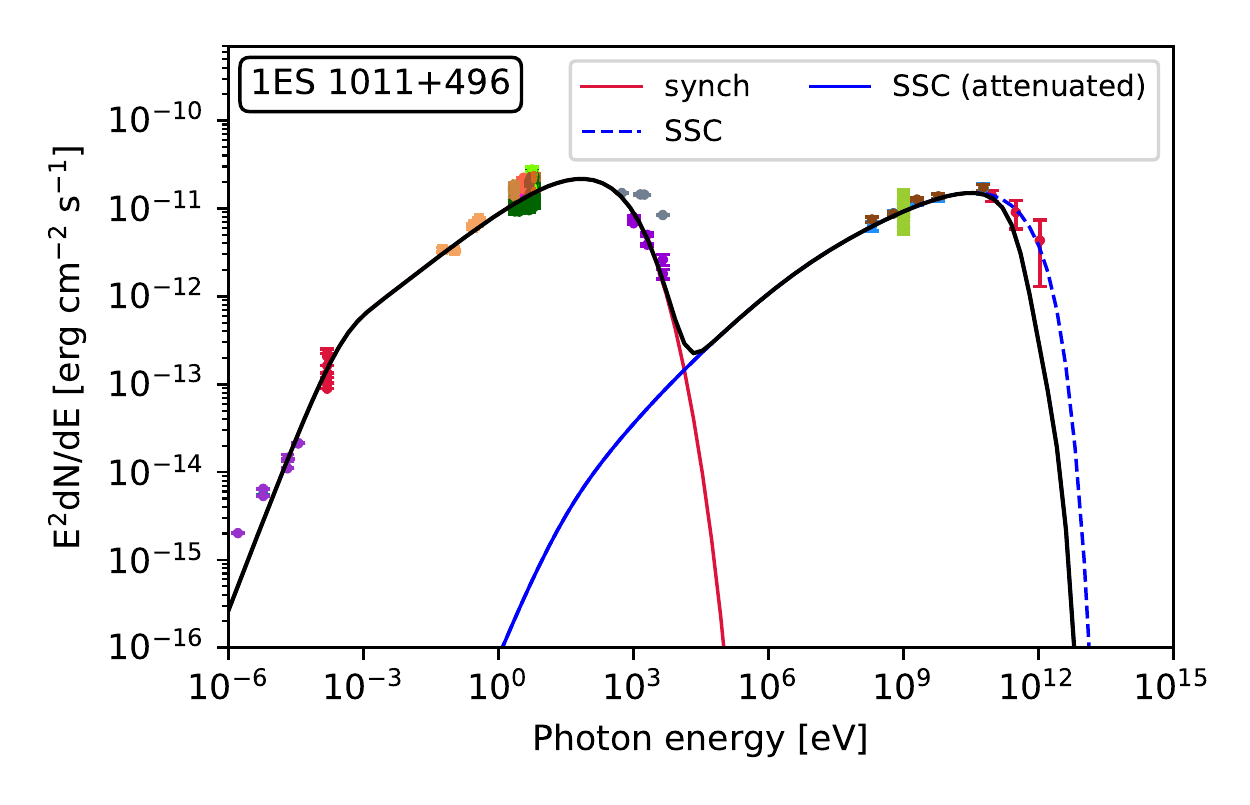}
\includegraphics[width = 0.49\textwidth]{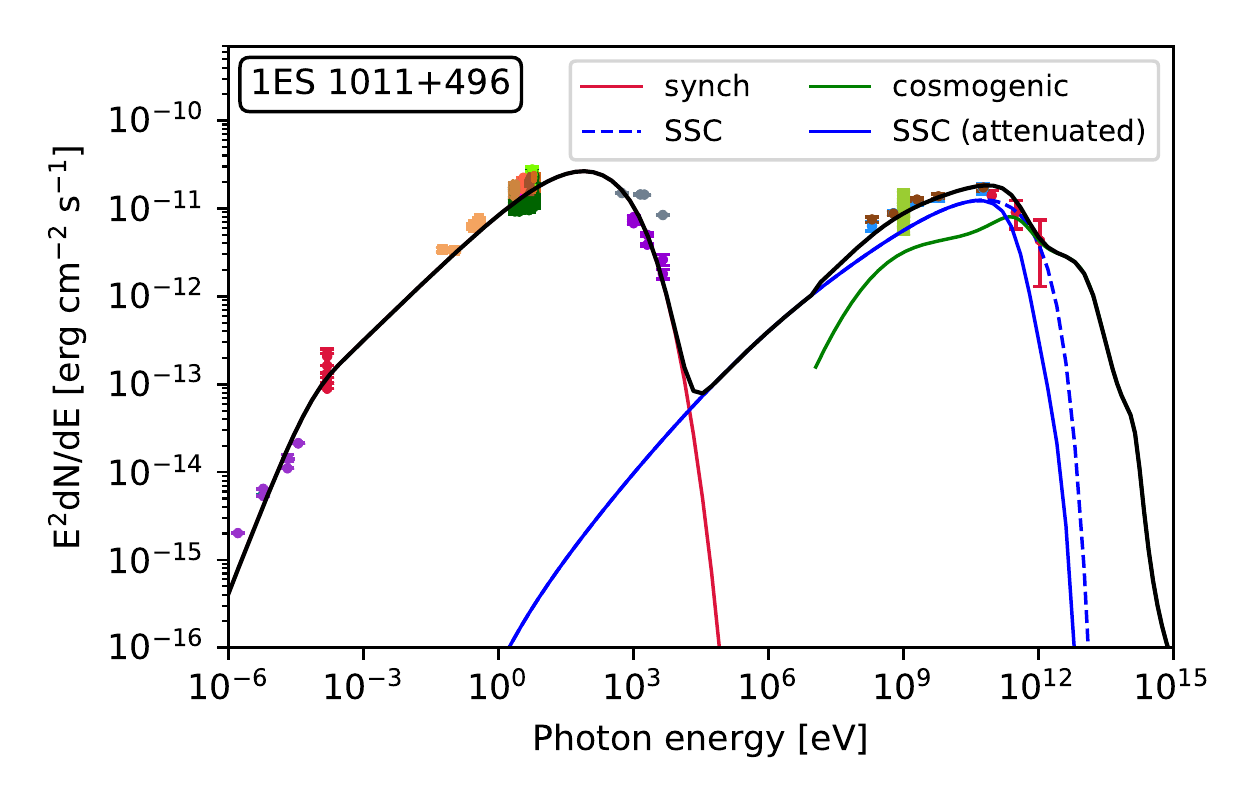}
\includegraphics[width = 0.49\textwidth]{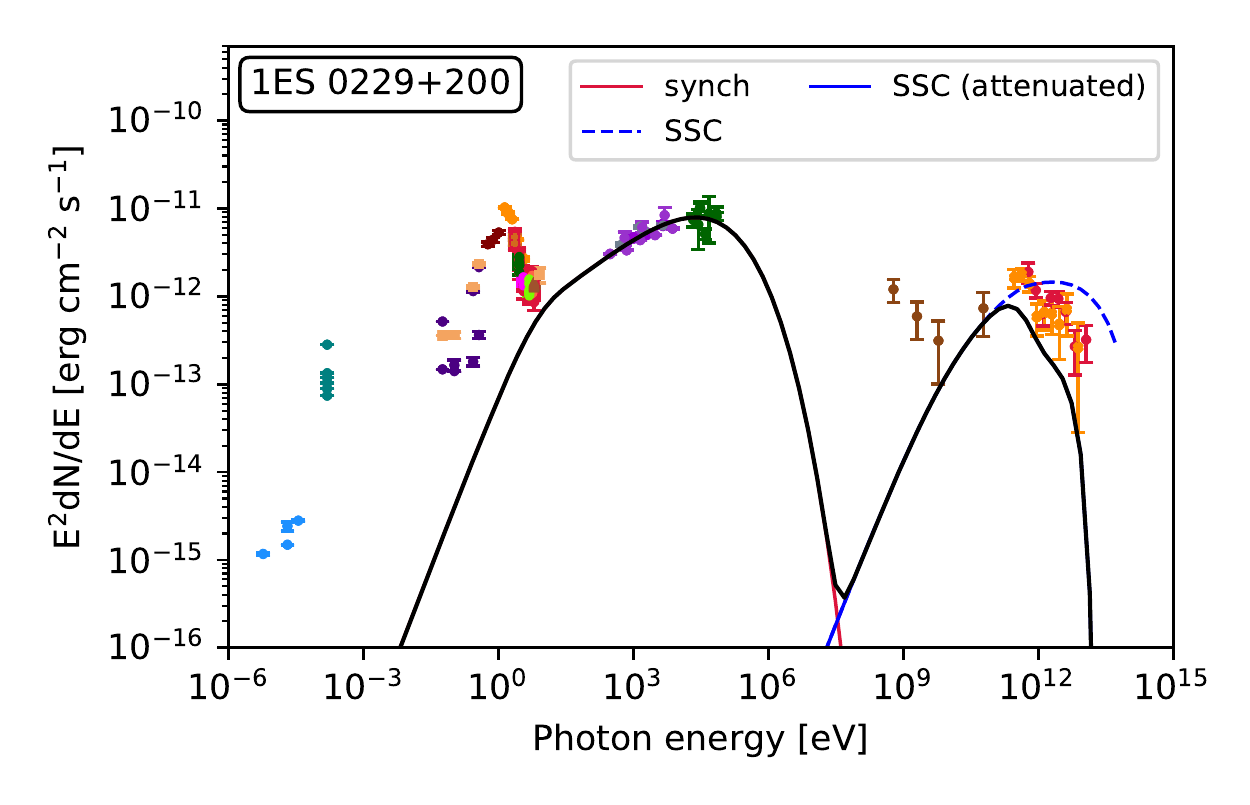}
\includegraphics[width = 0.49\textwidth]{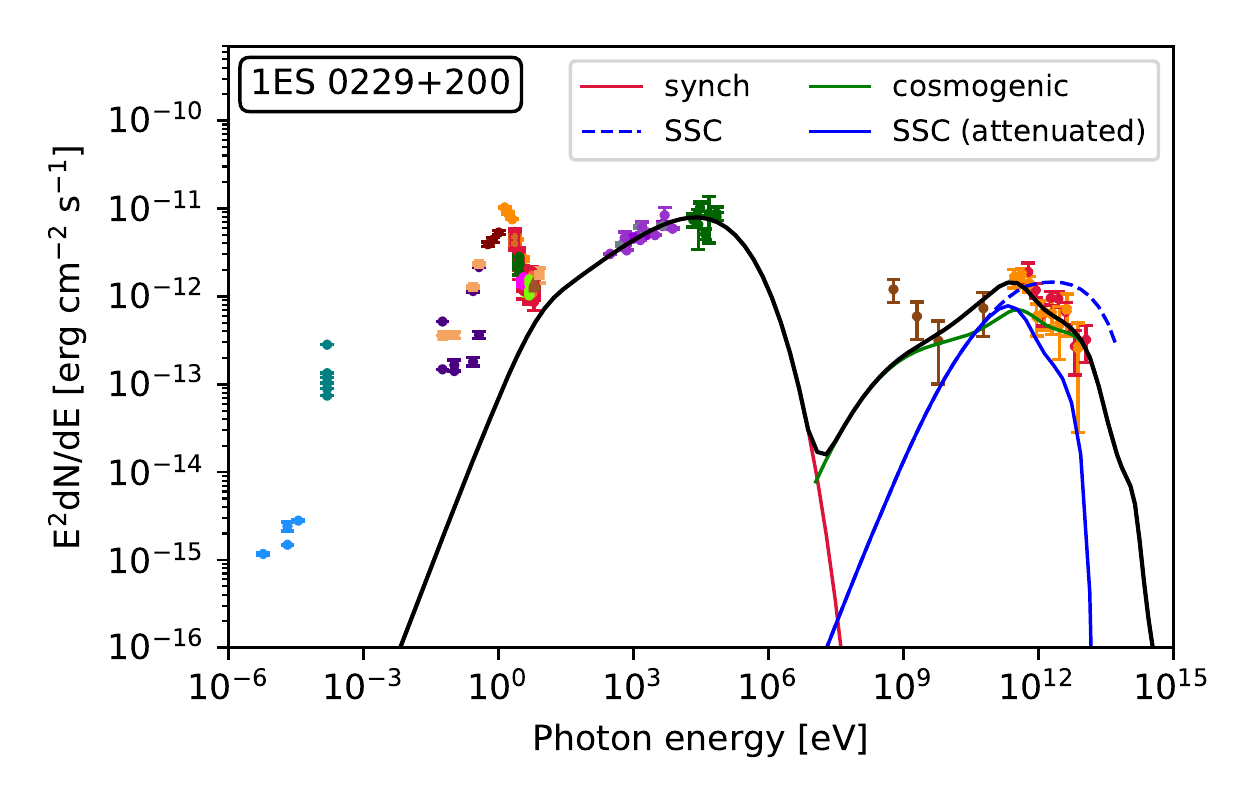}
\includegraphics[width = 0.49\textwidth]{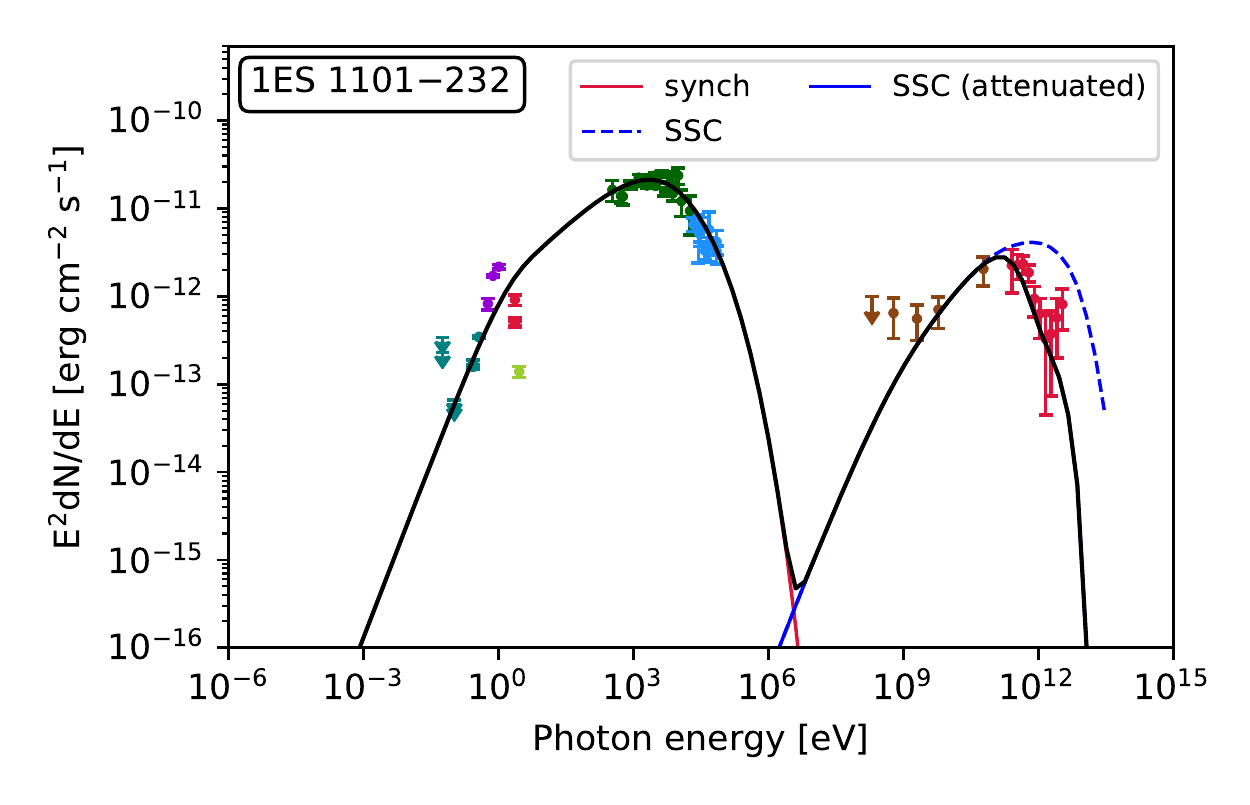}
\includegraphics[width = 0.49\textwidth]{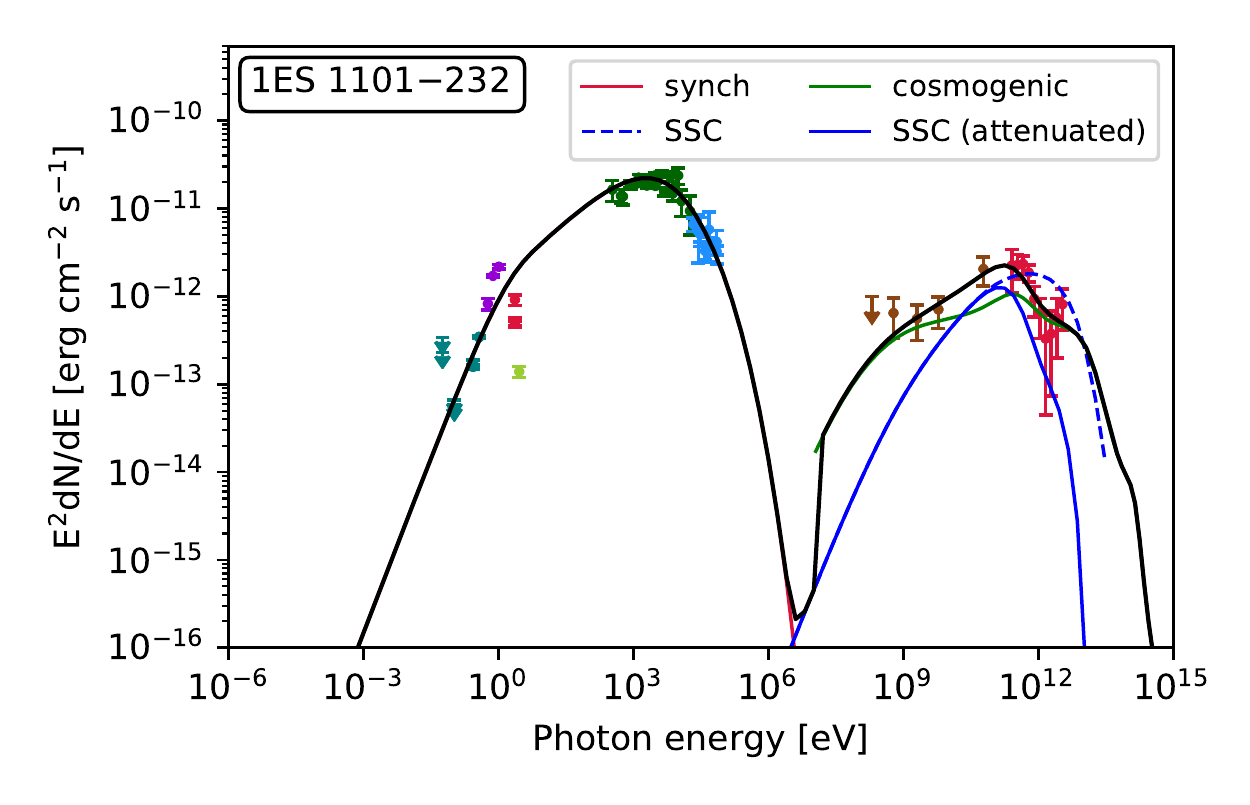}
\includegraphics[width = 0.49\textwidth]{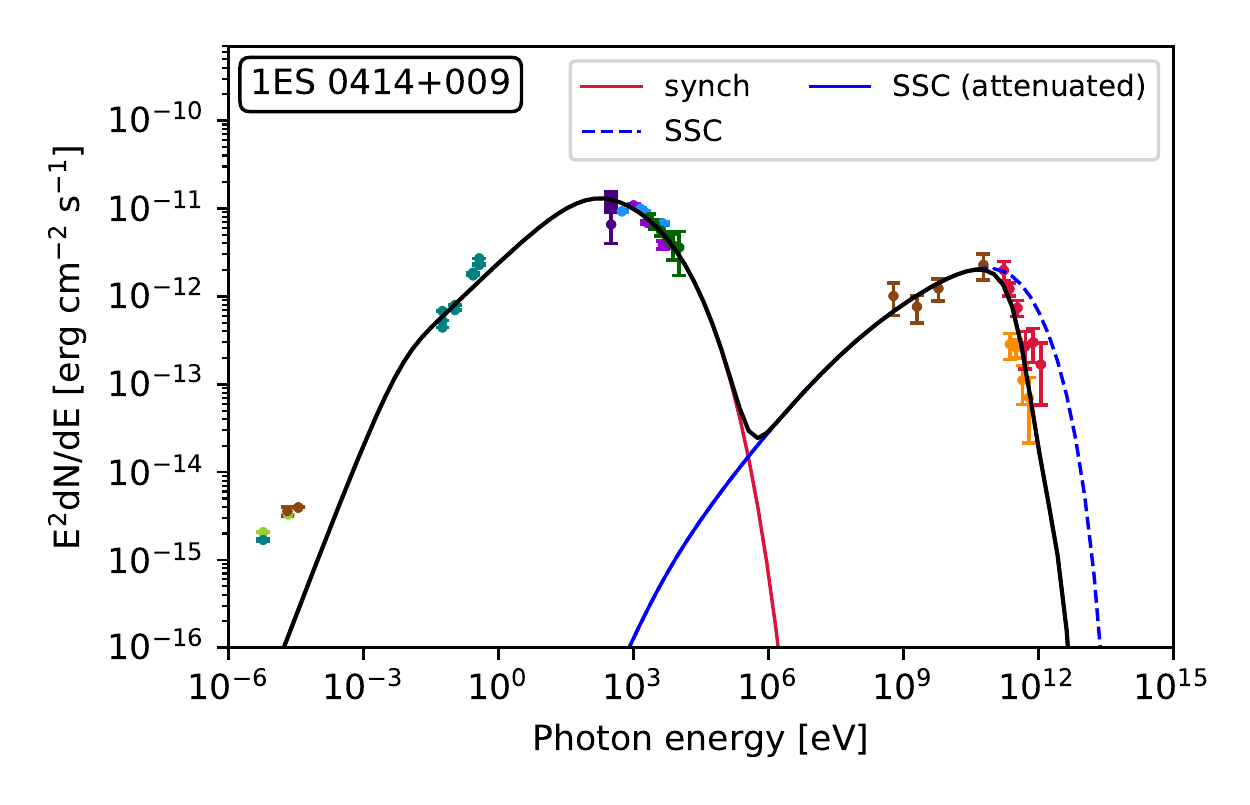}
\includegraphics[width = 0.49\textwidth]{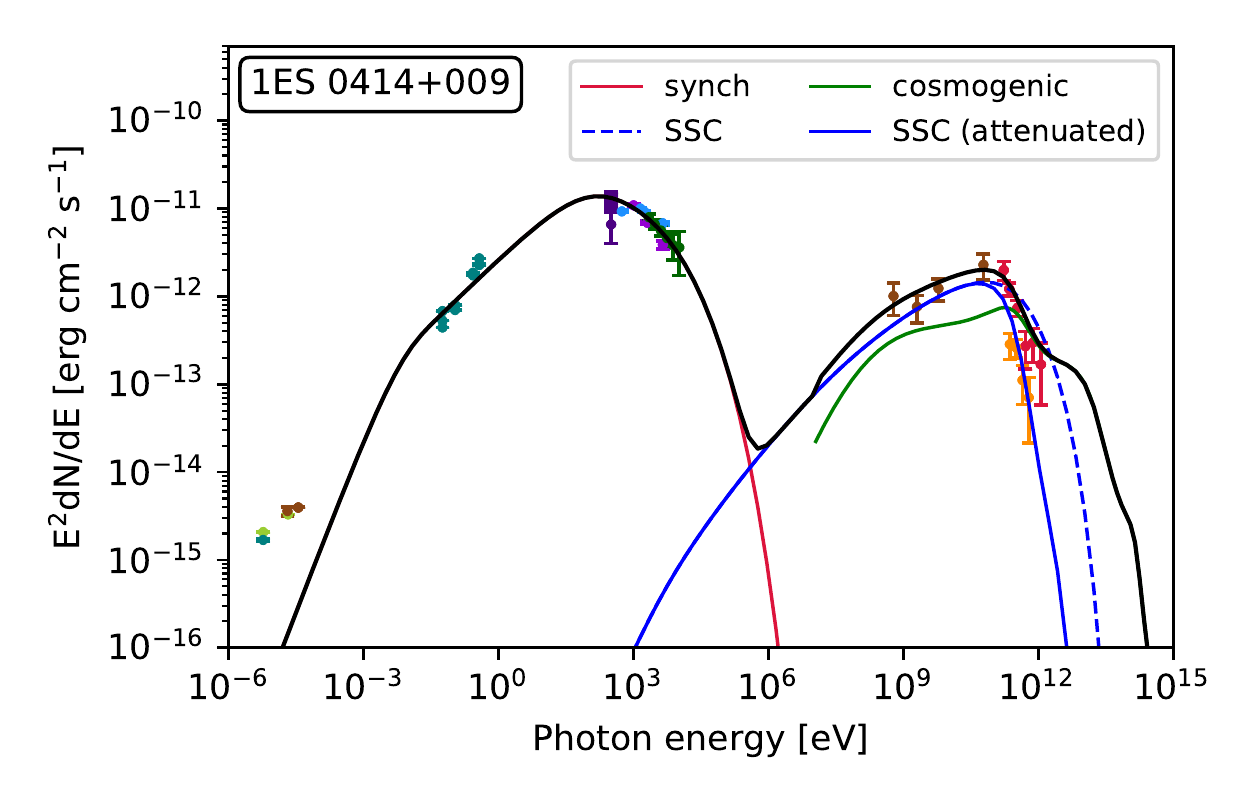}
\caption{\small{Multiwavelength spectrum of the HBLs, modeled using a pure leptonic model (\textit{left}) and a leptonic + hadronic model (\textit{right}).}}
\label{fig:rad_spectra}
\end{figure*}

The observed photon flux from inside the jet will be attenuated due to absorption in EBL and is taken into account using Gilmore et al. model. The optical depth for a gamma-ray observed at energy $E_\gamma$ is calculated by the following integral along the line of sight to the target at redshift $z$ \citep{Gilmore_12},
\begin{align}
\tau(E_\gamma , z_0) = & \dfrac{1}{2}\int_0^{z_0} dz\dfrac{dl}{dz}\int_{-1}^{+1}du (1-u) \nonumber \\
&\times\int_{E_{\rm min}}^\infty dE_{\rm bg} \ n(E_{\rm bg}, z)\sigma(E_{\gamma}(1+z), E_{\rm bg}, \theta)
\end{align}
where $E_{\rm min}$ is the redshifted threshold energy $E_{\rm th}$ for a background photon to interact with a $\gamma-$ray of energy $E_\gamma$.
\begin{equation}
E_{\rm min}=E_{\rm th} (1+z)^{-1} = \dfrac{2m_e^2c^4}{E_\gamma(1+z)(1-\cos\theta)}
\end{equation}
$n(E_{\rm bg}, z)$ is the proper density of target background photons as a function of energy $E_{\rm bg}$ and redshift $z$, and $u=\cos\theta$. $dl/dz$ is the cosmological line element given by,
\begin{equation}
\dfrac{dl}{dz} = \dfrac{c}{(1+z)H_0}\dfrac{1}{\sqrt{\Omega_m(1+z)^3+\Omega_\Lambda}}
\end{equation}
The attenuation affects only the high energy end of the SSC spectrum, as can be seen in Fig.~\ref{fig:rad_spectra} producing a flux $F_\gamma^{\rm obs}(E, z) = F_\gamma^{\rm int}(E, z)\exp[-\tau(E_\gamma, z_0)]$. An intuitive treatment of $\gamma\gamma_{bg}$ attenuation is also done in \cite{Razzaque_09} (see Eqn. 17 there).
The attenuation effect on the cascade photons due to the EBL is incorporated in the DINT code. We implement the Gilmore et al. EBL model in the DINT for propagation and cascade initiated by secondary EM particles. Our results do not depend significantly on the choice of other recent EBL models, e.g. \cite{Finke_10, Dominguez_11}.

\begin{figure*}
\centering
\includegraphics[width = 0.49\textwidth]{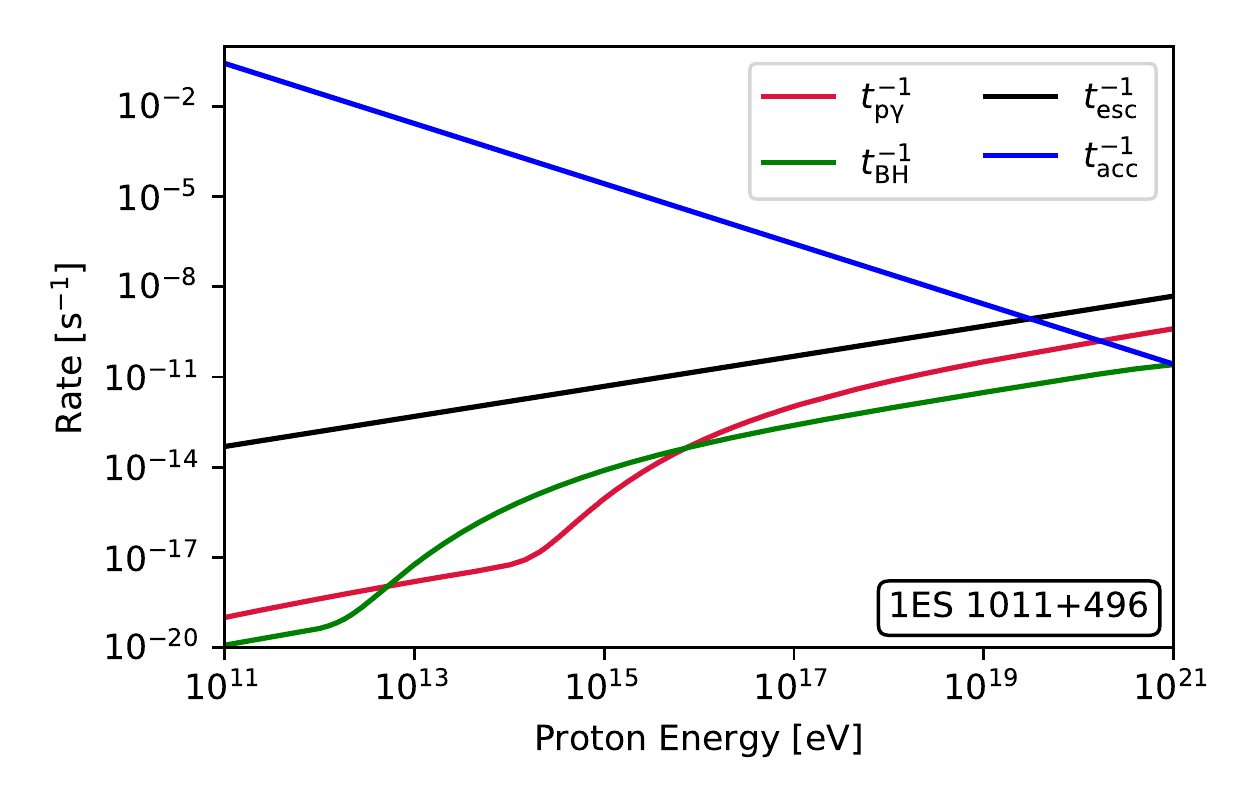}
\includegraphics[width = 0.49\textwidth]{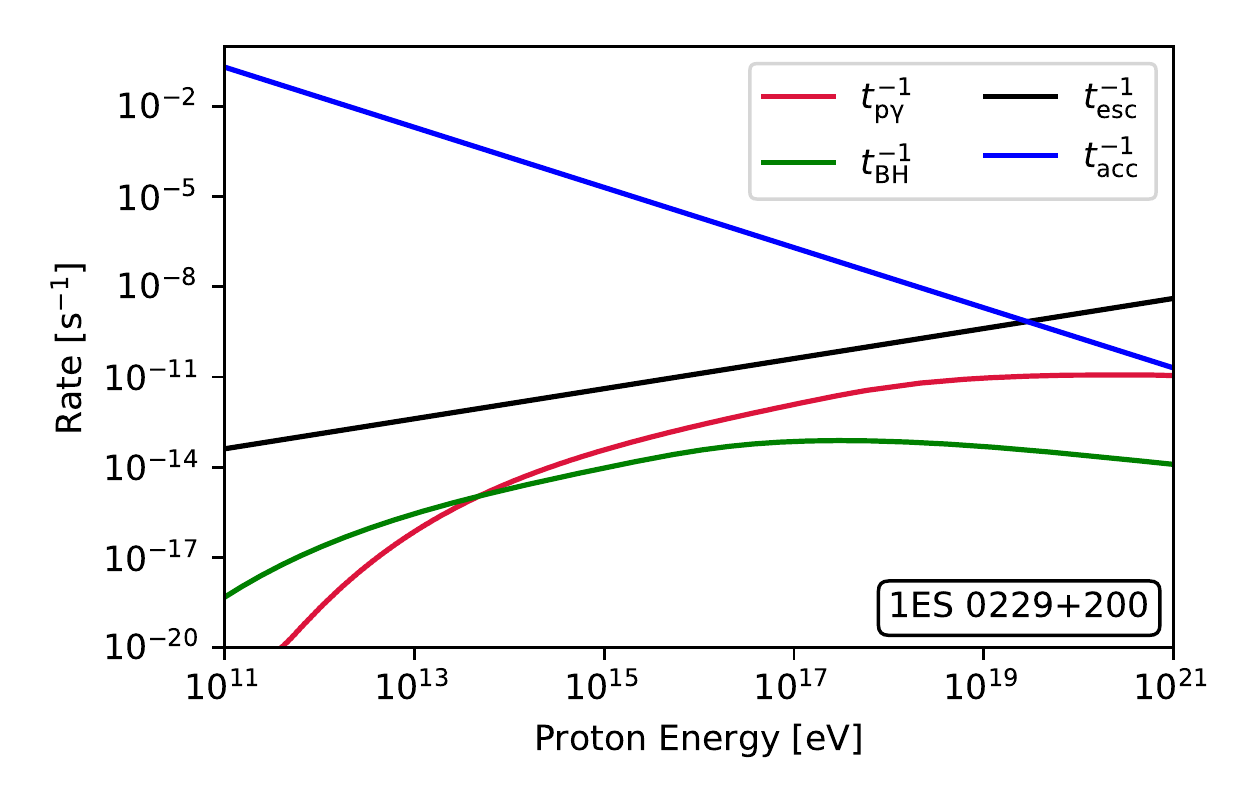}
\includegraphics[width = 0.49\textwidth]{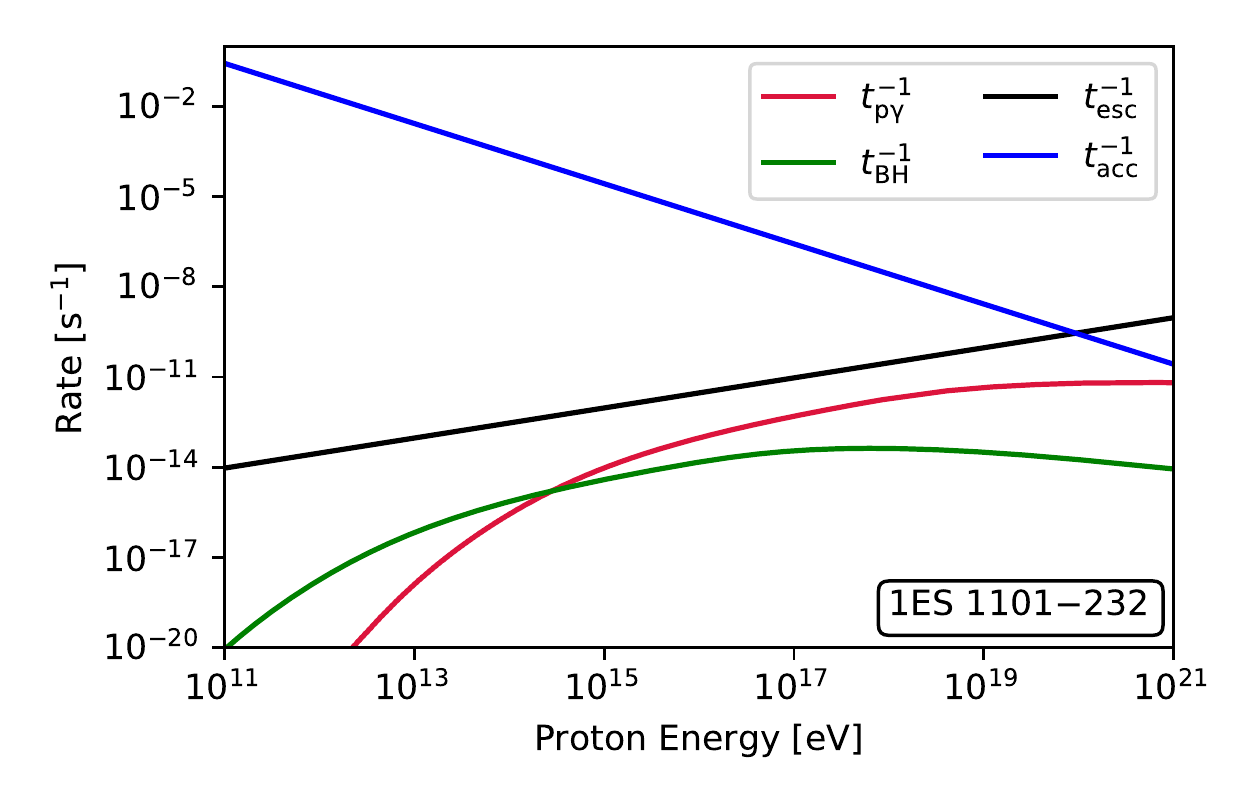}
\includegraphics[width = 0.49\textwidth]{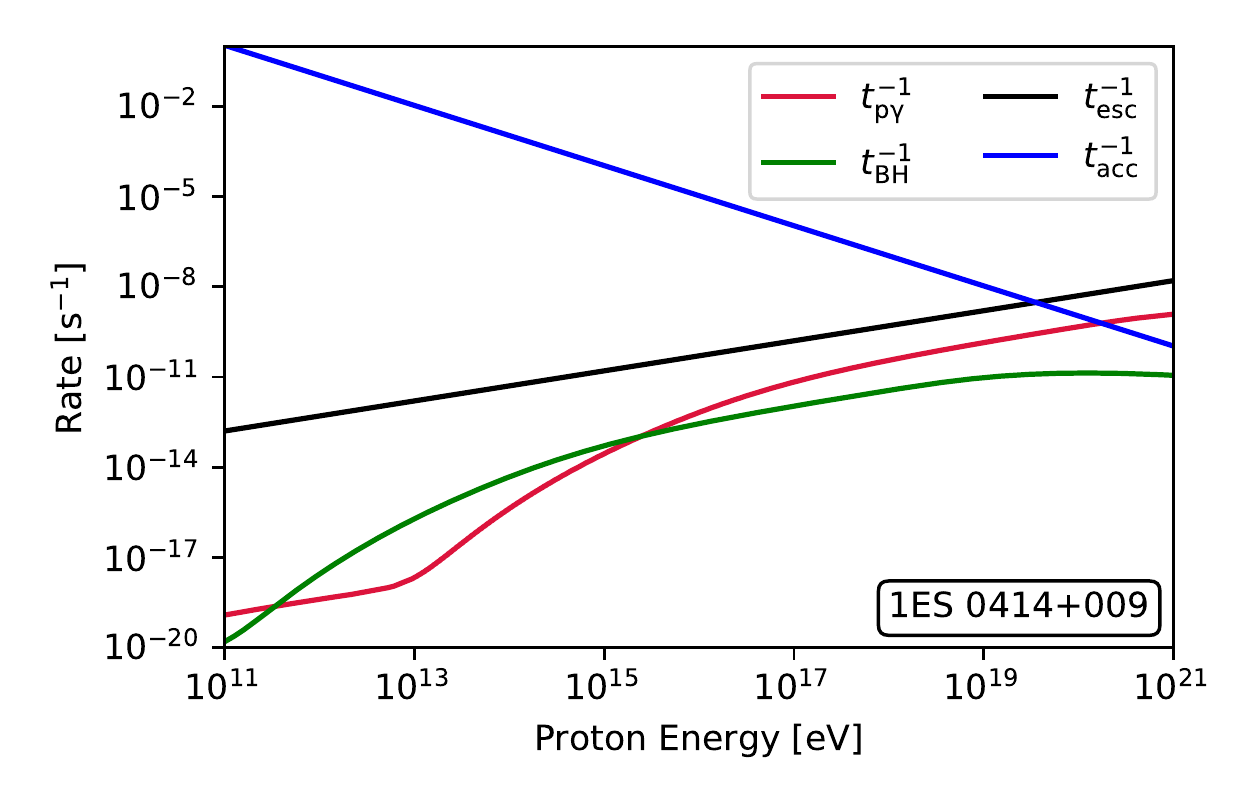}
\caption{\small{Timescale of photohadronic interactions inside the jet, with target photons from synchrotron and IC emission, calculated using Eqn.~(\ref{eqn:efficiency_ph}). The acceleration timescale ($t_{\rm acc}$) and escape timescale ($t_{\rm esc}$) are also shown for comparison, calculated from Eqn.~(\ref{eqn:escape_timescale}) and Eqn.~(\ref{eqn:acceleration}) respectively. The Bohm condition gives the minimum diffusion leading to a lower value of $t_{esc}$. Particles can be more diffusive than this and thus, $t_{\rm esc}$ is adjusted by varying the diffusion coefficient $D_0$ such that acceleration dominates up to $10^{19}$ eV (see text for more details). The $p\gamma$ and Bethe-Heitler interaction rates are found to be orders of magnitude less than escape rate. The photon spectrum from $\pi^0$ decay inside the jet is calculated and found to be $\sim 10$ orders of magnitude less than the peak VHE flux, for the same normalization as required for contribution from UHECR interactions.}}
\label{fig:timescales}
\end{figure*}

\begin{table*}
\caption{\label{tab:bestfit} Fit parameters for the multiwavelength SED modeling in Fig.~\ref{fig:rad_spectra}}
 \begin{ruledtabular}
 \begin{tabular}{lrrccccccccccc}
 HBL & $E_{\rm e, min}$  & $E_{\rm e, cut}$ & $\alpha$ & $R$ & $B$ & $\delta_D$ & $L_{\rm e}$ & $L_{\rm B}$ & $L_{\rm UHECR}$ & $L_{Edd}$\\
 & [GeV] & [GeV] & & [cm] & [Gauss] & & [erg/s] & [erg/s] & [erg/s] & [erg/s]\\
 \hline
 \multicolumn{11}{c}{Pure leptonic model}\\
 \hline
 1ES 1011+496 & 0.08 & 75.0 & 2.2 & 1.5$\times10^{17}$ & 0.024 & 20 & $5.8\times10^{38}$ & $1.9\times10^{43}$ & -- & \\
 1ES 0229+200 & 10.00 & 1500.0 & 2.2 & $1.0\times10^{16}$ & 0.015 & 40 & $1.3\times10^{38}$ & $1.3\times10^{41}$ & -- &\\
 1ES 1101--232 & 5.70 & 550.0 & 2.0 & $8.4\times10^{16}$ & 0.020 & 22 & $6.0\times10^{37}$ & $5.1\times10^{42}$ & -- & \\
 1ES 0414+009 & 0.20 & 200.0 & 2.0 & $7.0\times10^{16}$ & 0.080 & 22 & $7.6\times10^{37}$ & $5.7\times10^{43}$ & -- &\\
 \hline
 \multicolumn{11}{c}{Leptonic + hadronic (UHECR) model}\\
 \hline
  1ES 1011+496 & 0.04 & 65.0 & 2.0 & $2.2\times10^{17}$ & 0.020 & 20 & $3.8\times10^{38}$ & $2.9\times10^{43}$ & $4.8\times10^{44}$ & $5.1\times10^{46}$\\
 1ES 0229+200 & 10.00 & 1500.0 & 2.2 & $1.0\times10^{16}$ & 0.015 & 40 & $1.3\times10^{38}$ & $1.3\times10^{41}$ & $2.6\times10^{43}$ & $1.7\times10^{47}$\\
 1ES 1101--232 & 5.70 & 500.0 & 2.0 & $1.4\times10^{17}$ & 0.020 & 22 & $3.5\times10^{37}$ & $1.4\times10^{43}$ & $3.0\times10^{43}$ & $1.0\times10^{47}$\\
 1ES 0414+009 & 0.20 & 200.0 & 2.0 & $9.0\times10^{16}$ & 0.080 & 22 & $5.9\times10^{37}$ & $9.4\times10^{43}$ & $1.0\times10^{44}$ & $2.0\times10^{47}$\\

 \end{tabular}
 \end{ruledtabular}
\end{table*}

We consider the sources: 1ES 1011+496, 1ES 0229+200, 1ES 1101--232, and 1ES 0414+009. These representative class of HBLs covers a wide range in redshift (indicated in Table~\ref{tab:hadronic}) and hence provides a laboratory for testing the plausibility of $\gamma-$ray production from UHECR interactions. First, we calculate the timescale of photopion and Bethe-Heitler interactions $t_{p\gamma}$, $t_{\rm BH}$, inside the jet and compare it with the escape timescale $t_{\rm esc}$, using Eqn.~(\ref{eqn:efficiency_ph}) and Eqn.~(\ref{eqn:escape_timescale}). The seed photons are considered to be synchrotron as well as IC photons. We also calculate the acceleration timescale $t_{acc}$ using Eqn.~(\ref{eqn:acceleration}). The various timescales and interaction rates are shown in Fig.~\ref{fig:timescales}. In the Bohm diffusion approximation, the difffusion co-efficient can be written as $D=\eta r_Lc/3$. However, particles can be more diffusive than this and we consider the Kraichnan model of diffusion with the diffusion co-efficient $D_0$ adjusted between $10^{27}$ to $10^{30}$ cm$^2/$s, such that the acceleration rate dominates over escape rate at least up to $E=10^{19}$ eV. The maximum acceleration energy of protons inside the blob is found to be a few EeV from Eqn.~(\ref{eqn:Hillas}), for the sources studied. In view of the dominance of acceleration over escape and uncertainties in fit parameters, we set $E_{\rm p,max}=10^{19}$ eV for all the sources considered. The Eddington luminosity of the blazars can be calculated using the expression $L_{\rm Edd}=10^{47}(M_{\rm BH}/10^9 M_\odot)$ erg/s. In the absence of an estimated BH mass, we consider $M_{\rm BH}=10^{9}$ $M_\odot$ for 1ES 1101--232. For the other sources, the masses of the SMBHs are taken from \cite{Falomo_03}. We assume $\delta_D \simeq \Gamma$ in our calculations, which is valid for a viewing angle of the order of few degrees.

The synchrotron spectrum is modeled using a one-zone leptonic emission in GAMERA. The jet parameters obtained from the fit is used to calculate the SSC spectrum and is adjusted to extend up to the highest energy possible. Beyond this energy, the hadronic contribution dominates. The photon spectrum produced in UHECR interactions, peaking at $\sim$ TeV energies, is incorporated to well explain the data points in the entire VHE range. The synchrotron and SSC luminosities are doppler boosted in the observer frame by $\delta^4$. Here the multiwavelength data is taken for the quiescent state. The fit to photon SEDs obtained from a pure leptonic origin are presented in the left panels of Fig.~\ref{fig:rad_spectra}. The corresponding parameter values are given in top section of Table~\ref{tab:bestfit}. \citet{Petropoulou_15b} have done a detailed analysis of hadronic losses inside the jet resulting in photon spectrum in the high-energy regime. The proton jet power is much higher in \citet{Petropoulou_15b} compared to ours. We have calculated the efficiency of $p\gamma$ interactions inside the jet using the formalism in \cite{Kelner_08} and found that the photon spectrum resulting from $\pi^0$ decay is insignificant in comparison to the observed flux. Thus, we consider the interactions of UHECR protons only, during propagation over cosmological distances. The SSC flux depends on the radius of the spherical blob of emitting zone inside the jet and is adjusted suitably in case of a lepto-hadronic fit. The fit parameters used to model the observed SEDs using a combined leptonic + hadronic (UHECR) scenario are listed in the bottom section of Table~\ref{tab:bestfit} and the multiwavelength fits are presented in the right panels of Fig.~\ref{fig:rad_spectra}.

\textbf{1ES 1011+496}: The optical data, X-ray data and radio-to-X-ray flux ratio show typical properties of an HBL \citep{Padovani_95, Ahnen_16}. It is situated at a redshift $z=0.212$ and the VHE $\gamma-$ray emission was first discovered by MAGIC observations triggered by an optical outburst in March 2007 \citep{Albert_07}. The source has been well observed in 0.1 $-$ 300 GeV band by Fermi-LAT \citep{Fermi_3FGL} and 0.3-10 keV band by Swift-XRT \citep{Abdo_10}. The source is also listed in the second catalog of hard Femi-LAT sources. A photohadronic scenario to explain the high-energy SED is employed in \cite{Sahu_17}. The flux value of the low- and high-energy peak are comparable. We find, the pure-leptonic model is unable to cover the highest energy data points due to EBL attenuation. Increasing $E_{\rm e,cut}$ worsens the fit for synchrotron spectrum. In the lepto-hadronic fit, photon spectrum from UHECR interactions can indeed explain the highest energy data points.

\textbf{1ES 0229+200}: This BL Lac object at redshift 0.140 \citep{Woo_05} shows an extremely hard intrinsic TeV spectrum with the synchrotron spectrum peaking at exceptionally high energies near hard X-ray regime \citep{Costamante_18}. It has one of the highest inverse Compton (IC) peak frequency and the narrowest electron distribution among the extreme blazars known \citep{Kaufmann_11}. The source was first discovered by HESS in 2004 \citep{Aharonian_07a}. The X-ray-to-radio flux ratio classifies it as an HBL \citep{Giommi_95}. The SED in the high-energy (HE) band has been modeled using reprocessed GeV emission from pair production on EBL and subsequent cascade \citep{Tavecchio_10}, using a similar method applied to $\gamma-$ray bursts in \cite{Razzaque_04}. The HE data is obtained from Fermi-LAT \citep{Ackermann_13} and the VHE data is obtained from HESS \citep{Abramowski_14} and VERITAS collaboration \citep{Cerruti_13, Aliu_14}. In our pure-leptonic modeling, there is no significant change in the SSC spectrum on increasing $E_{\rm e,cut}$ beyond the value considered due to KN suppression. We show, the necessity of an additional component is inevitable and the spectrum arising from UHECR interactions well explains the highest energy data points.

\textbf{1ES 1101--232}: This HBL resides in a elliptical host galaxy at redshift $z=0.186$. It was first detected by Ariel-5 X-ray satellite and was misidentified with the Abell 1146 galaxy cluster at redshift $z=0.139$ \citep{Maccagni_78, McHardy_81}. The optical and radio data led to the correct identification as a BL Lac object \citep{Buckley_85, Remillard_89}. The highest energy data points exhibit very hard TeV spectra, which has been explained earlier using hadronic origin inside the jet emission region \citep{Cerruti_15}. The HE and VHE $\gamma-$ray data is obtained from Fermi-LAT and HESS observations \citep{Aharonian_07b, Abramowski_14}. It can be seen that that the highest energy data points are not well-covered by the SSC spectrum alone in a pure-leptonic fit. The fit improves considerably in the entire VHE range with the addition of the hadronic component originating in UHECRs.

\textbf{1ES 0414+009}: The optical spectrum of this HBL at redshift $z=0.287$ is described by the sum of the emission due to a standard elliptical galaxy and a relatively flat power law \citep{Halpern_91}. It was first detected by HEAO 1 satellite \citep{Gursky_78} in the energy range 0.2 keV $-$ 10 MeV. It is one of the furthest VHE blazar with very hard TeV $\gamma-$ray spectrum and well-determined redshift. The supermassive black hole (SMBH) at the center has a mass $2\times10^{9}M_\odot$ \citep{Falomo_03}. The HE (100 MeV $-$ 100 GeV) data is obtained from Fermi-LAT and the source is listed in the Fermi 4FGL catalog. The VHE ($E>100$ GeV) $\gamma-$ray data is taken from observations by HESS \citep{Abramowski_12} and VERITAS \citep{Aliu_12} collaborations. Despite the attenuation due to EBL background, a good fit to the observed SED is achieved using a pure-leptonic model alone. However, the fit to data points at the highest energy improves on adding the photon spectrum from UHECR interactions.

The broadband SED of the sources is plotted from the archival data as retrieved from ASDC SED builder \citep{Stratta_11}. The data points in the radio band are obtained from GBT \citep{Gregory_96}, CLASSCAT \citep{Myers_03}, NIEPPOCAT \citep{Nieppola_07}, NVSS \citep{Condon_98}, Planck \citep{Ade_14}, and WMAP \citep{Wright_09} observations. WISE catalog gives the flux values in infrared energies \citep{Wright_10}. The data for optical-UV wavebands are found from the UV and optical telescope (UVOT) of the Swift observatory \citep{Giommi_12} and the GALEX catalog \citep{Bianchi_11}. BeppoSax \citep{Beckmann_02}, XMM-Newton \citep{Watson_09}, and Swift-XRT \citep{D'Elia_13} provides the soft X-ray data, while the hard X-ray data is obtained from Swift-BAT observations \citep{Baumgartner_13}.

\begin{figure*}
\centering
\includegraphics[width = 0.49\textwidth]{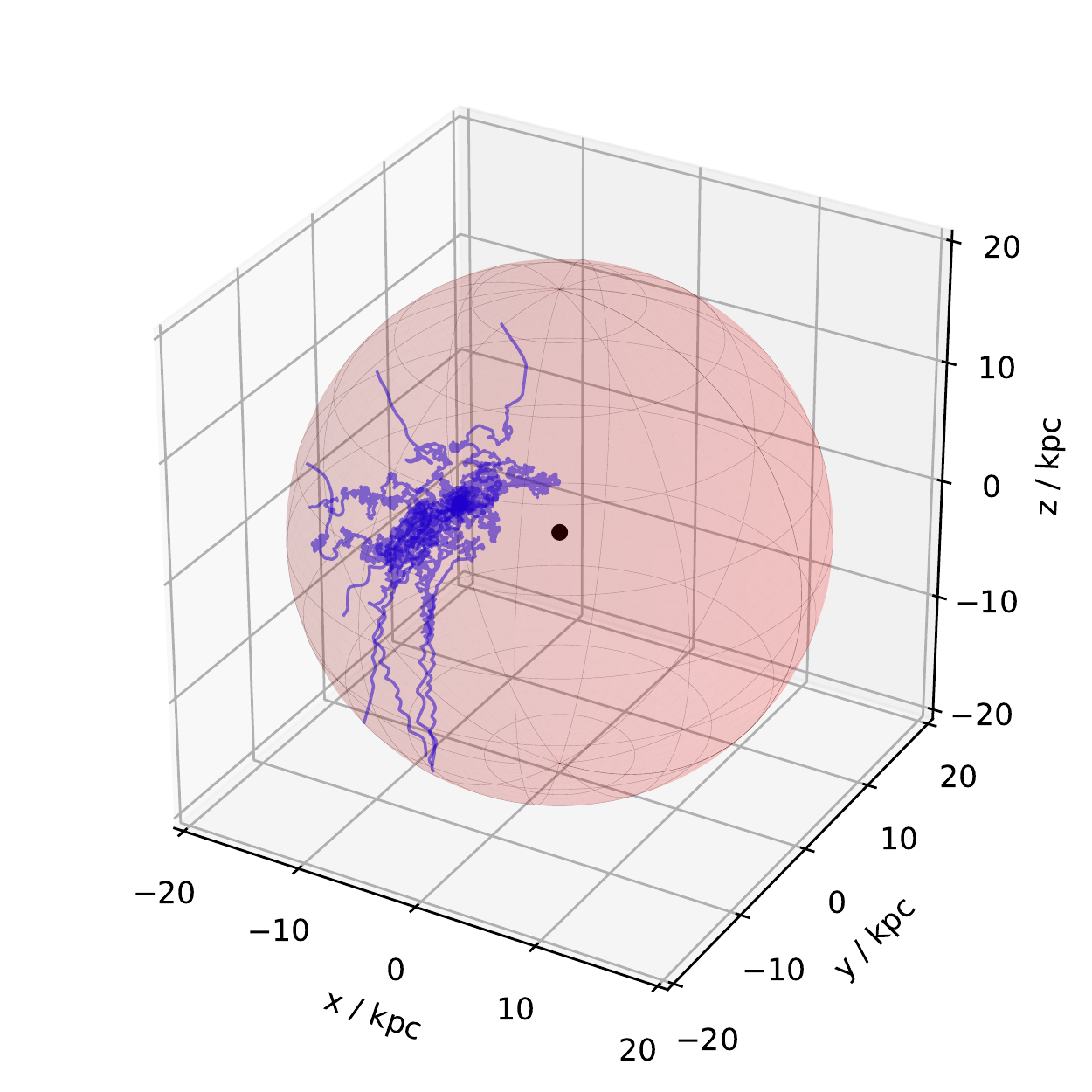}
\includegraphics[width = 0.49\textwidth]{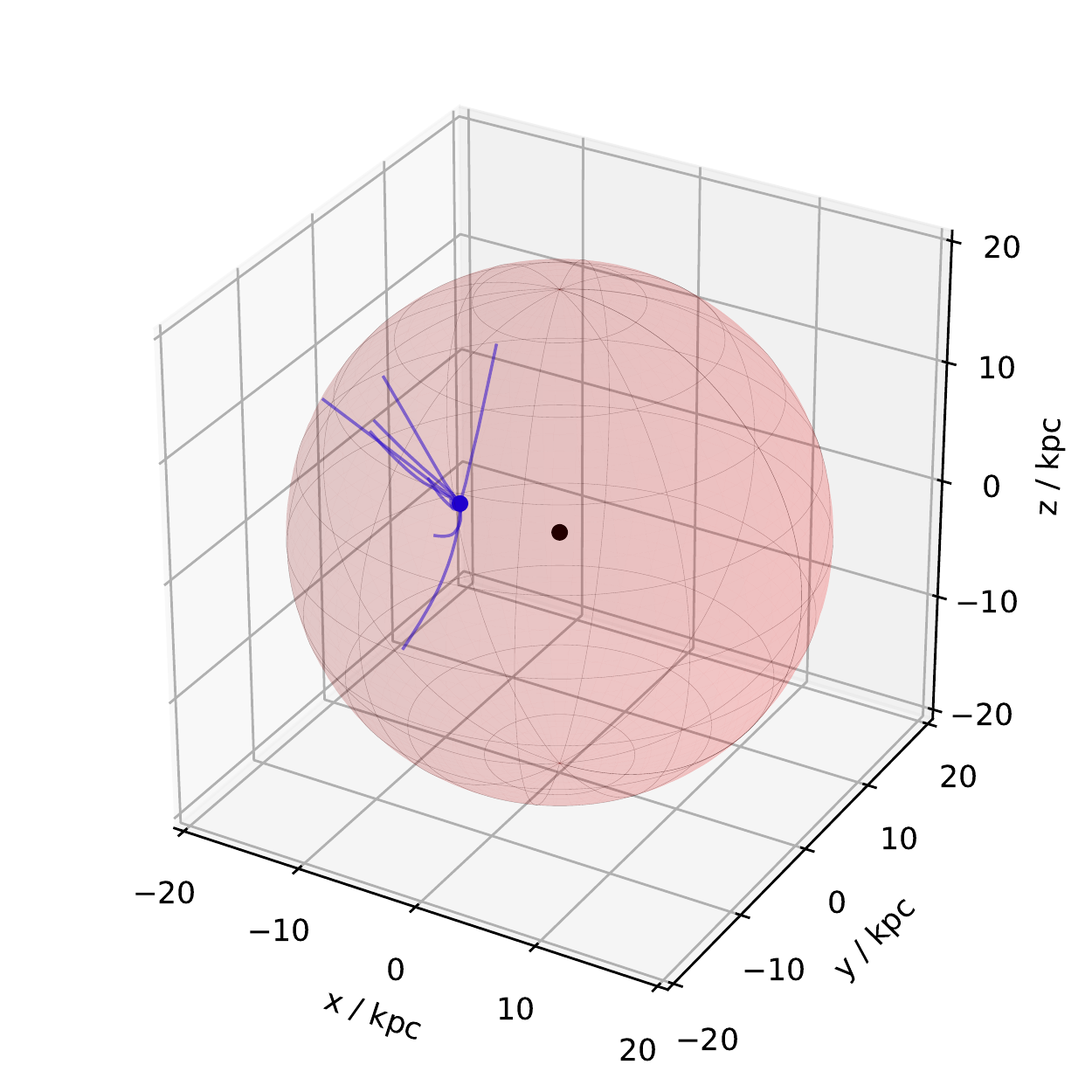}
\caption{\small{3D trajectories of 10 UHECR anti-protons emitted isotropically and backtracked from the Earth in Janson \& Farrar magnetic field of the Galaxy upto a halo radius of 20 kpc. The black dot indicates the Galactic center. See text for more details. \textit{Left:} For $E=0.1$ EeV, the deflections are high and no directionality information can be retained. \textit{Right:} For $E=10$ EeV, the deflections are small and the anti-protons travel in almost straight lines, escaping the Galaxy.}}
\label{fig:gal_traj}
\end{figure*}

Now, since the survival rate of UHECRs within $0.1^\circ$ of the propagation direction towards the observer is found to vary between 39\% and 48\% at a distance of 1 Mpc from the observer and the turbulent correlation length is taken to be $l_c=1$ Mpc, UHECR events from the source should arrive at Earth. Although $\gamma-$rays travel mostly undeflected once produced, the Galactic magnetic field (GMF) being many orders of magnitude stronger than the EGMF, can deflect the UHECRs significantly. The deflection suffered by UHECRs in the GMF can be approximated as \citep{Dermer&Menon},
\begin{equation}
\theta_{\rm def, MW} \approx \dfrac{0.9^\circ}{\sin b} \bigg(\dfrac{60\text{ EeV}}{E/Z}\bigg) \bigg(\dfrac{B}{10^{-9}\text{ G}}\bigg) \bigg(\dfrac{h_{\rm disk}}{1\text{ kpc}}\bigg)
\end{equation}
where $b$ is the galactic latitude of the source and $h_{\rm disk}$ is the height of the Galactic disk. The GMF may create a strong shadowing effect on the true location of the sources \citep{Fraija_18}. This makes a direct estimation of the observed event rate difficult. The computational efficiency of forward propagation accounting for magnetic field effects over large distances with a point observer is very low in \textsc{CRPropa 3}. We do backtracking simulations of cosmic rays with opposite charge from the observer to the edge of the Galaxy for showing a graphical representation of UHECR trajectories in GMF. We consider a realistic magnetic field, such as that of Jansson \& Farrar field model (JF12) with both random striated and random turbulent components \citep{Jansson_12}. For backtracking, 10 UHECR anti-protons are injected isotropically from the Earth and is similar to protons traveling towards Earth in forward simulations. In JF12 model the field is set to zero for $r > 20$ kpc and in a 1 kpc radius sphere centered on the Galactic center. The trajectories originating from Earth terminates at the boundary of the 20 kpc radius sphere. We consider two such cases when the UHECR protons are observed at Earth with $E_{\rm obs}=$ 0.1 EeV and 10 EeV. The resulting trajectories are shown in the left and right panels of Fig.~\ref{fig:gal_traj}, respectively. The black and blue dots represent the Galactic center and the Earth at a distance of 8.5 kpc from the Galactic center, respectively. 


An estimate of the number of UHECR events that can be expected at the Pierre Auger observatory (PAO) in Malarg\"ue, Argentina can be calculated from,
\begin{equation}
N_{\rm evt, p} = \dfrac{1}{\xi_B}\dfrac{\Xi\omega(\delta)}{\Omega}\int_{E_{\rm th}^{\rm obs}}^{{E_{\rm max}^{\rm obs}}} \dfrac{dN}{dE}dE \label{eqn:UHECR_events}
\end{equation}
where $\omega(\delta)$ is the relative exposure at a point source in the sky compared to the largest exposure on the sky \citep{Sommers_01} and $\Xi$ is the total integrated exposure over the detector's field of view. Here the additional factor $\xi_B$ accounts for UHECRs surviving within $0.1^\circ$ of the initial propagation direction, at a distance of 1 Mpc from Earth. Since the maximum UHECR energy $E_{\rm p,max}$ is taken to be only 10 EeV for all the sources, the UHECR events are subjected to large deflections in the GMF. The value of $E_{\rm max}^{\rm obs}$ being smaller compared to the GZK cutoff energy \citep{Greisen_66, Zatsepin_66} or, the energy threshold for AGN correlation analysis by PAO \citep{PAO_08}, an intermediate-scale anisotropy study is difficult to do. Also in large-scale anisotropic studies, the most significant signal found is the dipolar modulation in right ascension at energies $E>8$ EeV \citep{Aab_17, Aab_18}. In the latter case, a statistically significant oversampling at any grid point in the sky, compared to the background events is unexpected owing to the deflection and huge number of observed events at these low energies. Therefore, if the energy of the UHECRs from the HBLs we considered is restricted to below 10 EeV as required by our model, then it will be difficult to detect them directly as discussed in \cite{Razzaque_12}.

\begin{figure}
\centering
\includegraphics[width = 0.49\textwidth]{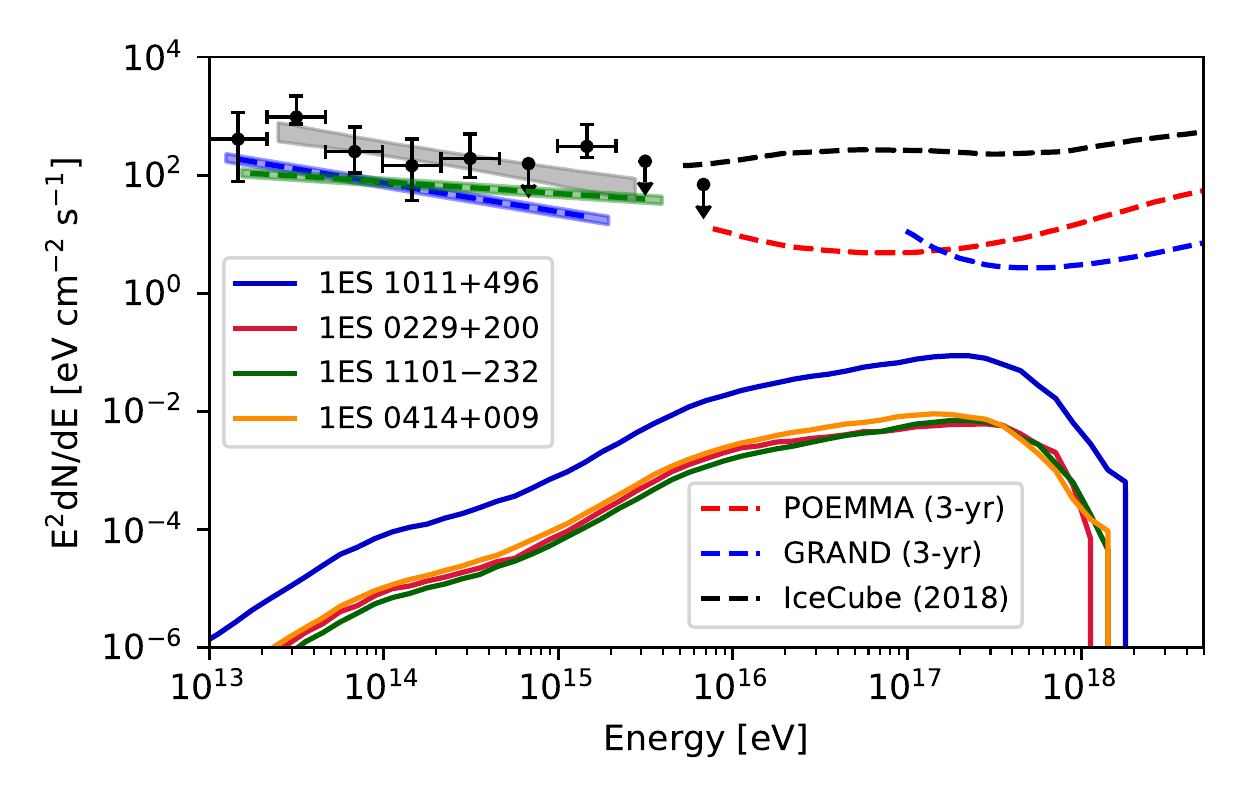}
\caption{\small{All-flavor neutrino flux at Earth produced in the same UHECR interactions as producing EM particles. The POEMMA and GRAND-200K sensitivities are multiplied by $4\pi$ steradians to compare with the neutrino fluxes in same units. The differential upper limit on extremely-high-energy cosmic neutrino flux by IceCube is also shown \citep{IceCube_18}. The black data points show the observed astrophysical diffuse neutrino flux \citep{IceCube_15}. The flux upper limit from 2LAC blazars is shown using equal weighting for power-law with spectral index 2.5 (blue shaded region) and 2.2 (green shaded region) \citep{IceCube_17}.}}
\label{fig:neu_spec}
\end{figure}

Since neutrinos travel unhindered by interactions and undeflected by magnetic fields, the neutrinos produced from the sources near to the line of sight direction are expected to arrive at Earth. The obtained luminosity in neutrinos is constrained from the luminosity requirement in UHECRs to explain the VHE $\gamma-$ray flux. Thus, we can write,
\begin{equation}
L_\nu = L_{\rm UHECR} \times f_{\rm CR\rightarrow\nu} \times \xi_B
\end{equation} 
where $f_{\rm CR\rightarrow\nu}$ is the ratio of the power in produced secondary neutrinos $L_{\nu,p}$ due to propagation of UHECRs, to the injected UHECR power $L_{\rm UHECR}$. This gives the normalization of secondary neutrino flux arriving within 0.1 degrees of the direction of propagation in which the observer is situated. The resulting all-flavor neutrino fluxes are shown in Fig.~\ref{fig:neu_spec}. The extrapolated 3-year sensitivities for the proposed future detectors POEMMA \citep{poemma1, poemma2} and the 200,000 antenna array GRAND-200K \citep{grand1, grand2} are also shown. The detector sensitivities are multiplied by $4\pi$ steradians to obtain the isotropic sensitivity that can be compared with the calculated neutrino flux in units of eV cm$^{-2}$ s$^{-1}$. A 3-years full operation by these detctors will not be sufficient to constrain the neutrino flux from the sources studied.

The secondary neutrino flux from 1ES 1011+496 is higher compared to the other sources. This is expected because of its higher flux value of the high energy bump in the multiwavelength SED, leading to a higher $\gamma-$ray flux required from hadronic interaction channels and hence a higher proton luminosity. The neutrino flux obtained peaks at $2.2\times10^{17}$ eV. The IceCube 8-year differential flux upper limit \citep{IceCube_18}, isotropized at this peak-energy, is of the order of $10^{3}$ times higher than the peak flux from 1ES 1011+496, as shown in figure. Thus a possibility of detection of the neutrino flux obtained from the BL Lacs in our study is unfavorable. The lower energy peak in the neutrino spectrum is not well pronounced because of the proton injection spectral index value $\alpha\sim2$ considered for the sources. The observed astrophysical diffuse neutrino flux is shown in Fig.~\ref{fig:neu_spec}. The grey shaded region indicates the best-fit power law with spectral index $\sim$2.5 \citep{IceCube_15}. Assuming the neutrino flux follows the measured $\gamma-$ray energy flux exactly, the TeV--PeV upper limit from 2LAC blazars is also shown for two spectral indices viz., 2.2 (shown in green shaded region) and 2.5 (shown in blue shaded region). This yields a maximum of 19\%--27\% contribution of the total 2LAC blazar sample to the observed best-fit value of the astrophysical neutrino flux \citep{IceCube_17}.

%
%

\section{Discussions} \label{sec:discussions}

The one-zone leptonic emission model is often found to be inadequate in explaining the high-energy $\gamma-$ray spectrum of a number of BL Lacs. These high-energy BL Lac objects (HBLs) exhibit a hard intrinsic TeV spectrum, which requires an alternate explanation. In many studies, multi-zone emission is employed to fit the broadband spectrum up to the highest energies observed \citep[see, eg.,][]{Prince_19, Xue_19b}. Among hadronic origin $\gamma-$ray models, those which invoke proton synchrotron emission requires extremely high kinetic power and very-high values of doppler factor or magnetic field \citep{Petropoulou_12}. If protons are cooled efficiently by synchrotron photons produced from accelelerated electrons, the contribution from photopion production, Bethe-Heitler interactions and muon synchrotron emission becomes important \citep{Manheim_92, Mucke_03}. 

In this study, we exploit yet another hadronic scenraio where the UHECRs escaping from the jet can interact with cosmic background photons (CMB and EBL), to produce secondary electromagnetic particles. These particles can initiate EM cascades, during propagation over cosmological distances, leading to the production of VHE $\gamma-$ray spectrum near the high-energy bump in the blazar SEDs \citep{Essey_10a}. Hadronic losses inside the blazar jet is found to be insignificant for the jet parameters considered in our model, facilitating the escape of UHECRs. For simplicity, we consider protons as the only UHECRs injected by these sources. We explain the multiwavelength SED of the HBLs by a suitable utilization of this lepto-hadronic model. The parameters for leptonic contribution is adjusted to extend the spectrum up to the highest energies possible and to simultaneously fit the synchrotron spectrum. Beyond this, the contribution from UHECR interactions dominate. The jet power required in such a scenario is calculated and compared with the Eddington luminosity.

We find, this model is successful in fitting the broadband emission spectrum of selected HBLs without exceeding the luminosity budget. The total proton luminosity $L_{\rm p}$, considering relativistic protons down to $\sim10$ GeV energies, will be approximately $5-10$ times of the $L_{\rm UHECR}$ value calculated. Taking this into account doesn't affects the credibility of our model, and the total jet power, in our case, still remains lower than $L_{\rm Edd}$. It is shown in \cite{Razzaque_12}, that required $L_{\rm UHECR}$ increases with increasing values of the injection spectral index $\alpha$ and decreasing values of $E_{\rm p,min}$. They have found that the lower limit on the jet power exceeds the Eddington luminosity for injection spectral index $\alpha>2.2$ for all the sources considered. In our analysis, we restrict ourselves to $\alpha\leqslant2.2$. 

The value of $\xi_B$ considered in the analysis is the survival rate of UHECRs within $0.1^\circ$ of the initial propagation direction. This is not the same as $0.1^\circ$ within the line of sight of the observer because the origin of angle is different in the two cases. Such a restriction provides increased constraints, which decreases the observed photon flux and thus increases the required UHECR power. The two has a one-to-one correspondence depending on the source distance. Hence, this factor is important to constrict the propagation within a narrow cone leading to an increased probability of interception by the observer. This is not considered in earlier works, thus reducing the luminosity requirement in UHECRs. If we calculate the survival rate of UHECRs within a smaller deflection angle than 0.1 degrees, a lower value of EGMF will be necessary for a substantial contribution to photon flux from UHECR interactions. The value of $B_{\rm rms}$ considered in our study is $10^{-5}$ nG, which is higher than the lower limit estimated in \cite{Essey_11b}.

Secondary charged EM particles produced within 0.1 degrees of the direction of propagation can still get deflected by the EGMF or GMF, reducing the observed cascade flux. This is not accounted for in the study. But it is also possible that contribution from shower development initiated outside of 0.1 degrees of propagation direction is intercepted along the line of sight. But the energy loss timescale of electrons/positrons being very low compared to protons, the fraction of events lost by such EM deflection should be negligible. Also, calculating $\xi_B$ only up to 1 Mpc introduces very negligible error, because only an infinitesimal fraction of total EM particles are produced nearer to 1 Mpc from the Earth and the coherence length is set to $l_c=1$ Mpc. We checked the fraction for 1ES 1011+496 at $d_c=$ 895 Mpc from 1D simulation, accepting 100\% of the produced events along the propagation direction. The fraction of EM events produced at distances less than 1 Mpc is found to be 0.0006. 

Since the acceleration rate dominates over the escape rate up to $\gtrsim10^{19}$ eV and the maximum energy obtained from the Hillas condition comes out to be a few EeV, we consider $E_{\rm p,max}=10^{19}$ eV in our analysis. This also accounts for the uncertainty in fit parameters arising because electrons lose energy much faster than protons and as a result they are restricted to a smaller region, while protons can travel larger distances without significant energy loss. Thus, the confinement regions of electrons and protons are most likely different. This can result in slightly different blob radii as viewed by electrons and protons. $E_{\rm p,max}$ is less than the threshold of photopion production on CMB, allowing the UHECR protons to travel cosmological distances and interact with EBL relatively close to Earth \citep{Essey_10a}. Hence, photopion interactions with EBL photons is dominant. Also, the proton injection is modeled to be a simple power law instead of an exponential cutoff power law. This makes no difference as the observed spectrum is not sensitive to the intrinsic source spectrum owing to the dominance of secondary photons from line of sight UHECR interactions \citep{Essey_11b}. However, the choice of the injection spectral index ($\alpha$) does have significant impact on the secondary neutrino spectrum. The lower energy peak becomes more and more prominent with higher values of $\alpha$ \citep[see, eg.,][]{Das_19}.

The total kinetic power in UHECRs required to explain the VHE spectrum also depends on the redshift of the sources. For higher values of $z$, the conversion of UHECR energy to $\gamma-$ray energy will be higher and the value of $f_{\rm CR}$ will be higher. This will decrease the value of $L_{\rm UHECR}$. Again, with increasing $z$, the survival rate of UHECRs along the line of sight decreases. The HBLs selected for the study spans over a wide range in redshift and is highest for 1ES 0414+009 ($z=0.287$). In this case also, the jet power required is less than that of a $10^{9}$ $M_\odot$ SMBH. It is predicted in \cite{Essey_11a} that a hard intrinsic TeV spectrum of distant blazars showing no or little attenuation can be attributed to the fact that production of secondary $\gamma-$rays, occuring near to the observer compared to the source distance, dominates the observed $\gamma-$ray signal at the VHE regime. This justifies the competency of UHECR interaction model, implemented in explaining the high-energy $\gamma-$ray spectrum from distant AGN.

In modeling the multiwavelength spectrum of blazar SEDs, an equipartition of energy density between magnetic fields and radiating particles is assumed in many studies to reduce the number of free parameters. In our pure-leptonic analysis, the emission region is far from equipartition. The ratio $u'_{\rm e}/u'_{\rm B}$ is very small in our case. The values of the magnetic field considered inside the jet is already low and decreasing them further to lower $u'_{\rm B}$ will result in a poor fit for the synchrotron spectrum. A departure from equipartition results in more luminosity requirement than minimum, as is also obtained in \cite{Zacharias_16}. But this doesn't pose a theoretical difficulty as long as $L_{\rm jet}\lesssim L_{\rm Edd}$, which is true in our case. In lepto-hadronic fits, the VHE emission being dominated by the hadronic component, the values of $(u'_{\rm e}+u'_{\rm p})/u'_{\rm B}$ are near to equipartition. The range of values obtained are between 16 and 200 and similar to that obtained for photohadronic interactions inside the jet in \cite{Cerruti_15}. However, the energy density in electrons in the lepto-hadronic case is still low, resulting in low values of $u'_{\rm e}/u'_{\rm B}$. Lowering the magnetic field value further disallows the Hillas condition from achieving the required $E_{\rm p,max}$ value \cite[see][for a detailed discussion]{Basumallick_17}. In \cite{Petropoulou_15a}, a departure from equipartition results from extremely high and dominant values of $u'_{\rm p}$ due to photohadronic interactions inside the jet.

The neutrino flux from individual BL Lacs obtained in our analysis is too low to be detected by currently operating and upcoming future detectors. A stacked flux from all BL Lacs with similar UHECR production mechanism we discussed could be interesting, however. The spectral shape is in accordance with that found in \cite{Essey_10b}. However, the neutrino flux they obtained for 1ES 0229+200 is many orders of magntude higher than that obtained in our calculations. The reason for this is the low luminosity requirement in UHECRs, as the VHE spectrum is modeled in our analysis using a comparable contribution from leptonic and hadronic counterparts at the peak. This hybrid contribution results in low $L_{\rm UHECR}$ for all the sources we have studied. The flux of secondary neutrinos is inversely proportional to redshift and the scaling is valid as long as the UHECR protons remain within the angular resolution of the detector \citep{Essey_11a}. Such a pattern is not seen in our results because of different peak VHE $\gamma-$ray flux of the sources, resulting in different UHECR luminosity requirements and hence distinct values of $L_\nu$. Given the low maximum proton energy (10 EeV) required in our model, and deflections in the EGMF and GMF, identifying UHECRs coming from individual BL Lacs we discussed will be difficult.

\section{Conclusions} \label{sec:conclusions}

The hard intrinsic TeV spectrum of high-energy peaked BL Lacs (HBLs) showing very less attenuation is an intriguing mystery in astroparticle physics. A fit to the multiwavelength SEDs of selected HBLs, over a wide redshift range, is obtained by invoking contributions from hadronic channels arising in UHECR interactions on cosmic background photons. The resulting lepto-hadronic spectrum is well equipped to explain the observed SED in the VHE regime, even for the sources at the highest redshift considered. AGN are extremely energetic astrophysical objects that can produce both $\gamma-$rays and UHECRs. However, energies beyond 10 EeV are not easily produced inside the jet. Thus, UHECR interactions on EBL dominates during cosmological propagation. For a substantial contribution from UHECR interactions, the protons must be collimated along the line of sight of the observer. The secondary particles produced from interactions of UHECRs with the EBL photons initiate EM cascades. With detailed modeling of these cascades and their propagation, we have shown that these contributions from UHECRs can fit VHE gamma-ray data beyond the applicability of the leptonic emission from the jets of a number of BL Lacs.

\section{Acknowledgments}

The work of S.R. was partially supported by the National Research Foundation (South Africa) with Grant No. 111749 (CPRR) and by the University of Johannesburg Research Council grant.

\software{\textsc{CRPropa 3} \citep{Batista_16}, DINT \citep{Lee_98, Heiter_18}, GAMERA (\url{http://libgamera.github.io/GAMERA/docs/main_page.html}).}

\appendix

\section{Effects of maximum UHECR energy $E_{\rm p,max}$} \label{app:best}

\begin{figure*}[h]
\centering
\includegraphics[width = 0.33\textwidth]{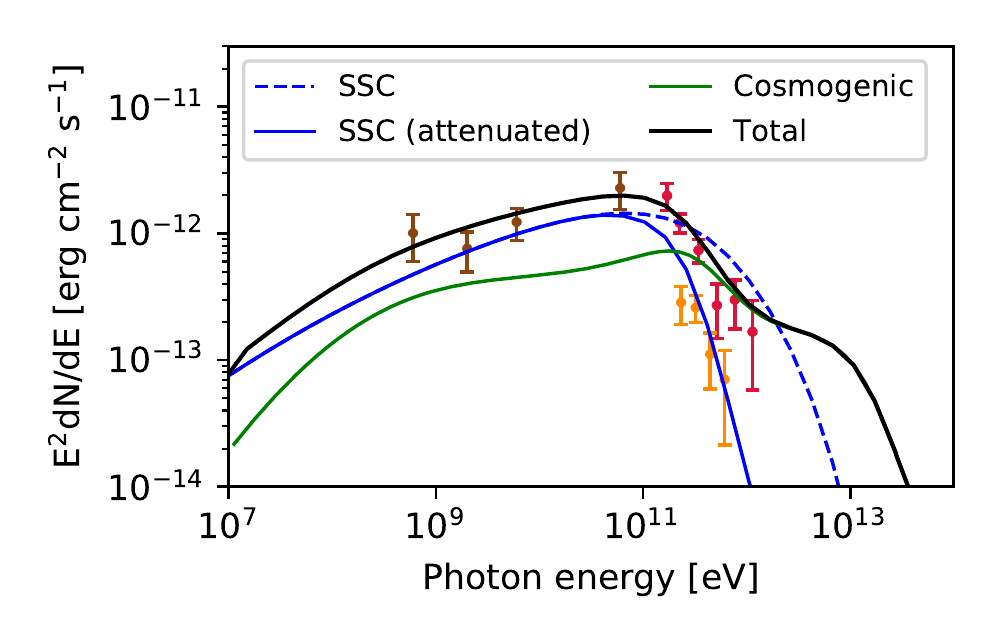}
\includegraphics[width = 0.33\textwidth]{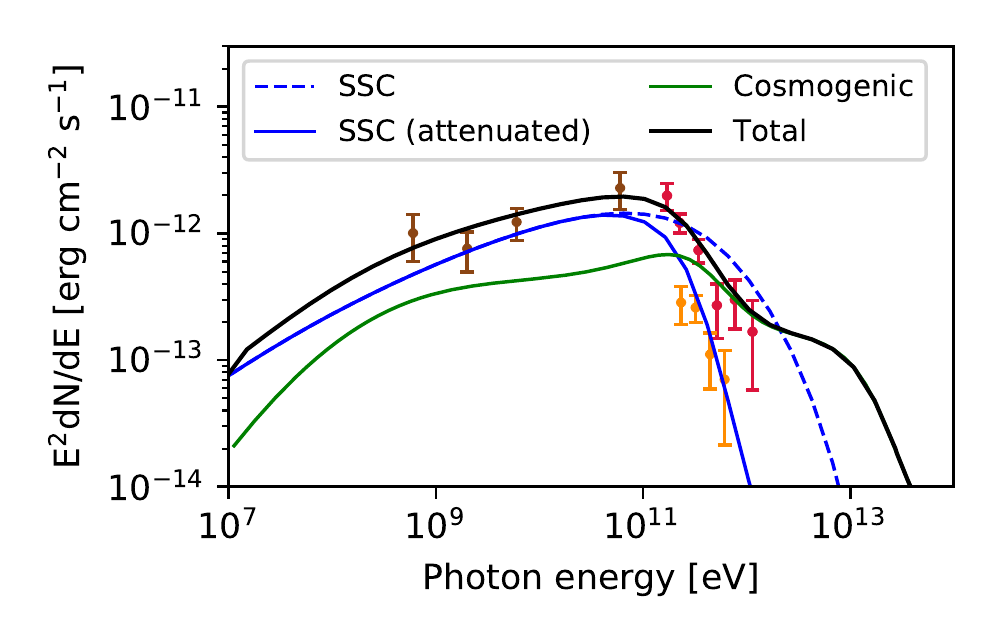}
\includegraphics[width = 0.33\textwidth]{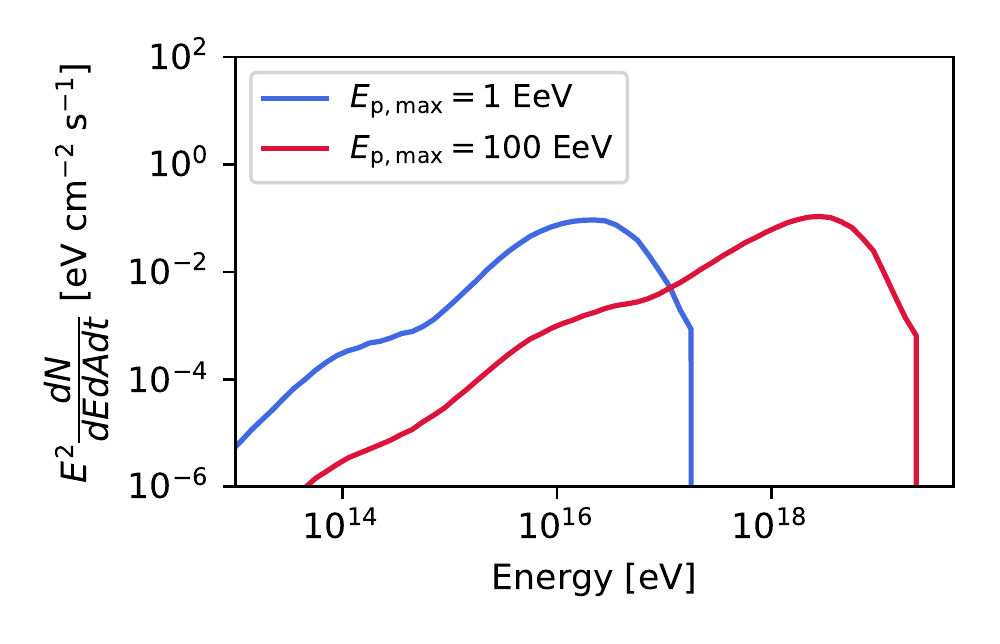}
\caption{\small{The high-energy spectrum of the HBL 1ES 0414+009, fitted with varying maximum proton energy at the source. The left and middle panels show the fit for $E_{\rm p,max}=1$ and 100 EeV, respectively. Not much variation can be seen in the spectral features. However, due to different survival rates in the EGMF and different energy conversion efficiency from UHECRs to $\gamma-$rays and neutrinos, the luminosity requirement at the source is different. The right panel shows the neutrino fluxes for both cases. See the main text for details.}}
\label{fig:epmax}
\end{figure*}

The spectrum of the observed $\gamma-$rays is governed primarily by the EM cascade of secondary particles produced from UHECR interactions on the CMB and EBL. The intrinsic source parameters have little or no effect on the VHE spectrum. However, the luminosity requirements can vary widely with source parameters, particularly $E_{\rm p,max}$, provided the UHECRs propagate for distances long enough to traverse multiple mean interaction lengths. In our work, we consider $E_{\rm p, max}=10$ EeV based on the Hillas criterion and the comparison between acceleration and escape timescales (cf. Fig.~\ref{fig:timescales}). But a different value obtained from different modeling can have a significant effect on $L_{\rm UHECR}$ due to increased or decreased interactions on CMB and EBL. We check the effects due to the choice of maximum proton energy $E_{\rm p,max}$ on luminosity requirements. For this, we consider two values, $E_{\rm p,max}=1$ and 100 EeV and fit the high-energy peak of the HBL 1ES 0414+009. This source has the highest redshift among those studied here. As a result, the effect of $E_{\rm p, max}$ variation on the survival rate of UHECRs along the line of sight, after suffering deflections in the EGMF, is expected to be the most prominent. The value of $\xi_B$ is found to be 0.259 and 0.397 for $E_{\rm p, max}$ values of 1 and 100 EeV, respectively. The UHECR injection index is taken to be $\alpha=2$, and all other parameters are kept the same as in Sec.~\ref{sec:result}. The ratio of power in produced  $\gamma-$ photons to the injected UHECR power, $f_{\rm CR}$ is also calculated. The value of $f_{\rm CR}$ comes out to be 0.0052 and 0.2559, differing by two orders of magnitude, for $E_{\rm p, max}$ values 1 and 100 EeV respectively. The resultant fits are shown in the left and middle panel of Fig.~\ref{fig:epmax}. The differences are negligible. The lower energy part of the spectrum remains unchanged and hence is not shown.

\begin{figure*}[h]
\centering
\includegraphics[width = 0.35\textwidth]{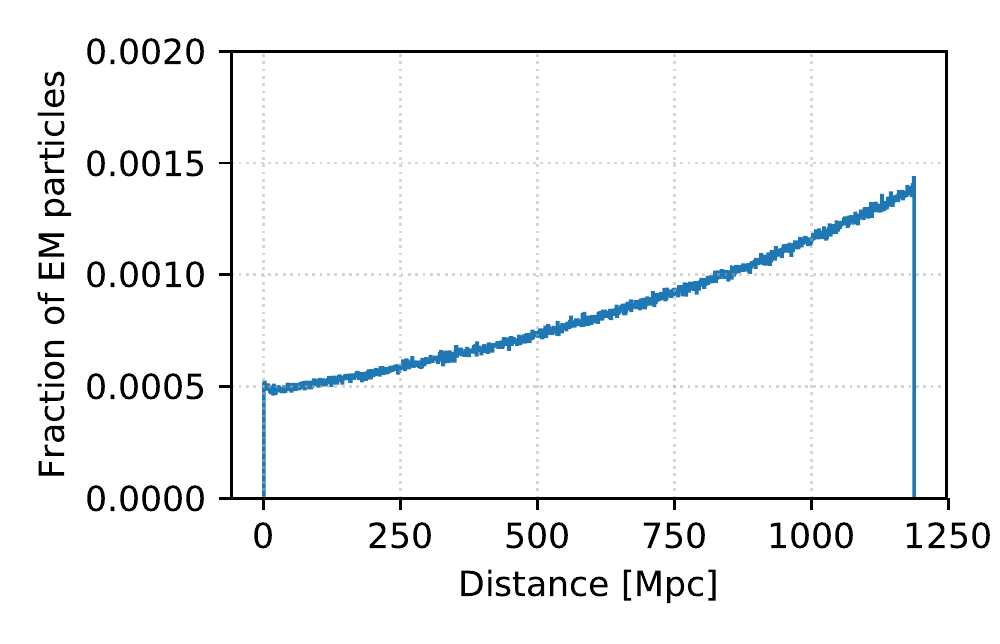}
\includegraphics[width = 0.35\textwidth]{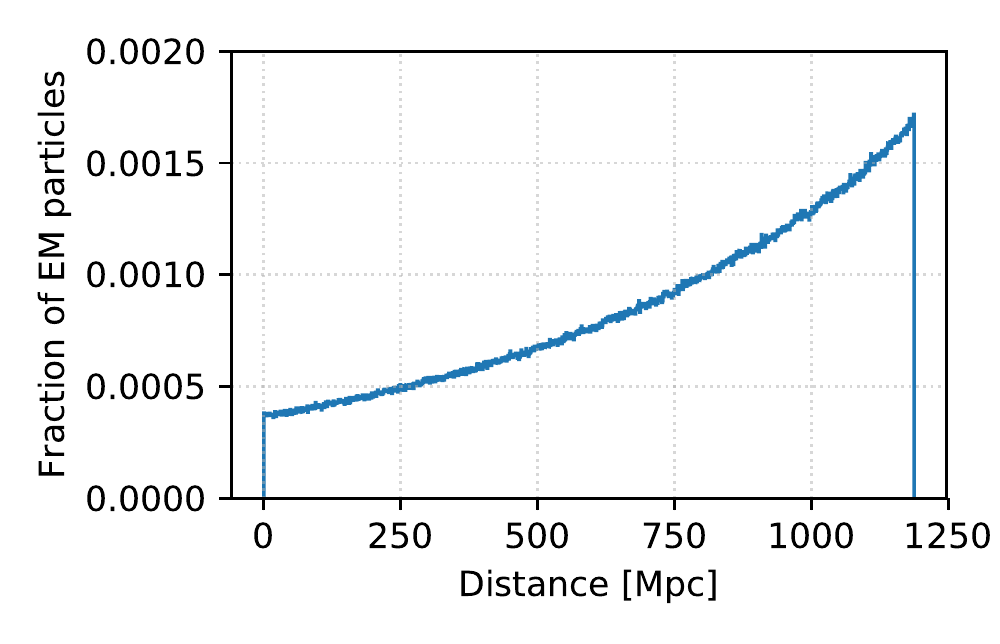}
\caption{\small{Fraction of total EM particles ($\mathrm{e^+}$, $\mathrm{e^-}$, $\gamma$) produced from UHECR interactions, binned over distance from the observer, for proton energy cutoff $E_{\rm p,max}$=1} EeV (\textit{left}) and 100 EeV (\textit{right}). The source is at $d_c\sim1189$ Mpc.}
\label{fig:em_dist}
\end{figure*}

The required luminosity is also found to differ by two orders of magnitude, being $2.47\times10^{45}$ erg/s for $E_{\rm p,max}=1$ EeV and  $3.05\times10^{43}$ erg/s for $E_{\rm p,max}=100$ EeV. The distribution of EM particles produced from UHECR interactions at distance steps of 1 Mpc from the observer to the source is shown in Fig.~\ref{fig:em_dist}. For higher $E_{\rm p,max}$, more secondaries will be produced near the sources, tending to reduce the observed $\gamma-$rays along the line of sight. But, the value of $f_{\rm CR}$ also increases for higher $E_{\rm p,max}$ due to increased photopion production on CMB, resulting in $\pi^0$ decay photons, and thus reducing the required value of $L_{\rm UHECR}$. While for lower $E_{\rm p,max}$, the main contribution to the observed $\gamma-$ray signals comes from EM cascade of $\mathrm{e^+e^-}$ pairs produced in Bethe-Heitler interactions on CMB and EBL. Thus the relative dominance of various processes changes due to varying maximum energy. This also leaves an imprint on the subsequent neutrino spectrum. Resonant photopion production on CMB with UHECRs of energy $\sim 50$ EeV produces the so-called GZK neutrinos and shifts the peak flux at higher energy. This can be seen on the right panel of Fig.~\ref{fig:epmax}.



\bibliography{blazar}{}

\providecommand{\noopsort}[1]{}\providecommand{\singleletter}[1]{#1}%
\begin{thebibliography}{}
\expandafter\ifx\csname natexlab\endcsname\relax\def\natexlab#1{#1}\fi
\providecommand{\url}[1]{\href{#1}{#1}}
\providecommand{\dodoi}[1]{doi:~\href{http://doi.org/#1}{\nolinkurl{#1}}}
\providecommand{\doeprint}[1]{\href{http://ascl.net/#1}{\nolinkurl{http://ascl.net/#1}}}
\providecommand{\doarXiv}[1]{\href{https://arxiv.org/abs/#1}{\nolinkurl{https://arxiv.org/abs/#1}}}

\bibitem[{Aab {et~al.}(2015)Aab, Abreu, Aglietta, Ahn, Al~Samarai,
  {et~al.}}]{PAO_15}
Aab, A., Abreu, P., Aglietta, M., {et~al.} 2015, Phys. Rev. D, 91, 092008,
  \dodoi{10.1103/PhysRevD.91.092008}

\bibitem[{Aartsen {et~al.}(2018)Aartsen, Ackermann, Adams, Aguilar, Ahlers, \&
  et~al.}]{IceCube_18}
Aartsen, M.~G., Ackermann, M., Adams, J., {et~al.} 2018, Phys. Rev. D, 98,
  062003, \dodoi{10.1103/PhysRevD.98.062003}

\bibitem[{{Aartsen} {et~al.}(2015){Aartsen}, {Abraham}, {Ackermann}, {Adams},
  {Aguilar}, {Ahlers}, {Ahrens}, {Altmann}, {Anderson}, {Archinger},
  {Arguelles}, {Arlen}, {Auffenberg}, {Bai}, {Barwick}, {Baum}, {Bay},
  {Beatty}, {Becker Tjus}, {Becker}, {Beiser}, {BenZvi}, {Berghaus}, {Berley},
  {Bernardini}, {Bernhard}, {Besson}, {Binder}, {Bindig}, {Bissok}, {Blaufuss},
  {Blumenthal}, {Boersma}, {Bohm}, {B{\"o}rner}, {Bos}, {Bose}, {B{\"o}ser},
  {Botner}, {Braun}, {Brayeur}, {Bretz}, {Brown}, {Buzinsky}, {Casey},
  {Casier}, {Cheung}, {Chirkin}, {Christov}, {Christy}, {Clark}, {Classen},
  {Coenders}, {Cowen}, {Cruz Silva}, {Daughhetee}, {Davis}, {Day}, {de
  Andr{\'e}}, {De Clercq}, {Dembinski}, {De Ridder}, {Desiati}, {de Vries}, {de
  Wasseige}, {de With}, {DeYoung}, {D{\'\i}az-V{\'e}lez}, {Dumm}, {Dunkman},
  {Eagan}, {Eberhardt}, {Ehrhardt}, {Eichmann}, {Euler}, {Evenson}, {Fadiran},
  {Fahey}, {Fazely}, {Fedynitch}, {Feintzeig}, {Felde}, {Filimonov}, {Finley},
  {Fischer-Wasels}, {Flis}, {Fuchs}, {Gaisser}, {Gaior}, {Gallagher},
  {Gerhardt}, {Ghorbani}, {Gier}, {Gladstone}, {Glagla}, {Gl{\"u}senkamp},
  {Goldschmidt}, {Golup}, {Gonzalez}, {Goodman}, {G{\'o}ra}, {Grant},
  {Gretskov}, {Groh}, {Gross}, {Ha}, {Haack}, {Haj Ismail}, {Hallgren},
  {Halzen}, {Hansmann}, {Hanson}, {Hebecker}, {Heereman}, {Helbing},
  {Hellauer}, {Hellwig}, {Hickford}, {Hignight}, {Hill}, {Hoffman}, {Hoffmann},
  {Holzapfel}, {Homeier}, {Hoshina}, {Huang}, {Huber}, {Huelsnitz}, {Hulth},
  {Hultqvist}, {In}, {Ishihara}, {Jacobi}, {Japaridze}, {Jero}, {Jurkovic},
  {Kaminsky}, {Kappes}, {Karg}, {Karle}, {Kauer}, {Keivani}, {Kelley}, {Kemp},
  {Kheirandish}, {Kiryluk}, {Kl{\"a}s}, {Klein}, {Kohnen}, {Kolanoski},
  {Konietz}, {Koob}, {K{\"o}pke}, {Kopper}, {Kopper}, {Koskinen}, {Kowalski},
  {Krings}, {Kroll}, {Kroll}, {Kunnen}, {Kurahashi}, {Kuwabara}, {Labare},
  {Lanfranchi}, {Larson}, {Lesiak-Bzdak}, {Leuermann}, {Leuner},
  {L{\"u}nemann}, {Madsen}, {Maggi}, {Mahn}, {Maruyama}, {Mase}, {Matis},
  {Maunu}, {McNally}, {Meagher}, {Medici}, {Meli}, {Menne}, {Merino}, {Meures},
  {Miarecki}, {Middell}, {Middlemas}, {Miller}, {Mohrmann}, {Montaruli},
  {Morse}, {Nahnhauer}, {Naumann}, {Niederhausen}, {Nowicki}, {Nygren},
  {Obertacke}, {Olivas}, {Omairat}, {O'Murchadha}, {Palczewski}, {Paul},
  {Pepper}, {P{\'e}rez de los Heros}, {Pfendner}, {Pieloth}, {Pinat},
  {Posselt}, {Price}, {Przybylski}, {P{\"u}tz}, {Quinnan}, {R{\"a}del},
  {Rameez}, {Rawlins}, {Redl}, {Reimann}, {Relich}, {Resconi}, {Rhode},
  {Richman}, {Richter}, {Riedel}, {Robertson}, {Rongen}, {Rott}, {Ruhe},
  {Ruzybayev}, {Ryckbosch}, {Saba}, {Sabbatini}, {Sand er}, {Sandrock},
  {Sandroos}, {Sarkar}, {Schatto}, {Scheriau}, {Schimp}, {Schmidt}, {Schmitz},
  {Schoenen}, {Sch{\"o}neberg}, {Sch{\"o}nwald}, {Schukraft}, {Schulte},
  {Seckel}, {Seunarine}, {Shanidze}, {Smith}, {Soldin}, {Spiczak}, {Spiering},
  {Stahlberg}, {Stamatikos}, {Stanev}, {Stanisha}, {Stasik}, {Stezelberger},
  {Stokstad}, {St{\"o}ssl}, {Strahler}, {Str{\"o}m}, {Strotjohann}, {Sullivan},
  {Sutherland}, {Taavola}, {Taboada}, {Ter-Antonyan}, {Terliuk},
  {Te{\v{s}}i{\'c}}, {Tilav}, {Toale}, {Tobin}, {Tosi}, {Tselengidou}, {Unger},
  {Usner}, {Vallecorsa}, {Vandenbroucke}, {van Eijndhoven}, {Vanheule}, {van
  Santen}, {Veenkamp}, {Vehring}, {Voge}, {Vraeghe}, {Walck}, {Wallace},
  {Wallraff}, {Wandkowsky}, {Weaver}, {Wendt}, {Westerhoff}, {Whelan},
  {Whitehorn}, {Wichary}, {Wiebe}, {Wiebusch}, {Wille}, {Williams}, {Wissing},
  {Wolf}, {Wood}, {Woschnagg}, {Xu}, {Xu}, {Xu}, {Yanez}, {Yodh}, {Yoshida},
  {Zarzhitsky}, {Zoll}, \& {IceCube Collaboration}}]{IceCube_15}
{Aartsen}, M.~G., {Abraham}, K., {Ackermann}, M., {et~al.} 2015, \apj, 809, 98,
  \dodoi{10.1088/0004-637X/809/1/98}

\bibitem[{{Aartsen} {et~al.}(2017){Aartsen}, {Abraham}, {Ackermann}, {Adams},
  {Aguilar}, {Ahlers}, {Ahrens}, {Altmann}, {Andeen}, {Anderson}, {Ansseau},
  {Anton}, {Archinger}, {Arguelles}, {Arlen}, {Auffenberg}, {Axani}, {Bai},
  {Barwick}, {Baum}, {Bay}, {Beatty}, {Becker Tjus}, {Becker}, {BenZvi},
  {Berghaus}, {Berley}, {Bernardini}, {Bernhard}, {Besson}, {Binder}, {Bindig},
  {Bissok}, {Blaufuss}, {Blot}, {Boersma}, {Bohm}, {B{\"o}rner}, {Bos}, {Bose},
  {B{\"o}ser}, {Botner}, {Braun}, {Brayeur}, {Bretz}, {Burgman}, {Casey},
  {Casier}, {Cheung}, {Chirkin}, {Christov}, {Clark}, {Classen}, {Coenders},
  {Collin}, {Conrad}, {Cowen}, {Cruz Silva}, {Daughhetee}, {Davis}, {Day}, {de
  Andr{\'e}}, {De Clercq}, {del Pino Rosendo}, {Dembinski}, {De Ridder},
  {Desiati}, {de Vries}, {de Wasseige}, {de With}, {DeYoung},
  {D{\'\i}az-V{\'e}lez}, {di Lorenzo}, {Dujmovic}, {Dumm}, {Dunkman},
  {Eberhardt}, {Ehrhardt}, {Eichmann}, {Euler}, {Evenson}, {Fahey}, {Fazely},
  {Feintzeig}, {Felde}, {Filimonov}, {Finley}, {Flis}, {F{\"o}sig},
  {Franckowiak}, {Fuchs}, {Gaisser}, {Gaior}, {Gallagher}, {Gerhardt},
  {Ghorbani}, {Giang}, {Gladstone}, {Glagla}, {Gl{\"u}senkamp}, {Goldschmidt},
  {Golup}, {Gonzalez}, {G{\'o}ra}, {Grant}, {Griffith}, {Haack}, {Haj Ismail},
  {Hallgren}, {Halzen}, {Hansen}, {Hansmann}, {Hansmann}, {Hanson}, {Hebecker},
  {Heereman}, {Helbing}, {Hellauer}, {Hickford}, {Hignight}, {Hill}, {Hoffman},
  {Hoffmann}, {Holzapfel}, {Homeier}, {Hoshina}, {Huang}, {Huber}, {Huelsnitz},
  {Hultqvist}, {In}, {Ishihara}, {Jacobi}, {Japaridze}, {Jeong}, {Jero},
  {Jones}, {Jurkovic}, {Kappes}, {Karg}, {Karle}, {Katz}, {Kauer}, {Keivani},
  {Kelley}, {Kemp}, {Kheirandish}, {Kim}, {Kintscher}, {Kiryluk}, {Kittler},
  {Klein}, {Kohnen}, {Koirala}, {Kolanoski}, {Konietz}, {K{\"o}pke}, {Kopper},
  {Kopper}, {Koskinen}, {Kowalski}, {Krings}, {Kroll}, {Kr{\"u}ckl},
  {Kr{\"u}ger}, {Kunnen}, {Kunwar}, {Kurahashi}, {Kuwabara}, {Labare},
  {Lanfranchi}, {Larson}, {Lennarz}, {Lesiak-Bzdak}, {Leuermann}, {Leuner},
  {Lu}, {L{\"u}nemann}, {Madsen}, {Maggi}, {Mahn}, {Mancina}, {Mandelartz},
  {Maruyama}, {Mase}, {Maunu}, {McNally}, {Meagher}, {Medici}, {Meier}, {Meli},
  {Menne}, {Merino}, {Meures}, {Miarecki}, {Middell}, {Mohrmann}, {Montaruli},
  {Moulai}, {Nahnhauer}, {Naumann}, {Neer}, {Niederhausen}, {Nowicki},
  {Nygren}, {Obertacke Pollmann}, {Olivas}, {Omairat}, {O'Murchadha},
  {Palczewski}, {Pandya}, {Pankova}, {Penek}, {Pepper}, {P{\'e}rez de los
  Heros}, {Pfendner}, {Pieloth}, {Pinat}, {Posselt}, {Price}, {Przybylski},
  {Quinnan}, {Raab}, {R{\"a}del}, {Rameez}, {Rawlins}, {Reimann}, {Relich},
  {Resconi}, {Rhode}, {Richman}, {Riedel}, {Robertson}, {Rongen}, {Rott},
  {Ruhe}, {Ryckbosch}, {Rysewyk}, {Sabbatini}, {Sanchez Herrera}, {Sandrock},
  {Sandroos}, {Sarkar}, {Satalecka}, {Schimp}, {Schlunder}, {Schmidt},
  {Schoenen}, {Sch{\"o}neberg}, {Sch{\"o}nwald}, {Schumacher}, {Seckel},
  {Seunarine}, {Soldin}, {Song}, {Spiczak}, {Spiering}, {Stahlberg},
  {Stamatikos}, {Stanev}, {Stasik}, {Steuer}, {Stezelberger}, {Stokstad},
  {St{\"o}{\ss}l}, {Str{\"o}m}, {Strotjohann}, {Sullivan}, {Sutherland},
  {Taavola}, {Taboada}, {Tatar}, {Ter-Antonyan}, {Terliuk}, {Te{\v{s}}i{\'c}},
  {Tilav}, {Toale}, {Tobin}, {Toscano}, {Tosi}, {Tselengidou}, {Turcati},
  {Unger}, {Usner}, {Vallecorsa}, {Vandenbroucke}, {van Eijndhoven},
  {Vanheule}, {van Rossem}, {van Santen}, {Veenkamp}, {Vehring}, {Voge},
  {Vraeghe}, {Walck}, {Wallace}, {Wallraff}, {Wandkowsky}, {Weaver}, {Wendt},
  {Westerhoff}, {Whelan}, {Wickmann}, {Wiebe}, {Wiebusch}, {Wille}, {Williams},
  {Wills}, {Wissing}, {Wolf}, {Wood}, {Woolsey}, {Woschnagg}, {Xu}, {Xu}, {Xu},
  {Yanez}, {Yodh}, {Yoshida}, {Zoll}, \& {IceCube Collaboration}}]{IceCube_17}
---. 2017, \apj, 835, 45, \dodoi{10.3847/1538-4357/835/1/45}

\bibitem[{{Abdo} {et~al.}(2010){Abdo}, {Ackermann}, {Agudo}, {Ajello}, {Aller},
  {Aller}, {Angelakis}, {Arkharov}, {Axelsson}, {Bach}, \& et~al.}]{Abdo_10}
{Abdo}, A.~A., {Ackermann}, M., {Agudo}, I., {et~al.} 2010, \apj, 716, 30,
  \dodoi{10.1088/0004-637X/716/1/30}

\bibitem[{{Abeysekara} {et~al.}(2017{\natexlab{a}})}]{HAWC_17a}
{Abeysekara}, A.~U., {et~al.} 2017{\natexlab{a}}, \apj, 843, 116,
  \dodoi{10.3847/1538-4357/aa789f}

\bibitem[{{Abeysekara} {et~al.}(2017{\natexlab{b}})}]{HAWC_17b}
---. 2017{\natexlab{b}}, \apj, 843, 40, \dodoi{10.3847/1538-4357/aa7556}

\bibitem[{{Acciari} {et~al.}(2019)}]{MAGIC_19}
{Acciari}, V.~A., {et~al.} 2019, Mon. Not. R. Astron. Soc, 486, 4233,
  \dodoi{10.1093/mnras/stz943}

\bibitem[{Ackermann {et~al.}(2013)Ackermann, Ajello, Allafort, Asano, Atwood,
  Baldini, Ballet, Barbiellini, Bastieri, Bechtol, Bellazzini, Bloom,
  Bonamente, Borgland, Bottacini, Brandt, Bregeon, Brigida, Bruel, Buehler,
  Burnett, Busetto, Buson, Caliandro, Cameron, Caraveo, Casandjian, Cecchi,
  Charles, Chaty, Chekhtman, Cheung, Chiang, Cillis, Ciprini, Claus,
  Cohen-Tanugi, Colafrancesco, Conrad, Cutini, {et~al.}}]{Ackermann_13}
Ackermann, M., Ajello, M., Allafort, A., {et~al.} 2013, \apj, 765, 54,
  \dodoi{10.1088/0004-637x/765/1/54}

\bibitem[{{Ackermann} {et~al.}(2015)}]{Fermi_3LAC}
{Ackermann}, M., {et~al.} 2015, \apj, 810, 14,
  \dodoi{10.1088/0004-637X/810/1/14}

\bibitem[{{Adams} {et~al.}(2017)}]{poemma2}
{Adams}, J., {et~al.} 2017, ArXiv e-prints.
\newblock \doarXiv{1703.04513}

\bibitem[{{Aharonian} {et~al.}(2006)}]{HESS_06}
{Aharonian}, F., {et~al.} 2006, \nat, 440, 1018, \dodoi{10.1038/nature04680}

\bibitem[{{Aharonian} {et~al.}(2007{\natexlab{a}}){Aharonian}, {Akhperjanian},
  {Barres de Almeida}, {Bazer-Bachi}, {Behera}, {Beilicke}, {Benbow},
  {Bernl{\"o}hr}, {Boisson}, {Bolz}, {Borrel}, {Braun}, {Brion}, {Brown},
  {B{\"u}hler}, {Bulik}, {B{\"u}sching}, {Boutelier}, {Carrigan}, {Chadwick},
  {Chounet}, {Clapson}, {Coignet}, {Cornils}, {Costamante}, {Dalton},
  {Degrange}, {Dickinson}, {Djannati-Ata{\"i}}, {Domainko}, {O'C.~Drury},
  {Dubois}, {Dubus}, {Dyks}, {Egberts}, {Emmanoulopoulos}, {Espigat},
  {Farnier}, {Feinstein}, {Fiasson}, {F{\"o}rster}, {Fontaine}, {Funk},
  {F{\"u}{\ss}ling}, {Gallant}, {Giebels}, {Glicenstein}, {Gl{\"u}ck}, {Goret},
  {Hadjichristidis}, {Hauser}, {Hauser}, {Heinzelmann}, {Henri}, {Hermann},
  {Hinton}, {Hoffmann}, {Hofmann}, {Holleran}, {Hoppe}, {Horns},
  {Jacholkowska}, {de Jager}, {Jung}, {Katarzy{\'n}ski}, {Kendziorra},
  {Kerschhaggl}, {Kh{\'e}lifi}, {Keogh}, {Komin}, {Kosack}, {Lamanna},
  {Latham}, {Lemi{\`e}re}, {Lemoine-Goumard}, {Lenain}, {Lohse}, {Martin},
  {Martineau-Huynh}, {Marcowith}, {Masterson}, {Maurin}, {Maurin}, {McComb},
  {Moderski}, {Moulin}, {de Naurois}, {Nedbal}, {Nolan}, {Ohm}, {Olive}, {de
  O{\~n}a Wilhelmi}, {Orford}, {Osborne}, {Ostrowski}, {Panter}, {Pedaletti},
  {Pelletier}, {Petrucci}, {Pita}, {P{\"u}hlhofer}, {Punch}, {Ranchon},
  {Raubenheimer}, {Raue}, {Rayner}, {Renaud}, {Ripken}, {Rob}, {Rolland},
  {Rosier-Lees}, {Rowell}, {Rudak}, {Ruppel}, {Sahakian}, {Santangelo},
  {Schlickeiser}, {Sch{\"o}ck}, {Schr{\"o}der}, {Schwanke}, {Schwarzburg},
  {Schwemmer}, {Shalchi}, {Sol}, {Spangler}, {Stawarz}, {Steenkamp},
  {Stegmann}, {Superina}, {Tam}, {Tavernet}, {Terrier}, {van Eldik},
  {Vasileiadis}, {Venter}, {Vialle}, {Vincent}, {Vivier}, {V{\"o}lk}, {Volpe},
  {Wagner}, {Ward}, {Zdziarski}, \& {Zech}}]{Aharonian_07a}
{Aharonian}, F., {Akhperjanian}, A.~G., {Barres de Almeida}, U., {et~al.}
  2007{\natexlab{a}}, \aap, 475, L9, \dodoi{10.1051/0004-6361:20078462}

\bibitem[{{Aharonian} {et~al.}(2007{\natexlab{b}}){Aharonian}, {Akhperjanian},
  {Bazer-Bachi}, {Beilicke}, {Benbow}, {Berge}, {Bernl{\"o}hr}, {Boisson},
  {Bolz}, {Borrel}, {Braun}, {Brion}, {Brown}, {B{\"u}hler}, {B{\"u}sching},
  {Boutelier}, {Carrigan}, {Chadwick}, {Chounet}, {Coignet}, {Cornils},
  {Costamante}, {Degrange}, {Dickinson}, {Djannati-Ata{\"i}}, {O'C.~Drury},
  {Dubus}, {Egberts}, {Emmanoulopoulos}, {Espigat}, {Farnier}, {Feinstein},
  {Ferrero}, {Fiasson}, {Fontaine}, {Funk}, {Funk}, {F{\"u}{\ss}ling},
  {Gallant}, {Giebels}, {Glicenstein}, {Gl{\"u}ck}, {Goret}, {Hadjichristidis},
  {Hauser}, {Hauser}, {Heinzelmann}, {Henri}, {Hermann}, {Hinton}, {Hoffmann},
  {Hofmann}, {Holleran}, {Hoppe}, {Horns}, {Jacholkowska}, {de Jager},
  {Kendziorra}, {Kerschhaggl}, {Kh{\'e}lifi}, {Komin}, {Kosack}, {Lamanna},
  {Latham}, {Le Gallou}, {Lemi{\`e}re}, {Lemoine-Goumard}, {Lohse}, {Martin},
  {Martineau-Huynh}, {Marcowith}, {Masterson}, {Maurin}, {McComb}, {Moulin},
  {de Naurois}, {Nedbal}, {Nolan}, {Noutsos}, {Olive}, {Orford}, {Osborne},
  {Panter}, {Pelletier}, {Petrucci}, {Pita}, {P{\"u}hlhofer}, {Punch},
  {Ranchon}, {Raubenheimer}, {Raue}, {Rayner}, {Ripken}, {Rob}, {Rolland},
  {Rosier-Lees}, {Rowell}, {Sahakian}, {Santangelo}, {Saug{\'e}}, {Schlenker},
  {Schlickeiser}, {Schr{\"o}der}, {Schwanke}, {Schwarzburg}, {Schwemmer},
  {Shalchi}, {Sol}, {Spangler}, {Spanier}, {Steenkamp}, {Stegmann}, {Superina},
  {Tam}, {Tavernet}, {Terrier}, {Tluczykont}, {van Eldik}, {Vasileiadis},
  {Venter}, {Vialle}, {Vincent}, {V{\"o}lk}, {Wagner}, \&
  {Ward}}]{Aharonian_07b}
{Aharonian}, F., {Akhperjanian}, A.~G., {Bazer-Bachi}, A.~R., {et~al.}
  2007{\natexlab{b}}, \aap, 470, 475, \dodoi{10.1051/0004-6361:20077057}

\bibitem[{{Aharonian}(2002)}]{Aharonian_02}
{Aharonian}, F.~A. 2002, \mnras, 332, 215,
  \dodoi{10.1046/j.1365-8711.2002.05292.x}

\bibitem[{{Ahnen} {et~al.}(2016){Ahnen}, {Ansoldi}, {Antonelli}, {Antoranz},
  {Babic}, {Banerjee}, {Bangale}, {Barres de Almeida}, {Barrio}, {Becerra
  Gonz{\'a}lez}, {Bednarek}, {Bernardini}, {Biasuzzi}, {Biland}, {Blanch},
  {Bonnefoy}, {Bonnoli}, {Borracci}, {Bretz}, {Carmona}, {Carosi},
  {Chatterjee}, {Clavero}, {Colin}, {Colombo}, {Contreras}, {Cortina},
  {Covino}, {Da Vela}, {Dazzi}, {De Angelis}, {De Caneva}, {De Lotto}, {de
  O{\~n}a Wilhelmi}, {Delgado Mendez}, {Di Pierro}, {Dominis Prester},
  {Dorner}, {Doro}, {Einecke}, {Elsaesser}, {Fern{\'a}ndez-Barral}, {Fidalgo},
  {Fonseca}, {Font}, {Frantzen}, {Fruck}, {Galindo}, {Garc{\'{\i}}a L{\'o}pez},
  {Garczarczyk}, {Garrido Terrats}, {Gaug}, {Giammaria}, {Eisenacher Glawion},
  {Godinovi{\'c}}, {Gonz{\'a}lez Mu{\~n}oz}, {Guberman}, {Hanabata},
  {Hayashida}, {Herrera}, {Hose}, {Hrupec}, {Hughes}, {Idec}, {Kodani},
  {Konno}, {Kubo}, {Kushida}, {La Barbera}, {Lelas}, {Lindfors}, {Lombardi},
  {Longo}, {L{\'o}pez}, {L{\'o}pez-Coto}, {L{\'o}pez-Oramas}, {Lorenz},
  {Majumdar}, {Makariev}, {Mallot}, {Maneva}, {Manganaro}, {Mannheim},
  {Maraschi}, {Marcote}, {Mariotti}, {Mart{\'{\i}}nez}, {Mazin}, {Menzel},
  {Miranda}, {Mirzoyan}, {Moralejo}, {Nakajima}, {Neustroev}, {Niedzwiecki},
  {Nievas Rosillo}, {Nilsson}, {Nishijima}, {Noda}, {Orito}, {Overkemping},
  {Paiano}, {Palacio}, {Palatiello}, {Paneque}, {Paoletti}, {Paredes},
  {Paredes-Fortuny}, {Persic}, {Poutanen}, {Prada Moroni}, {Prandini},
  {Puljak}, {Reinthal}, {Rhode}, {Rib{\'o}}, {Rico}, {Rodriguez Garcia},
  {R{\"u}gamer}, {Saito}, {Satalecka}, {Scapin}, {Schultz}, {Schweizer},
  {Shore}, {Sillanp{\"a}{\"a}}, {Sitarek}, {Snidaric}, {Sobczynska},
  {Stamerra}, {Steinbring}, {Strzys}, {Takalo}, {Takami}, {Tavecchio},
  {Temnikov}, {Terzi{\'c}}, {Tescaro}, {Teshima}, {Thaele}, {Torres}, {Toyama},
  {Treves}, {Verguilov}, {Vovk}, {Ward}, {Will}, {Wu}, {Zanin}, {Lucarelli},
  {Pittori}, {Vercellone}, {Berdyugin}, {Carini}, {L{\"a}hteenm{\"a}ki},
  {Pasanen}, {Pease}, {Sainio}, {Tornikoski}, \& {Walters}}]{Ahnen_16}
{Ahnen}, M.~L., {Ansoldi}, S., {Antonelli}, L.~A., {et~al.} 2016, \mnras, 459,
  2286, \dodoi{10.1093/mnras/stw710}

\bibitem[{Albert {et~al.}(2007)Albert, Aliu, Anderhub, Antoranz, Armada,
  Baixeras, Barrio, Bartko, Bastieri, Becker, Bednarek, Berger, Bigongiari,
  Biland, Bock, Bordas, Bosch-Ramon, Bretz, Britvitch, Camara, Carmona,
  Chilingarian, Coarasa, Commichau, Contreras, Cortina, Costado, Curtef,
  Danielyan, Dazzi, Angelis, Delgado, de~los Reyes, Lotto,
  Domingo-Santamar{\'{\i}}a, Dorner, Doro, Errando, Fagiolini, Ferenc,
  Fern{\'{a}}ndez, Firpo, Flix, Fonseca, Font, Fuchs, Galante,
  Garc{\'{\i}}a-L{\'{o}}pez, Garczarczyk, Gaug, Giller, Goebel, Hakobyan,
  Hayashida, Hengstebeck, Herrero, Höhne, Hose, Hsu, Jacon, Jogler, Kosyra,
  Kranich, Kritzer, Laille, Lindfors, Lombardi, Longo, L{\'{o}}pez,
  L{\'{o}}pez, Lorenz, Majumdar, Maneva, Mannheim, Mansutti, Mariotti,
  Mart{\'{\i}}nez, Mazin, Merck, Meucci, Meyer, Miranda, Mirzoyan, Mizobuchi,
  Moralejo, Nieto, Nilsson, Ninkovic, O{\~{n}}a-Wilhelmi, Otte, Oya, Paneque,
  Panniello, Paoletti, Paredes, Pasanen, Pascoli, Pauss, Pegna, Perlman,
  Persic, Peruzzo, Piccioli, Prandini, Puchades, Raymers, Rhode, Rib{\'{o}},
  Rico, Rissi, Robert, Rügamer, Saggion, Saito, S{\'{a}}nchez, Sartori,
  Scalzotto, Scapin, Schmitt, Schweizer, Shayduk, Shinozaki, Shore, Sidro,
  Sillanpää, Sobczynska, Stamerra, Stark, Takalo, Tavecchio, Temnikov,
  Tescaro, Teshima, Torres, Turini, Vankov, Vitale, Wagner, Wibig, Wittek,
  Zandanel, Zanin, \& Zapatero}]{Albert_07}
Albert, J., Aliu, E., Anderhub, H., {et~al.} 2007, \apj, 667, L21,
  \dodoi{10.1086/521982}

\bibitem[{{Aliu} {et~al.}(2012){Aliu}, {Archambault}, {Arlen}, {Aune},
  {Beilicke}, {Benbow}, {B{\"o}ttcher}, {Bouvier}, {Bugaev}, {Cannon},
  {Cesarini}, {Ciupik}, {Collins-Hughes}, {Connolly}, {Cui}, {Dickherber},
  {Dumm}, {Errando}, {Falcone}, {Federici}, {Feng}, {Finley}, {Finnegan},
  {Fortson}, {Furniss}, {Galante}, {Gall}, {Godambe}, {Griffin}, {Grube},
  {Gyuk}, {Hanna}, {Holder}, {Huan}, {Hughes}, {Hui}, {Imran}, {Jameil},
  {Kaaret}, {Karlsson}, {Kertzman}, {Kerr}, {Khassen}, {Kieda}, {Krawczynski},
  {Krennrich}, {Lang}, {Lee}, {Madhavan}, {Majumdar}, {McArthur}, {McCann},
  {Moriarty}, {Mukherjee}, {Nelson}, {O'Faol{\'a}in de Bhr{\'o}ithe}, {Ong},
  {Orr}, {Otte}, {Park}, {Perkins}, {Pichel}, {Pohl}, {Quinn}, {Ragan},
  {Reynolds}, {Roache}, {Ruppel}, {Saxon}, {Schroedter}, {Sembroski},
  {{\c{S}}ent{\"u}rk}, {Smith}, {Staszak}, {Stroh}, {Telezhinsky},
  {Te{\v{s}}i{\'c}}, {Theiling}, {Thibadeau}, {Tsurusaki}, {Varlotta},
  {Vassiliev}, {Vivier}, {Wakely}, {Ward}, {Weinstein}, {Welsing}, {Williams},
  \& {Zitzer}}]{Aliu_12}
{Aliu}, E., {Archambault}, S., {Arlen}, T., {et~al.} 2012, \apj, 755, 118,
  \dodoi{10.1088/0004-637X/755/2/118}

\bibitem[{{Aliu} {et~al.}(2014){Aliu}, {Archambault}, {Arlen}, {Aune},
  {Behera}, {Beilicke}, {Benbow}, {Berger}, {Bird}, {Bouvier}, {Buckley},
  {Bugaev}, {Byrum}, {Cerruti}, {Chen}, {Ciupik}, {Connolly}, {Cui}, {Duke},
  {Dumm}, {Errand o}, {Falcone}, {Federici}, {Feng}, {Finley}, {Fleischhack},
  {Fortin}, {Fortson}, {Furniss}, {Galante}, {Gilland ers}, {Griffin},
  {Griffiths}, {Grube}, {Gyuk}, {Hanna}, {Holder}, {Hughes}, {Humensky},
  {Johnson}, {Kaaret}, {Kertzman}, {Khassen}, {Kieda}, {Krawczynski},
  {Krennrich}, {Lang}, {Madhavan}, {Maier}, {Majumdar}, {McArthur}, {McCann},
  {Meagher}, {Millis}, {Moriarty}, {Mukherjee}, {Nieto}, {O'Faol{\'a}in de
  Bhr{\'o}ithe}, {Ong}, {Otte}, {Park}, {Perkins}, {Pohl}, {Popkow}, {Prokoph},
  {Quinn}, {Ragan}, {Reyes}, {Reynolds}, {Richards}, {Roache}, {Sembroski},
  {Smith}, {Staszak}, {Telezhinsky}, {Theiling}, {Varlotta}, {Vassiliev},
  {Vincent}, {Wakely}, {Weekes}, {Weinstein}, {Welsing}, {Williams}, {Zajczyk},
  \& {Zitzer}}]{Aliu_14}
---. 2014, \apj, 782, 13, \dodoi{10.1088/0004-637X/782/1/13}

\bibitem[{{Alves Batista} {et~al.}(2016){Alves Batista}, {Dundovic}, {Erdmann},
  {Kampert}, {Kuempel}, {M{\"u}ller}, {Sigl}, {van Vliet}, {Walz}, \&
  {Winchen}}]{Batista_16}
{Alves Batista}, R., {Dundovic}, A., {Erdmann}, M., {et~al.} 2016, Journal of
  Cosmology and Astroparticle Physics, 2016, 038,
  \dodoi{10.1088/1475-7516/2016/05/038}

\bibitem[{Basumallick \& Gupta(2017)}]{Basumallick_17}
Basumallick, P.~P., \& Gupta, N. 2017, \apj, 844, 58,
  \dodoi{10.3847/1538-4357/aa7a12}

\bibitem[{Baumgartner {et~al.}(2013)Baumgartner, Tueller, Markwardt, Skinner,
  Barthelmy, Mushotzky, Evans, \& Gehrels}]{Baumgartner_13}
Baumgartner, W.~H., Tueller, J., Markwardt, C.~B., {et~al.} 2013, The
  Astrophysical Journal Supplement Series, 207, 19,
  \dodoi{10.1088/0067-0049/207/2/19}

\bibitem[{{Beckmann, V.} {et~al.}(2002){Beckmann, V.}, {Wolter, A.}, {Celotti,
  A.}, {Costamante, L.}, {Ghisellini, G.}, {Maccacaro, T.}, \& {Tagliaferri,
  G.}}]{Beckmann_02}
{Beckmann, V.}, {Wolter, A.}, {Celotti, A.}, {et~al.} 2002, A\&A, 383, 410,
  \dodoi{10.1051/0004-6361:20011752}

\bibitem[{{Berezinskii} \& {Grigor'eva}(1988)}]{Berezinskii_88}
{Berezinskii}, V.~S., \& {Grigor'eva}, S.~I. 1988, \aap, 199, 1

\bibitem[{{Berezinsky} {et~al.}(2006){Berezinsky}, {Gazizov}, \&
  {Grigorieva}}]{Berezinsky_06}
{Berezinsky}, V., {Gazizov}, A., \& {Grigorieva}, S. 2006, \prd, 74, 043005,
  \dodoi{10.1103/PhysRevD.74.043005}

\bibitem[{{Bianchi} {et~al.}(2011){Bianchi}, {Efremova}, {Herald}, {Girardi},
  {Zabot}, {Marigo}, \& {Martin}}]{Bianchi_11}
{Bianchi}, L., {Efremova}, B., {Herald}, J., {et~al.} 2011, \mnras, 411, 2770,
  \dodoi{10.1111/j.1365-2966.2010.17890.x}

\bibitem[{{B{\"o}ttcher} \& {Els}(2016)}]{Bottcher_16}
{B{\"o}ttcher}, M., \& {Els}, P. 2016, \apj, 821, 102,
  \dodoi{10.3847/0004-637X/821/2/102}

\bibitem[{{B{\"o}ttcher} {et~al.}(2013){B{\"o}ttcher}, {Reimer}, {Sweeney}, \&
  {Prakash}}]{Bottcher_13}
{B{\"o}ttcher}, M., {Reimer}, A., {Sweeney}, K., \& {Prakash}, A. 2013, \apj,
  768, 54, \dodoi{10.1088/0004-637X/768/1/54}

\bibitem[{{Buckley} {et~al.}(1985){Buckley}, {Tuohy}, \&
  {Remillard}}]{Buckley_85}
{Buckley}, D.~A.~H., {Tuohy}, I.~R., \& {Remillard}, R.~A. 1985, Proceedings of
  the Astronomical Society of Australia, 6, 147

\bibitem[{{Celotti} \& {Ghisellini}(2008)}]{Celotti_08}
{Celotti}, A., \& {Ghisellini}, G. 2008, Mon. Not. R. Astron. Soc, 385, 283,
  \dodoi{10.1111/j.1365-2966.2007.12758.x}

\bibitem[{{Cerruti}(2013)}]{Cerruti_13}
{Cerruti}, M. 2013, arXiv e-prints, arXiv:1307.8091.
\newblock \doarXiv{1307.8091}

\bibitem[{{Cerruti} {et~al.}(2015){Cerruti}, {Zech}, {Boisson}, \&
  {Inoue}}]{Cerruti_15}
{Cerruti}, M., {Zech}, A., {Boisson}, C., \& {Inoue}, S. 2015, \mnras, 448,
  910, \dodoi{10.1093/mnras/stu2691}

\bibitem[{{Chodorowski} {et~al.}(1992){Chodorowski}, {Zdziarski}, \&
  {Sikora}}]{Chodorowski_92}
{Chodorowski}, M.~J., {Zdziarski}, A.~A., \& {Sikora}, M. 1992, \apj, 400, 181,
  \dodoi{10.1086/171984}

\bibitem[{{Condon} {et~al.}(1998){Condon}, {Cotton}, {Greisen}, {Yin},
  {Perley}, {Taylor}, \& {Broderick}}]{Condon_98}
{Condon}, J.~J., {Cotton}, W.~D., {Greisen}, E.~W., {et~al.} 1998, \aj, 115,
  1693, \dodoi{10.1086/300337}

\bibitem[{{Costamante} {et~al.}(2018){Costamante}, {Bonnoli}, {Tavecchio},
  {Ghisellini}, {Tagliaferri}, \& {Khangulyan}}]{Costamante_18}
{Costamante}, L., {Bonnoli}, G., {Tavecchio}, F., {et~al.} 2018, \mnras, 477,
  4257, \dodoi{10.1093/mnras/sty857}

\bibitem[{{Das} {et~al.}(2019){Das}, {Razzaque}, \& {Gupta}}]{Das_19}
{Das}, S., {Razzaque}, S., \& {Gupta}, N. 2019, \prd, 99, 083015,
  \dodoi{10.1103/PhysRevD.99.083015}

\bibitem[{{D'Elia} {et~al.}(2013){D'Elia}, {Perri}, {Puccetti}, {Capalbi},
  {Giommi}, {Burrows}, {Campana}, {Tagliaferri}, {Cusumano}, {Evans},
  {Gehrels}, {Kennea}, {Moretti}, {Nousek}, {Osborne}, {Romano}, \&
  {Stratta}}]{D'Elia_13}
{D'Elia}, V., {Perri}, M., {Puccetti}, S., {et~al.} 2013, \aap, 551, A142,
  \dodoi{10.1051/0004-6361/201220863}

\bibitem[{Dermer \& Menon(2009)}]{Dermer&Menon}
Dermer, C.~D., \& Menon, G. 2009, {High energy radiation from black holes:
  gamma rays, cosmic rays, and neutrinos}, Princeton series in astrophysics
  (Princeton, NJ: Princeton Univ. Press).
\newblock \url{https://cds.cern.ch/record/1225453}

\bibitem[{{Dermer} {et~al.}(2009){Dermer}, {Razzaque}, {Finke}, \&
  {Atoyan}}]{Dermer_09}
{Dermer}, C.~D., {Razzaque}, S., {Finke}, J.~D., \& {Atoyan}, A. 2009, New
  Journal of Physics, 11, 065016, \dodoi{10.1088/1367-2630/11/6/065016}

\bibitem[{{Dom{\'\i}nguez} {et~al.}(2011){Dom{\'\i}nguez}, {Primack},
  {Rosario}, {Prada}, {Gilmore}, {Faber}, {Koo}, {Somerville},
  {P{\'e}rez-Torres}, {P{\'e}rez-Gonz{\'a}lez}, {Huang}, {Davis},
  {Guhathakurta}, {Barmby}, {Conselice}, {Lozano}, {Newman}, \&
  {Cooper}}]{Dominguez_11}
{Dom{\'\i}nguez}, A., {Primack}, J.~R., {Rosario}, D.~J., {et~al.} 2011,
  \mnras, 410, 2556, \dodoi{10.1111/j.1365-2966.2010.17631.x}

\bibitem[{Essey {et~al.}(2011{\natexlab{a}})Essey, Ando, \&
  Kusenko}]{Essey_11b}
Essey, W., Ando, S., \& Kusenko, A. 2011{\natexlab{a}}, Astroparticle Physics,
  35, 135 , \dodoi{https://doi.org/10.1016/j.astropartphys.2011.06.010}

\bibitem[{Essey {et~al.}(2011{\natexlab{b}})Essey, Kalashev, Kusenko, \&
  Beacom}]{Essey_11a}
Essey, W., Kalashev, O., Kusenko, A., \& Beacom, J.~F. 2011{\natexlab{b}},
  \apj, 731, 51, \dodoi{10.1088/0004-637x/731/1/51}

\bibitem[{Essey {et~al.}(2010)Essey, Kalashev, Kusenko, \& Beacom}]{Essey_10b}
Essey, W., Kalashev, O.~E., Kusenko, A., \& Beacom, J.~F. 2010, Phys. Rev.
  Lett., 104, 141102, \dodoi{10.1103/PhysRevLett.104.141102}

\bibitem[{{Essey} \& {Kusenko}(2010)}]{Essey_10a}
{Essey}, W., \& {Kusenko}, A. 2010, Astroparticle Physics, 33, 81,
  \dodoi{10.1016/j.astropartphys.2009.11.007}

\bibitem[{{Falomo} {et~al.}(2003){Falomo}, {Carangelo}, \&
  {Treves}}]{Falomo_03}
{Falomo}, R., {Carangelo}, N., \& {Treves}, A. 2003, \mnras, 343, 505,
  \dodoi{10.1046/j.1365-8711.2003.06690.x}

\bibitem[{{Fang} {et~al.}(2017)}]{grand2}
{Fang}, K., {et~al.} 2017, ArXiv e-prints, ICRC2017, 996.
\newblock \doarXiv{1708.05128}

\bibitem[{{Fermi-LAT Collaboration} {et~al.}(2015){Fermi-LAT Collaboration},
  {Acero}, {et~al.}}]{Fermi_3FGL}
{Fermi-LAT Collaboration}, {Acero}, F., {et~al.} 2015, 218, 23,
  \dodoi{10.1088/0067-0049/218/2/23}

\bibitem[{{Finke}(2019)}]{Finke_19}
{Finke}, J.~D. 2019, \apj, 870, 28, \dodoi{10.3847/1538-4357/aaf00c}

\bibitem[{{Finke} {et~al.}(2010){Finke}, {Razzaque}, \& {Dermer}}]{Finke_10}
{Finke}, J.~D., {Razzaque}, S., \& {Dermer}, C.~D. 2010, \apj, 712, 238,
  \dodoi{10.1088/0004-637X/712/1/238}

\bibitem[{Fraija {et~al.}(2018)Fraija, Aguilar-Ruiz, Galván-Gámez, Marinelli,
  \& de Diego}]{Fraija_18}
Fraija, N., Aguilar-Ruiz, E., Galván-Gámez, A., Marinelli, A., \& de Diego,
  J.~A. 2018, Mon. Not. R. Astron. Soc, 481, 4461,
  \dodoi{10.1093/mnras/sty2561}

\bibitem[{{Ghisellini} {et~al.}(1993){Ghisellini}, {Padovani}, {Celotti}, \&
  {Maraschi}}]{Ghisellini_93}
{Ghisellini}, G., {Padovani}, P., {Celotti}, A., \& {Maraschi}, L. 1993, \apj,
  407, 65, \dodoi{10.1086/172493}

\bibitem[{{Gilmore} {et~al.}(2012){Gilmore}, {Somerville}, {Primack}, \&
  {Dom{\'\i}nguez}}]{Gilmore_12}
{Gilmore}, R.~C., {Somerville}, R.~S., {Primack}, J.~R., \& {Dom{\'\i}nguez},
  A. 2012, Mon. Not. R. Astron. Soc, 422, 3189,
  \dodoi{10.1111/j.1365-2966.2012.20841.x}

\bibitem[{{Giommi} {et~al.}(1995){Giommi}, {Ansari}, \& {Micol}}]{Giommi_95}
{Giommi}, P., {Ansari}, S.~G., \& {Micol}, A. 1995, \aaps, 109, 267

\bibitem[{{Giommi} {et~al.}(2012){Giommi}, {Polenta}, {L{\"a}hteenm{\"a}ki},
  {Thompson}, {Capalbi}, {Cutini}, {Gasparrini}, {Gonz{\'a}lez-Nuevo},
  {Le{\'o}n-Tavares}, {L{\'o}pez-Caniego}, {Mazziotta}, {Monte}, {Perri},
  {Rain{\`o}}, {Tosti}, {Tramacere}, {Verrecchia}, {Aller}, {Aller},
  {Angelakis}, {Bastieri}, {Berdyugin}, {Bonaldi}, {Bonavera}, {Burigana},
  {Burrows}, {Buson}, {Cavazzuti}, {Chincarini}, {Colafrancesco}, {Costamante},
  {Cuttaia}, {D'Ammando}, {de Zotti}, {Frailis}, {Fuhrmann}, {Galeotta},
  {Gargano}, {Gehrels}, {Giglietto}, {Giordano}, {Giroletti}, {Keih{\"a}nen},
  {King}, {Krichbaum}, {Lasenby}, {Lavonen}, {Lawrence}, {Leto}, {Lindfors},
  {Mandolesi}, {Massardi}, {Max-Moerbeck}, {Michelson}, {Mingaliev}, {Natoli},
  {Nestoras}, {Nieppola}, {Nilsson}, {Partridge}, {Pavlidou}, {Pearson},
  {Procopio}, {Rachen}, {Readhead}, {Reeves}, {Reimer}, {Reinthal},
  {Ricciardi}, {Richards}, {Riquelme}, {Saarinen}, {Sajina}, {Sandri},
  {Savolainen}, {Sievers}, {Sillanp{\"a}{\"a}}, {Sotnikova}, {Stevenson},
  {Tagliaferri}, {Takalo}, {Tammi}, {Tavagnacco}, {Terenzi}, {Toffolatti},
  {Tornikoski}, {Trigilio}, {Turunen}, {Umana}, {Ungerechts}, {Villa}, {Wu},
  {Zacchei}, {Zensus}, \& {Zhou}}]{Giommi_12}
{Giommi}, P., {Polenta}, G., {L{\"a}hteenm{\"a}ki}, A., {et~al.} 2012, \aap,
  541, A160, \dodoi{10.1051/0004-6361/201117825}

\bibitem[{{Gregory} {et~al.}(1996){Gregory}, {Scott}, {Douglas}, \&
  {Condon}}]{Gregory_96}
{Gregory}, P.~C., {Scott}, W.~K., {Douglas}, K., \& {Condon}, J.~J. 1996,
  \apjs, 103, 427, \dodoi{10.1086/192282}

\bibitem[{Greisen(1966)}]{Greisen_66}
Greisen, K. 1966, Phys. Rev. Lett., 16, 748, \dodoi{10.1103/PhysRevLett.16.748}

\bibitem[{{Gursky} {et~al.}(1978){Gursky}, {Bradt}, {Doxsey}, {Schwartz},
  {Schwarz}, {Dower}, {Fabbiano}, {Griffiths}, {Johnston}, {Leach}, {Ramsey},
  \& {Spada}}]{Gursky_78}
{Gursky}, H., {Bradt}, H., {Doxsey}, R., {et~al.} 1978, \apj, 223, 973,
  \dodoi{10.1086/156329}

\bibitem[{{H.~E.~S.~S. Collaboration} {et~al.}(2014){H.~E.~S.~S.
  Collaboration}, {Abramowski}, {Aharonian}, {Ait Benkhali}, {Akhperjanian},
  {Ang{\"u}ner}, {Anton}, {Backes}, {Balenderan}, {Balzer}, {Barnacka},
  {Becherini}, {Becker Tjus}, {Bernl{\"o}hr}, {Birsin}, {Bissaldi}, {Biteau},
  {B{\"o}ttcher}, {Boisson}, {Bolmont}, {Bordas}, {Brucker}, {Brun}, {Brun},
  {Bulik}, {Carrigan}, {Casanova}, {Chadwick}, {Chalme-Calvet}, {Chaves},
  {Cheesebrough}, {Chr{\'e}tien}, {Colafrancesco}, {Cologna}, {Conrad},
  {Couturier}, {Cui}, {Dalton}, {Daniel}, {Davids}, {Degrange}, {Deil},
  {deWilt}, {Dickinson}, {Djannati-At{\"a}{\i}}, {Domainko}, {Drury}, {Dubus},
  {Dutson}, {Dyks}, {Dyrda}, {Edwards}, {Egberts}, {Eger}, {Espigat},
  {Farnier}, {Fegan}, {Feinstein}, {Fernand es}, {Fernandez}, {Fiasson},
  {Fontaine}, {F{\"o}rster}, {F{\"u}{\ss}ling}, {Gajdus}, {Gallant},
  {Garrigoux}, {Giavitto}, {Giebels}, {Glicenstein}, {Grondin},
  {Grudzi{\'n}ska}, {H{\"a}ffner}, {Hahn}, {Harris}, {Heinzelmann}, {Henri},
  {Hermann}, {Hervet}, {Hillert}, {Hinton}, {Hofmann}, {Hofverberg}, {Holler},
  {Horns}, {Jacholkowska}, {Jahn}, {Jamrozy}, {Janiak}, {Jankowsky}, {Jung},
  {Kastendieck}, {Katarzy{\'n}ski}, {Katz}, {Kaufmann}, {Kh{\'e}lifi},
  {Kieffer}, {Klepser}, {Klochkov}, {Klu{\'z}niak}, {Kneiske}, {Kolitzus},
  {Komin}, {Kosack}, {Krakau}, {Krayzel}, {Kr{\"u}ger}, {Laffon}, {Lamanna},
  {Lefaucheur}, {Lemie`re}, {Lemoine-Goumard}, {Lenain}, {Lohse}, {Lopatin},
  {Lu}, {Marandon}, {Marcowith}, {Marx}, {Maurin}, {Maxted}, {Mayer}, {McComb},
  {M{\'e}hault}, {Meintjes}, {Menzler}, {Meyer}, {Moderski}, {Mohamed},
  {Moulin}, {Murach}, {Naumann}, {de Naurois}, {Niemiec}, {Nolan}, {Oakes},
  {Odaka}, {Ohm}, {de O{\~n}a Wilhelmi}, {Opitz}, {Ostrowski}, {Oya}, {Panter},
  {Parsons}, {Paz Arribas}, {Pekeur}, {Pelletier}, {Perez}, {Petrucci},
  {Peyaud}, {Pita}, {Poon}, {P{\"u}hlhofer}, {Punch}, {Quirrenbach}, {Raab},
  {Raue}, {Reichardt}, {Reimer}, {Reimer}, {Renaud}, {de los Reyes}, {Rieger},
  {Rob}, {Romoli}, {Rosier-Lees}, {Rowell}, {Rudak}, {Rulten}, {Sahakian},
  {Sanchez}, {Santangelo}, {Schlickeiser}, {Sch{\"u}ssler}, {Schulz},
  {Schwanke}, {Schwarzburg}, {Schwemmer}, {Sol}, {Spengler}, {Spies},
  {Stawarz}, {Steenkamp}, {Stegmann}, {Stinzing}, {Stycz}, {Sushch},
  {Tavernet}, {Tavernier}, {Taylor}, {Terrier}, {Tluczykont}, {Trichard},
  {Valerius}, {van Eldik}, {van Soelen}, {Vasileiadis}, {Venter}, {Viana},
  {Vincent}, {V{\"o}lk}, {Volpe}, {Vorster}, {Vuillaume}, {Wagner}, {Wagner},
  {Wagner}, {Ward}, {Weidinger}, {Weitzel}, {White}, {Wierzcholska},
  {Willmann}, {W{\"o}rnlein}, {Wouters}, {Yang}, {Zabalza}, {Zacharias},
  {Zdziarski}, {Zech}, {Zechlin}, \& {Malyshev}}]{Abramowski_14}
{H.~E.~S.~S. Collaboration}, {Abramowski}, A., {Aharonian}, F., {et~al.} 2014,
  \aap, 562, A145, \dodoi{10.1051/0004-6361/201322510}

\bibitem[{{Halpern} {et~al.}(1991){Halpern}, {Chen}, {Madejski}, \&
  {Chanan}}]{Halpern_91}
{Halpern}, J.~P., {Chen}, V.~S., {Madejski}, G.~M., \& {Chanan}, G.~A. 1991,
  \aj, 101, 818, \dodoi{10.1086/115725}

\bibitem[{{Heiter} {et~al.}(2018){Heiter}, {Kuempel}, {Walz}, \&
  {Erdmann}}]{Heiter_18}
{Heiter}, C., {Kuempel}, D., {Walz}, D., \& {Erdmann}, M. 2018, Astroparticle
  Physics, 102, 39, \dodoi{10.1016/j.astropartphys.2018.05.003}

\bibitem[{{H.E.S.S.~Collaboration} {et~al.}(2012){H.E.S.S.~Collaboration},
  {Abramowski}, {Acero}, {Aharonian}, {Akhperjanian}, {Anton}, {Balzer},
  {Barnacka}, {Barres de Almeida}, {Becherini}, {Becker}, {Behera},
  {Bernloehr}, {Birsin}, {Biteau}, {Bochow}, {Boisson}, {Bolmont}, {Bordas},
  {Brucker}, {Brun}, {Brun}, {Bulik}, {Buesching}, {Carrigan}, {Casanova},
  {Cerruti}, {Chadwick}, {Charbonnier}, {Chaves}, {Cheesebrough}, {Chounet},
  {Clapson}, {Coignet}, {Cologna}, {Conrad}, {Dalton}, {Daniel}, {Davids},
  {Degrange}, {Deil}, {Dickinson}, {Djannati-Ataie}, {Domainko}, {Drury},
  {Dubois}, {Dubus}, {Dutson}, {Dyks}, {Dyrda}, {Egberts}, {Eger}, {Espigat},
  {Fallon}, {Farnier}, {Feinstein}, {Fernandes}, {Fiasson}, {Fontaine},
  {Foerster}, {Fuesling}, {Gallant}, {Gast}, {Gerard}, {Gerbig}, {Giebels},
  {Glicenstein}, {Glueck}, {Goret}, {Goering}, {Haeffner}, {Hague}, {Hampf},
  {Hauser}, {Heinz}, {Heinzelmann}, {Henri}, {Hermann}, {Hinton}, {Hoffmann},
  {Hofmann}, {Hofverberg}, {Holler}, {Horns}, {Jacholkowska}, {de Jager},
  {Jahn}, {Jamrozy}, {Jung}, {Kastendieck}, {Katarzynski}, {Katz}, {Kaufmann},
  {Keogh}, {Khangulyan}, {Khelifi}, {Klochkov}, {Kluzniak}, {Kneiske}, {Komin},
  {Kosack}, {Kossakowski}, {Laffon}, {Lamanna}, {Lennarz}, {Lohse}, {Lopatin},
  {Lu}, {Marandon}, {Marcowith}, {Masbou}, {Maurin}, {Maxted}, {Mayer},
  {McComb}, {Medina}, {Mehault}, {Moderski}, {Moulin}, {Naumann},
  {Naumann-Godo}, {de Naurois}, {Nedbal}, {Nekrassov}, {Nguyen}, {Nicholas},
  {Niemiec}, {Nolan}, {Ohm}, {de Ona Wilhelmi}, {Opitz}, {Ostrowski}, {Oya},
  {Panter}, {Paz Arribas}, {Pedaletti}, {Pelletier}, {Petrucci}, {Pita},
  {Puehlhofer}, {Punch}, {Quirrenbach}, {Raue}, {Rayner}, {Reimer}, {Reimer},
  {Renaud}, {de Los Reyes}, {Rieger}, {Ripken}, {Rob}, {Rosier-Lees}, {Rowell},
  {Rudak}, {Rulten}, {Ruppel}, {Sahakian}, {Sanchez}, {Santangelo},
  {Schlickeiser}, {Schoeck}, {Schulz}, {Schwanke}, {Schwarzburg}, {Schwemmer},
  {Sheidaei}, {Sikora}, {Skilton}, {Sol}, {Spengler}, {Stawarz}, {Steenkamp},
  {Stegmann}, {Stinzing}, {Stycz}, {Sushch}, {Szostek}, {Tavernet}, {Terrier},
  {Tluczykont}, {Valerius}, {van Eldik}, {Vasileiadis}, {Venter}, {Vialle},
  {Viana}, {Vincent}, {Voelk}, {Volpe}, {Vorobiov}, {Vorster}, {Wagner},
  {Ward}, {White}, {Wierzcholska}, {Zacharias}, {Zajczyk}, {Zdziarski}, {Zech},
  {Zechlin}, {Costamante}, {Fegan}, \& {Ajello}}]{Abramowski_12}
{H.E.S.S.~Collaboration}, {Abramowski}, A., {Acero}, F., {et~al.} 2012, \aap,
  538, A103, \dodoi{10.1051/0004-6361/201118406}

\bibitem[{Hillas(1984)}]{Hillas_84}
Hillas, A.~M. 1984, Annual Review of Astronomy and Astrophysics, 22, 425,
  \dodoi{10.1146/annurev.aa.22.090184.002233}

\bibitem[{{Jansson} \& {Farrar}(2012)}]{Jansson_12}
{Jansson}, R., \& {Farrar}, G.~R. 2012, \apj, 757, 14,
  \dodoi{10.1088/0004-637X/757/1/14}

\bibitem[{{Joshi} \& {Gupta}(2013)}]{Joshi_13}
{Joshi}, J.~C., \& {Gupta}, N. 2013, \prd, 87, 023002,
  \dodoi{10.1103/PhysRevD.87.023002}

\bibitem[{{Kalashev} {et~al.}(2013){Kalashev}, {Kusenko}, \&
  {Essey}}]{Kalashev_13}
{Kalashev}, O.~E., {Kusenko}, A., \& {Essey}, W. 2013, \prl, 111, 041103,
  \dodoi{10.1103/PhysRevLett.111.041103}

\bibitem[{{Kaufmann} {et~al.}(2011){Kaufmann}, {Wagner}, {Tibolla}, \&
  {Hauser}}]{Kaufmann_11}
{Kaufmann}, S., {Wagner}, S.~J., {Tibolla}, O., \& {Hauser}, M. 2011, \aap,
  534, A130, \dodoi{10.1051/0004-6361/201117215}

\bibitem[{{Kelner} \& {Aharonian}(2008)}]{Kelner_08}
{Kelner}, S.~R., \& {Aharonian}, F.~A. 2008, \prd, 78, 034013,
  \dodoi{10.1103/PhysRevD.78.034013}

\bibitem[{{Lagage} \& {Cesarsky}(1983)}]{Lagage_83}
{Lagage}, P.~O., \& {Cesarsky}, C.~J. 1983, 125, 249

\bibitem[{{Lee}(1998)}]{Lee_98}
{Lee}, S. 1998, \prd, 58, 043004, \dodoi{10.1103/PhysRevD.58.043004}

\bibitem[{{Maccagni} {et~al.}(1978){Maccagni}, {Tarenghi}, {Cooke},
  {Maccacaro}, {Pye}, {Ricketts}, \& {Chincarini}}]{Maccagni_78}
{Maccagni}, D., {Tarenghi}, M., {Cooke}, B.~A., {et~al.} 1978, \aap, 62, 127

\bibitem[{{Mannheim} \& {Biermann}(1992)}]{Manheim_92}
{Mannheim}, K., \& {Biermann}, P.~L. 1992, \aap, 253, L21

\bibitem[{{Martineau-Huynh} {et~al.}(2017)}]{grand1}
{Martineau-Huynh}, O., {et~al.} 2017, in European Physical Journal Web of
  Conferences, Vol. 135, European Physical Journal Web of Conferences, 02001,
  \dodoi{10.1051/epjconf/201713502001}

\bibitem[{{McHardy} {et~al.}(1981){McHardy}, {Lawrence}, {Pye}, \&
  {Pounds}}]{McHardy_81}
{McHardy}, I.~M., {Lawrence}, A., {Pye}, J.~P., \& {Pounds}, K.~A. 1981,
  \mnras, 197, 893, \dodoi{10.1093/mnras/197.4.893}

\bibitem[{{M{\"u}cke} {et~al.}(2000){M{\"u}cke}, {Engel}, {Rachen},
  {Protheroe}, \& {Stanev}}]{Mucke_00}
{M{\"u}cke}, A., {Engel}, R., {Rachen}, J.~P., {Protheroe}, R.~J., \& {Stanev},
  T. 2000, Computer Physics Communications, 124, 290,
  \dodoi{10.1016/S0010-4655(99)00446-4}

\bibitem[{{M{\"u}cke} {et~al.}(2003){M{\"u}cke}, {Protheroe}, {Engel},
  {Rachen}, \& {Stanev}}]{Mucke_03}
{M{\"u}cke}, A., {Protheroe}, R.~J., {Engel}, R., {Rachen}, J.~P., \& {Stanev},
  T. 2003, Astroparticle Physics, 18, 593,
  \dodoi{10.1016/S0927-6505(02)00185-8}

\bibitem[{Murase {et~al.}(2012)Murase, Dermer, Takami, \& Migliori}]{Murase_12}
Murase, K., Dermer, C.~D., Takami, H., \& Migliori, G. 2012, The Astrophysical
  Journal, 749, 63, \dodoi{10.1088/0004-637x/749/1/63}

\bibitem[{Murase {et~al.}(2018)Murase, Oikonomou, \& Petropoulou}]{Murase_18}
Murase, K., Oikonomou, F., \& Petropoulou, M. 2018, \apj, 865, 124,
  \dodoi{10.3847/1538-4357/aada00}

\bibitem[{{Myers} {et~al.}(2003){Myers}, {Jackson}, {Browne}, {de Bruyn},
  {Pearson}, {Readhead}, {Wilkinson}, {Biggs}, {Blandford}, {Fassnacht},
  {Koopmans}, {Marlow}, {McKean}, {Norbury}, {Phillips}, {Rusin}, {Shepherd},
  \& {Sykes}}]{Myers_03}
{Myers}, S.~T., {Jackson}, N.~J., {Browne}, I.~W.~A., {et~al.} 2003, \mnras,
  341, 1, \dodoi{10.1046/j.1365-8711.2003.06256.x}

\bibitem[{{Neronov} \& {Vovk}(2010)}]{Neronov_10}
{Neronov}, A., \& {Vovk}, I. 2010, Science, 328, 73,
  \dodoi{10.1126/science.1184192}

\bibitem[{{Nieppola} {et~al.}(2007){Nieppola}, {Tornikoski},
  {L{\"a}hteenm{\"a}ki}, {Valtaoja}, {Hakala}, {Hovatta}, {Kotiranta},
  {Nummila}, {Ojala}, {Parviainen}, {Ranta}, {Saloranta}, {Torniainen}, \&
  {Tr{\"o}ller}}]{Nieppola_07}
{Nieppola}, E., {Tornikoski}, M., {L{\"a}hteenm{\"a}ki}, A., {et~al.} 2007,
  \aj, 133, 1947, \dodoi{10.1086/512609}

\bibitem[{{Olinto} {et~al.}(2017)}]{poemma1}
{Olinto}, A.~V., {et~al.} 2017, ArXiv e-prints, ICRC2017, 542.
\newblock \doarXiv{1708.07599}

\bibitem[{{Padovani} \& {Giommi}(1995)}]{Padovani_95}
{Padovani}, P., \& {Giommi}, P. 1995, \apj, 444, 567, \dodoi{10.1086/175631}

\bibitem[{{Petropoulou} {et~al.}(2015){Petropoulou}, {Dimitrakoudis},
  {Padovani}, {Mastichiadis}, \& {Resconi}}]{Petropoulou_15b}
{Petropoulou}, M., {Dimitrakoudis}, S., {Padovani}, P., {Mastichiadis}, A., \&
  {Resconi}, E. 2015, \mnras, 448, 2412, \dodoi{10.1093/mnras/stv179}

\bibitem[{{Petropoulou} \& {Mastichiadis}(2012)}]{Petropoulou_12}
{Petropoulou}, M., \& {Mastichiadis}, A. 2012, \mnras, 426, 462,
  \dodoi{10.1111/j.1365-2966.2012.21720.x}

\bibitem[{{Petropoulou} \& {Mastichiadis}(2015)}]{Petropoulou_15a}
---. 2015, Mon. Not. R. Astron. Soc, 447, 36, \dodoi{10.1093/mnras/stu2364}

\bibitem[{{Pierre Auger Collaboration} {et~al.}(2008){Pierre Auger
  Collaboration}, {Abraham}, {et~al.}}]{PAO_08}
{Pierre Auger Collaboration}, {Abraham}, J., {et~al.} 2008, Astroparticle
  Physics, 29, 188, \dodoi{10.1016/j.astropartphys.2008.01.002}

\bibitem[{{Pierre Auger Collaboration} {et~al.}(2017){Pierre Auger
  Collaboration}, {Aab}, {Abreu}, {Aglietta}, {Samarai}, {Albuquerque},
  {Allekotte}, {Almela}, {Alvarez Castillo}, {Alvarez-Mu{\~n}iz}, {Anastasi},
  {Anchordoqui}, {Andrada}, {Andringa}, {Aramo}, {Arqueros}, {Arsene},
  {Asorey}, {Assis}, {Aublin}, {Avila}, {Badescu}, {Balaceanu}, {Barbato},
  {Barreira Luz}, {Beatty}, {Becker}, {Bellido}, {Berat}, {Bertaina}, {Bertou},
  {Biermann}, {Billoir}, {Biteau}, {Blaess}, {Blanco}, {Blazek}, {Bleve},
  {Boh{\'a}{\v{c}}ov{\'a}}, {Boncioli}, {Bonifazi}, {Borodai}, {Botti},
  {Brack}, {Brancus}, {Bretz}, {Bridgeman}, {Briechle}, {Buchholz}, {Bueno},
  {Buitink}, {Buscemi}, {Caballero-Mora}, {Caccianiga}, {Cancio}, {Canfora},
  {Caramete}, {Caruso}, {Castellina}, {Cataldi}, {Cazon}, {Chavez},
  {Chinellato}, {Chudoba}, {Clay}, {Cobos}, {Colalillo}, {Coleman}, {Collica},
  {Coluccia}, {Concei{\c{c}}{\~a}o}, {Consolati}, {Contreras}, {Cooper},
  {Coutu}, {Covault}, {Cronin}, {D'Amico}, {Daniel}, {Dasso}, {Daumiller},
  {Dawson}, {de Almeida}, {de Jong}, {De Mauro}, {de Mello Neto}, {De Mitri},
  {de Oliveira}, {de Souza}, {Debatin}, {Deligny}, {Di Giulio}, {Di Matteo},
  {D{\'\i}az Castro}, {Diogo}, {Dobrigkeit}, {D'Olivo}, {Dorosti}, {dos Anjos},
  {Dova}, {Dundovic}, {Ebr}, {Engel}, {Erdmann}, {Erfani}, {Escobar},
  {Espadanal}, {Etchegoyen}, {Falcke}, {Farrar}, {Fauth}, {Fazzini}, {Fenu},
  {Fick}, {Figueira}, {Filip{\v{c}}i{\v{c}}}, {Fratu}, {Freire}, {Fujii},
  {Fuster}, {Gaior}, {Garc{\'\i}a}, {Garcia-Pinto}, {Gat{\'e}}, {Gemmeke},
  {Gherghel-Lascu}, {Ghia}, {Giaccari}, {Giammarchi}, {Giller}, {G{\l}as},
  {Glaser}, {Golup}, {G{\'o}mez Berisso}, {G{\'o}mez Vitale}, {Gonz{\'a}lez},
  {Gorgi}, {Gorham}, {Grillo}, {Grubb}, {Guarino}, {Guedes}, {Hampel},
  {Hansen}, {Harari}, {Harrison}, {Harton}, {Haungs}, {Hebbeker}, {Heck},
  {Heimann}, {Herve}, {Hill}, {Hojvat}, {Holt}, {Homola}, {H{\"o}randel},
  {Horvath}, {Hrabovsk{\'y}}, {Huege}, {Hulsman}, {Insolia}, {Isar}, {Jandt},
  {Jansen}, {Johnsen}, {Josebachuili}, {Jurysek}, {K{\"a}{\"a}p{\"a}},
  {Kambeitz}, {Kampert}, {Katkov}, {Keilhauer}, {Kemmerich}, {Kemp}, {Kemp},
  {Kieckhafer}, {Klages}, {Kleifges}, {Kleinfeller}, {Krause}, {Krohm},
  {Kuempel}, {Kukec Mezek}, {Kunka}, {Kuotb Awad}, {LaHurd}, {Lauscher},
  {Legumina}, {Leigui de Oliveira}, {Letessier-Selvon}, {Lhenry-Yvon}, {Link},
  {Lo Presti}, {Lopes}, {L{\'o}pez}, {L{\'o}pez Casado}, {Luce}, {Lucero},
  {Malacari}, {Mallamaci}, {Mandat}, {Mantsch}, {Mariazzi},
  {Mari{\textcommabelow s}}, {Marsella}, {Martello}, {Martinez}, {Mart{\'\i}nez
  Bravo}, {Mas{\'\i}as Meza}, {Mathes}, {Mathys}, {Matthews}, {Matthews},
  {Matthiae}, {Mayotte}, {Mazur}, {Medina}, {Medina-Tanco}, {Melo},
  {Menshikov}, {Merenda}, {Michal}, {Micheletti}, {Middendorf}, {Miramonti},
  {Mitrica}, {Mockler}, {Mollerach}, {Montanet}, {Morello}, {Mostaf{\'a}},
  {M{\"u}ller}, {M{\"u}ller}, {Muller}, {M{\"u}ller}, {Mussa}, {Naranjo},
  {Nellen}, {Nguyen}, {Niculescu-Oglinzanu}, {Niechciol}, {Niemietz},
  {Niggemann}, {Nitz}, {Nosek}, {Novotny}, {No{\v{z}}ka}, {N{\'u}{\~n}ez},
  {Ochilo}, {Oikonomou}, {Olinto}, {Palatka}, {Pallotta}, {Papenbreer},
  {Parente}, {Parra}, {Paul}, {Pech}, {Pedreira}, {Pkala}, {Pelayo},
  {Pe{\~n}a-Rodriguez}, {Pereira}, {Perl{\'\i}n}, {Perrone}, {Peters},
  {Petrera}, {Phuntsok}, {Piegaia}, {Pierog}, {Pieroni}, {Pimenta},
  {Pirronello}, {Platino}, {Plum}, {Porowski}, {Prado}, {Privitera}, {Prouza},
  {Quel}, {Querchfeld}, {Quinn}, {Ramos-Pollan}, {Rautenberg}, {Ravignani},
  {Revenu}, {Ridky}, {Riehn}, {Risse}, {Ristori}, {Rizi}, {Rodrigues de
  Carvalho}, {Rodriguez Fernandez}, {Rodriguez Rojo}, {Rogozin}, {Roncoroni},
  {Roth}, {Roulet}, {Rovero}, {Ruehl}, {Saffi}, {Saftoiu}, {Salamida},
  {Salazar}, {Saleh}, {Salesa Greus}, {Salina}, {S{\'a}nchez}, {Sanchez-Lucas},
  {Santos}, {Santos}, {Sarazin}, {Sarmento}, {Sarmiento}, {Sato}, {Schauer},
  {Scherini}, {Schieler}, {Schimp}, {Schmidt}, {Scholten}, {Schov{\'a}nek},
  {Schr{\"o}der}, {Schulz}, {Schumacher}, {Sciutto}, {Segreto}, {Settimo},
  {Shadkam}, {Shellard}, {Sigl}, {Silli}, {Sima}, {{\'S}mia{\l}kowski},
  {{\v{S}}m{\'\i}da}, {Snow}, {Sommers}, {Sonntag}, {Sorokin}, {Squartini},
  {Stanca}, {Stani{\v{c}}}, {Stasielak}, {Stassi}, {Strafella}, {Suarez},
  {Suarez Dur{\'a}n}, {Sudholz}, {Suomij{\"a}rvi}, {Supanitsky},
  {{\v{S}}up{\'\i}k}, {Swain}, {Szadkowski}, {Taboada}, {Taborda}, {Tapia},
  {Theodoro}, {Timmermans}, {Todero Peixoto}, {Tomankova}, {Tom{\'e}},
  {Torralba Elipe}, {Travnicek}, {Trini}, {Ulrich}, {Unger}, {Urban},
  {Vald{\'e}s Galicia}, {Vali{\~n}o}, {Valore}, {van Aar}, {van Bodegom}, {van
  den Berg}, {van Vliet}, {Varela}, {Vargas C{\'a}rdenas}, {Varner},
  {V{\'a}zquez}, {Veberi{\v{c}}}, {Ventura}, {Vergara Quispe}, {Verzi},
  {Vicha}, {Villase{\~n}or}, {Vorobiov}, {Wahlberg}, {Wainberg}, {Walz},
  {Watson}, {Weber}, {Weindl}, {Wiencke}, {Wilczy{\'n}ski}, {Wirtz},
  {Wittkowski}, {Wundheiler}, {Yang}, {Yushkov}, {Zas}, {Zavrtanik},
  {Zavrtanik}, {Zepeda}, {Zimmermann}, {Ziolkowski}, {Zong}, \&
  {Zuccarello}}]{Aab_17}
{Pierre Auger Collaboration}, {Aab}, A., {Abreu}, P., {et~al.} 2017, Science,
  357, 1266, \dodoi{10.1126/science.aan4338}

\bibitem[{{Pierre Auger Collaboration} {et~al.}(2018){Pierre Auger
  Collaboration}, {Aab}, {Abreu}, {Aglietta}, {Albuquerque}, {Albury},
  {Allekotte}, {Almela}, {Alvarez Castillo}, {Alvarez-Mu{\~n}iz}, {Anastasi},
  {Anchordoqui}, {Andrada}, {Andringa}, {Aramo}, {Asorey}, {Assis}, {Avila},
  {Badescu}, {Balaceanu}, {Barbato}, {Barreira Luz}, {Baur}, {Becker},
  {Bellido}, {Berat}, {Bertaina}, {Bertou}, {Biermann}, {Biteau}, {Blaess},
  {Blanco}, {Blazek}, {Bleve}, {Boh{\'a}{\v{c}}ov{\'a}}, {Bonifazi}, {Borodai},
  {Botti}, {Brack}, {Bretz}, {Bridgeman}, {Briechle}, {Buchholz}, {Bueno},
  {Buitink}, {Buscemi}, {Caballero-Mora}, {Caccianiga}, {Calcagni}, {Cancio},
  {Canfora}, {Carceller}, {Caruso}, {Castellina}, {Catalani}, {Cataldi},
  {Cazon}, {Chinellato}, {Chudoba}, {Chytka}, {Clay}, {Cobos Cerutti},
  {Colalillo}, {Coleman}, {Coluccia}, {Concei{\c{c}}{\~a}o}, {Consolati},
  {Contreras}, {Cooper}, {Coutu}, {Covault}, {Daniel}, {Dasso}, {Daumiller},
  {Dawson}, {Day}, {de Almeida}, {de Jong}, {De Mauro}, {de Mello Neto}, {De
  Mitri}, {de Oliveira}, {de Souza}, {Debatin}, {Deligny}, {Dhital}, {D{\'\i}az
  Castro}, {Diogo}, {Dobrigkeit}, {D'Olivo}, {Dorosti}, {dos Anjos}, {Dova},
  {Dundovic}, {Ebr}, {Engel}, {Erdmann}, {Escobar}, {Etchegoyen}, {Falcke},
  {Farmer}, {Farrar}, {Fauth}, {Fazzini}, {Feldbusch}, {Fenu}, {Ferreyro},
  {Figueira}, {Filip{\v{c}}i{\v{c}}}, {Freire}, {Fujii}, {Fuster},
  {Garc{\'\i}a}, {Gemmeke}, {Gherghel-Lascu}, {Ghia}, {Giaccari}, {Giammarchi},
  {Giller}, {G{\l}as}, {Glombitza}, {Golup}, {G{\'o}mez Berisso}, {G{\'o}mez
  Vitale}, {Gonz{\'a}lez}, {Goos}, {G{\'o}ra}, {Gorgi}, {Gottowik}, {Grubb},
  {Guarino}, {Guedes}, {Guido}, {Halliday}, {Hampel}, {Hansen}, {Harari},
  {Harrison}, {Harvey}, {Haungs}, {Hebbeker}, {Heck}, {Heimann}, {Hill},
  {Hojvat}, {Holt}, {Homola}, {H{\"o}rand el}, {Horvath}, {Hrabovsk{\'y}},
  {Huege}, {Hulsman}, {Insolia}, {Isar}, {Jandt}, {Johnsen}, {Josebachuili},
  {Jurysek}, {K{\"a}{\"a}p{\"a}}, {Kampert}, {Keilhauer}, {Kemmerich}, {Kemp},
  {Klages}, {Kleifges}, {Kleinfeller}, {Krause}, {Kuempel}, {Kukec Mezek},
  {Kuotb Awad}, {Lago}, {LaHurd}, {Lang}, {Legumina}, {Leigui de Oliveira},
  {Lenok}, {Letessier-Selvon}, {Lhenry-Yvon}, {Lo Presti}, {Lopes},
  {L{\'o}pez}, {L{\'o}pez Casado}, {Lorek}, {Luce}, {Lucero}, {Malacari},
  {Mallamaci}, {Mancarella}, {Mandat}, {Mantsch}, {Mariazzi},
  {Mari{\textcommabelow s}}, {Marsella}, {Martello}, {Martinez}, {Mart{\'\i}nez
  Bravo}, {Mathes}, {Mathys}, {Matthews}, {Matthiae}, {Mayotte}, {Mazur},
  {Medina-Tanco}, {Melo}, {Menshikov}, {Merenda}, {Michal}, {Micheletti},
  {Middendorf}, {Miramonti}, {Mitrica}, {Mockler}, {Mollerach}, {Montanet},
  {Morello}, {Morlino}, {Mostaf{\'a}}, {M{\"u}ller}, {Muller}, {M{\"u}ller},
  {Mussa}, {Nellen}, {Nguyen}, {Niculescu-Oglinzanu}, {Niechciol}, {Nitz},
  {Nosek}, {Novotny}, {No{\v{z}}ka}, {Nucita}, {N{\'u}{\~n}ez}, {Olinto},
  {Palatka}, {Pallotta}, {Papenbreer}, {Parente}, {Parra}, {Pech}, {Pedreira},
  {P{\c{e}}kala}, {Pelayo}, {Pe{\~n}a-Rodriguez}, {Pereira}, {Perlin},
  {Perrone}, {Peters}, {Petrera}, {Phuntsok}, {Pierog}, {Pimenta},
  {Pirronello}, {Platino}, {Poh}, {Pont}, {Porowski}, {Prado}, {Privitera},
  {Prouza}, {Puyleart}, {Querchfeld}, {Quinn}, {Ramos-Pollan}, {Rautenberg},
  {Ravignani}, {Reininghaus}, {Ridky}, {Riehn}, {Risse}, {Ristori}, {Rizi},
  {Rodrigues de Carvalho}, {Rodriguez Rojo}, {Roncoroni}, {Roth}, {Roulet},
  {Rovero}, {Ruehl}, {Saffi}, {Saftoiu}, {Salamida}, {Salazar}, {Saleh},
  {Salina}, {S{\'a}nchez}, {Santos}, {Santos}, {Sarazin}, {Sarmento},
  {Sarmiento-Cano}, {Sato}, {Savina}, {Schauer}, {Scherini}, {Schieler},
  {Schimassek}, {Schimp}, {Schmidt}, {Scholten}, {Schov{\'a}nek},
  {Schr{\"o}der}, {Schr{\"o}der}, {Schumacher}, {Sciutto}, {Shellard}, {Sigl},
  {Silli}, {Sima}, {{\v{S}}m{\'\i}da}, {Snow}, {Sommers}, {Soriano},
  {Souchard}, {Squartini}, {Stanca}, {Stani{\v{c}}}, {Stasielak}, {Stassi},
  {Stolpovskiy}, {Streich}, {Suarez}, {Su{\'a}rez-Dur{\'a}n}, {Sudholz},
  {Suomij{\"a}rvi}, {Supanitsky}, {{\v{S}}up{\'\i}k}, {Szadkowski}, {Taboada},
  {Taborda}, {Tapia}, {Timmermans}, {Todero Peixoto}, {Tom{\'e}}, {Torralba
  Elipe}, {Travnicek}, {Trini}, {Tueros}, {Ulrich}, {Unger}, {Urban},
  {Vald{\'e}s Galicia}, {Vali{\~n}o}, {Valore}, {van Bodegom}, {van den Berg},
  {van Vliet}, {Varela}, {Vargas C{\'a}rdenas}, {V{\'a}zquez}, {Veberi{\v{c}}},
  {Ventura}, {Vergara Quispe}, {Verzi}, {Vicha}, {Villase{\~n}or}, {Vorobiov},
  {Wahlberg}, {Wainberg}, {Watson}, {Weber}, {Weindl}, {Wiede{\'n}ski},
  {Wiencke}, {Wilczy{\'n}ski}, {Wirtz}, {Wittkowski}, {Wundheiler}, {Yang},
  {Yushkov}, {Zas}, {Zavrtanik}, {Zavrtanik}, {Zehrer}, {Zepeda}, {Zimmermann},
  {Ziolkowski}, {Zong}, {Zuccarello}, \& {Pierre Auger Collaboration}}]{Aab_18}
---. 2018, \apj, 868, 4, \dodoi{10.3847/1538-4357/aae689}

\bibitem[{{Planck Collaboration} {et~al.}(2014){Planck Collaboration}, {Ade},
  {Aghanim}, {Arg{\"u}eso}, {Armitage-Caplan}, {Arnaud}, {Ashdown},
  {Atrio-Barandela}, {Aumont}, {Baccigalupi}, \& et~al.}]{Ade_14}
{Planck Collaboration}, {Ade}, P.~A.~R., {Aghanim}, N., {et~al.} 2014, \aap,
  571, A28, \dodoi{10.1051/0004-6361/201321524}

\bibitem[{{Prince} {et~al.}(2019){Prince}, {Gupta}, \& {Nalewajko}}]{Prince_19}
{Prince}, R., {Gupta}, N., \& {Nalewajko}, K. 2019, \apj, 883, 137,
  \dodoi{10.3847/1538-4357/ab3afa}

\bibitem[{{Pushkarev} {et~al.}(2009){Pushkarev}, {Kovalev}, {Lister}, \&
  {Savolainen}}]{Pushkarev_09}
{Pushkarev}, A.~B., {Kovalev}, Y.~Y., {Lister}, M.~L., \& {Savolainen}, T.
  2009, 507, L33, \dodoi{10.1051/0004-6361/200913422}

\bibitem[{Razzaque {et~al.}(2009)Razzaque, Dermer, \& Finke}]{Razzaque_09}
Razzaque, S., Dermer, C.~D., \& Finke, J.~D. 2009, The Astrophysical Journal,
  697, 483, \dodoi{10.1088/0004-637x/697/1/483}

\bibitem[{{Razzaque} {et~al.}(2012){Razzaque}, {Dermer}, \&
  {Finke}}]{Razzaque_12}
{Razzaque}, S., {Dermer}, C.~D., \& {Finke}, J.~D. 2012, \apj, 745, 196,
  \dodoi{10.1088/0004-637X/745/2/196}

\bibitem[{Razzaque {et~al.}(2004)Razzaque, Meszaros, \& Zhang}]{Razzaque_04}
Razzaque, S., Meszaros, P., \& Zhang, B. 2004, \apj, 613, 1072,
  \dodoi{10.1086/423166}

\bibitem[{{Remillard} {et~al.}(1989){Remillard}, {Tuohy}, {Brissenden},
  {Buckley}, {Schwartz}, {Feigelson}, \& {Tapia}}]{Remillard_89}
{Remillard}, R.~A., {Tuohy}, I.~R., {Brissenden}, R.~J.~V., {et~al.} 1989,
  \apj, 345, 140, \dodoi{10.1086/167888}

\bibitem[{{Sahu} {et~al.}(2017){Sahu}, {de Le{\'o}n}, \& {Mirand a}}]{Sahu_17}
{Sahu}, S., {de Le{\'o}n}, A.~R., \& {Mirand a}, L.~S. 2017, European Physical
  Journal C, 77, 741, \dodoi{10.1140/epjc/s10052-017-5335-2}

\bibitem[{{Sahu} {et~al.}(2019){Sahu}, {L{\'o}pez Fort{\'\i}n}, \&
  {Nagataki}}]{Sahu_19}
{Sahu}, S., {L{\'o}pez Fort{\'\i}n}, C.~E., \& {Nagataki}, S. 2019, \apjl, 884,
  L17, \dodoi{10.3847/2041-8213/ab43c7}

\bibitem[{{Sahu} {et~al.}(2012){Sahu}, {Zhang}, \& {Fraija}}]{Sahu_12}
{Sahu}, S., {Zhang}, B., \& {Fraija}, N. 2012, \prd, 85, 043012,
  \dodoi{10.1103/PhysRevD.85.043012}

\bibitem[{{Settimo} \& {De Domenico}(2015)}]{Settimo_15}
{Settimo}, M., \& {De Domenico}, M. 2015, Astroparticle Physics, 62, 92,
  \dodoi{10.1016/j.astropartphys.2014.07.011}

\bibitem[{Sommers(2001)}]{Sommers_01}
Sommers, P. 2001, Astroparticle Physics, 14, 271 ,
  \dodoi{https://doi.org/10.1016/S0927-6505(00)00130-4}

\bibitem[{{Stratta} {et~al.}(2011){Stratta}, {Capalbi}, {Giommi}, {Primavera},
  {Cutini}, \& {Gasparrini}}]{Stratta_11}
{Stratta}, G., {Capalbi}, M., {Giommi}, P., {et~al.} 2011, arXiv e-prints,
  arXiv:1103.0749.
\newblock \doarXiv{1103.0749}

\bibitem[{{Supanitsky} \& {de Souza}(2013)}]{Supanitsky_13}
{Supanitsky}, A.~D., \& {de Souza}, V. 2013, Journal of Cosmology and
  Astroparticle Physics, 2013, 023, \dodoi{10.1088/1475-7516/2013/12/023}

\bibitem[{{Takami} {et~al.}(2013){Takami}, {Murase}, \& {Dermer}}]{Takami_13}
{Takami}, H., {Murase}, K., \& {Dermer}, C.~D. 2013, \apjl, 771, L32,
  \dodoi{10.1088/2041-8205/771/2/L32}

\bibitem[{{Tavecchio}(2014)}]{Tavecchio_14}
{Tavecchio}, F. 2014, \mnras, 438, 3255, \dodoi{10.1093/mnras/stt2437}

\bibitem[{Tavecchio {et~al.}(2010)Tavecchio, Ghisellini, Foschini, Bonnoli,
  Ghirlanda, \& Coppi}]{Tavecchio_10}
Tavecchio, F., Ghisellini, G., Foschini, L., {et~al.} 2010, Monthly Notices of
  the Royal Astronomical Society: Letters, 406, L70,
  \dodoi{10.1111/j.1745-3933.2010.00884.x}

\bibitem[{{The Fermi-LAT collaboration}(2019{\natexlab{a}})}]{Fermi_4FGL}
{The Fermi-LAT collaboration}. 2019{\natexlab{a}}, arXiv e-prints,
  arXiv:1902.10045.
\newblock \doarXiv{1902.10045}

\bibitem[{{The Fermi-LAT collaboration}(2019{\natexlab{b}})}]{Fermi_4LAC}
---. 2019{\natexlab{b}}, arXiv e-prints, arXiv:1905.10771.
\newblock \doarXiv{1905.10771}

\bibitem[{{VERITAS Collaboration} {et~al.}(2018){VERITAS Collaboration},
  {Archer}, {et~al.}}]{VERITAS_18}
{VERITAS Collaboration}, {Archer}, A., {et~al.} 2018, \prd, 98, 062004,
  \dodoi{10.1103/PhysRevD.98.062004}

\bibitem[{{Watson} {et~al.}(2009){Watson}, {Schr{\"o}der}, {Fyfe}, {Page},
  {Lamer}, {Mateos}, {Pye}, {Sakano}, {Rosen}, {Ballet}, {Barcons}, {Barret},
  {Boller}, {Brunner}, {Brusa}, {Caccianiga}, {Carrera}, {Ceballos}, {Della
  Ceca}, {Denby}, {Denkinson}, {Dupuy}, {Farrell}, {Fraschetti}, {Freyberg},
  {Guillout}, {Hambaryan}, {Maccacaro}, {Mathiesen}, {McMahon}, {Michel},
  {Motch}, {Osborne}, {Page}, {Pakull}, {Pietsch}, {Saxton}, {Schwope},
  {Severgnini}, {Simpson}, {Sironi}, {Stewart}, {Stewart}, {Stobbart}, {Tedds},
  {Warwick}, {Webb}, {West}, {Worrall}, \& {Yuan}}]{Watson_09}
{Watson}, M.~G., {Schr{\"o}der}, A.~C., {Fyfe}, D., {et~al.} 2009, \aap, 493,
  339, \dodoi{10.1051/0004-6361:200810534}

\bibitem[{{Woo} {et~al.}(2005){Woo}, {Urry}, {van der Marel}, {Lira}, \&
  {Maza}}]{Woo_05}
{Woo}, J.-H., {Urry}, C.~M., {van der Marel}, R.~P., {Lira}, P., \& {Maza}, J.
  2005, \apj, 631, 762, \dodoi{10.1086/432681}

\bibitem[{{Wright} {et~al.}(2009){Wright}, {Chen}, {Odegard}, {Bennett},
  {Hill}, {Hinshaw}, {Jarosik}, {Komatsu}, {Nolta}, {Page}, {Spergel},
  {Weiland}, {Wollack}, {Dunkley}, {Gold}, {Halpern}, {Kogut}, {Larson},
  {Limon}, {Meyer}, \& {Tucker}}]{Wright_09}
{Wright}, E.~L., {Chen}, X., {Odegard}, N., {et~al.} 2009, \apjs, 180, 283,
  \dodoi{10.1088/0067-0049/180/2/283}

\bibitem[{{Wright} {et~al.}(2010){Wright}, {Eisenhardt}, {Mainzer}, {Ressler},
  {Cutri}, {Jarrett}, {Kirkpatrick}, {Padgett}, {McMillan}, {Skrutskie},
  {Stanford}, {Cohen}, {Walker}, {Mather}, {Leisawitz}, {Gautier}, {McLean},
  {Benford}, {Lonsdale}, {Blain}, {Mendez}, {Irace}, {Duval}, {Liu}, {Royer},
  {Heinrichsen}, {Howard}, {Shannon}, {Kendall}, {Walsh}, {Larsen}, {Cardon},
  {Schick}, {Schwalm}, {Abid}, {Fabinsky}, {Naes}, \& {Tsai}}]{Wright_10}
{Wright}, E.~L., {Eisenhardt}, P.~R.~M., {Mainzer}, A.~K., {et~al.} 2010, \aj,
  140, 1868, \dodoi{10.1088/0004-6256/140/6/1868}

\bibitem[{{Xue} {et~al.}(2019{\natexlab{a}}){Xue}, {Liu}, {Petropoulou},
  {Oikonomou}, {Wang}, {Wang}, \& {Wang}}]{Xue_19b}
{Xue}, R., {Liu}, R.-Y., {Petropoulou}, M., {et~al.} 2019{\natexlab{a}}, arXiv
  e-prints, arXiv:1908.10190.
\newblock \doarXiv{1908.10190}

\bibitem[{{Xue} {et~al.}(2019{\natexlab{b}}){Xue}, {Liu}, {Wang}, {Yan}, \&
  {B{\"o}ttcher}}]{Xue_19a}
{Xue}, R., {Liu}, R.-Y., {Wang}, X.-Y., {Yan}, H., \& {B{\"o}ttcher}, M.
  2019{\natexlab{b}}, \apj, 871, 81, \dodoi{10.3847/1538-4357/aaf720}

\bibitem[{{Zacharias} \& {Wagner}(2016)}]{Zacharias_16}
{Zacharias}, M., \& {Wagner}, S.~J. 2016, \aap, 588, A110,
  \dodoi{10.1051/0004-6361/201526698}

\bibitem[{Zatsepin \& Kuzmin(1966)}]{Zatsepin_66}
Zatsepin, G.~T., \& Kuzmin, V.~A. 1966, JETP Lett., 4, 78

\bibitem[{{Zdziarski} \& {B{\"o}ttcher}(2015)}]{Zdziarski_15}
{Zdziarski}, A.~A., \& {B{\"o}ttcher}, M. 2015, \mnras, 450, L21,
  \dodoi{10.1093/mnrasl/slv039}

\end{thebibliography}
\bibliographystyle{aasjournal}



\end{document}